\documentclass[10pt,a4paper]{IEEEtran}

\usepackage{amsmath,epsfig,amssymb,verbatim,amsopn,color}
\usepackage{cite,xspace}
\usepackage{calrsfs} 
\usepackage{subfig}
\usepackage{url}
\usepackage{ifpdf}

\ifCLASSOPTIONonecolumn
    
    \renewcommand{\baselinestretch}{1.9}
    \usepackage[margin=25mm]{geometry}
\else
    \renewcommand{\baselinestretch}{1}
    
\fi

\newcommand{\lsph}{L^2(\mathbb{S}^2)}
\newcommand{\bv}[1]{\boldsymbol{#1}}

\newcommand{\unit}[1]{\bv{\hat{#1}}}

\newcommand{\lsoo}{L^2(\textrm{SO(3)})}

\newcommand{\shc}[3]{\big({#1} \big)_{#2}^{#3}}
\newcommand{\sohc}[4]{\big({#1} \big)^{#2}_{#3,#4}}

\newcommand{\intsph}{\int_{\mathbb{S}^2}}
\newcommand{\intso}{\int_{\textrm{SO($3$)}}}
\newcommand{\thetac}{\theta_\textrm{c}}

\newcommand{\figref}[1]{Fig.\,\ref{#1}}

\newcommand{\appref}[1]{Appendix\,\ref{#1}}

\newtheorem{definition}{Definition}

\newtheorem{remark}{Remark}

\graphicspath{{figs/},{Figures/}, {Authors/}}

\begin{document}
\title{Fast Directional Spatially Localized \\Spherical Harmonic Transform}
\author{%
Zubair~Khalid,~\IEEEmembership{Student Member,~IEEE}, Rodney A. Kennedy,~\IEEEmembership{Fellow,~IEEE}, Salman Durrani,~\IEEEmembership{Senior Member,~IEEE}, Parastoo Sadeghi,~\IEEEmembership{Senior Member,~IEEE}, Yves~Wiaux,~\IEEEmembership{Member,~IEEE},
 and Jason~D.~McEwen,~\IEEEmembership{Member,~IEEE}

\thanks{ Manuscript received August 06, 2012; revised November 22, 2012; accepted
January 27, 2013. Date of publication February 15, 2013; date of
current version April 03, 2013. Z.~Khalid, R.~A.~Kennedy and
P.~Sadeghi are supported by the
  Australian Research Council's Discovery Projects funding scheme
  (Project No.~DP1094350). Y.~Wiaux is supported in part by the Center
  for Biomedical Imaging (CIBM) of the Geneva and Lausanne
  Universities, EPFL, and the Leenaards and Louis-Jeantet foundations,
  and in part by the SNSF by Grant PP00P2-123438.  J.~D.~McEwen is
  supported by a Newton International Fellowship from the Royal
  Society and the British Academy.  }

 \thanks{Z.~Khalid, R.~A.~Kennedy, S.~Durrani and P.~Sadeghi are with
   the Research School of Engineering, College of Engineering and
   Computer Science, The Australian National University, Canberra,
   Australia~(email:~zubair.khalid@anu.edu.au; rodney.kennedy@anu.edu.au; salman.durrani@anu.edu.au; parastoo.sadeghi@anu.edu.au).}

   \thanks{ Y.~Wiaux is with the Institute of Electrical Engineering
   and the Institute of Bioengineering, Ecole Polytechnique
   F{\'e}d{\'e}rale de Lausanne (EPFL), Lausanne, Switzerland.
   Y.~Wiaux is also with the Institute of Bioengineering, EPFL,
   \mbox{CH-1015} Lausanne, Switzerland, and the Department of
   Radiology and Medical Informatics, University of Geneva (UniGE),
   Geneva, Switzerland~(email:~yves.wiaux@epfl.ch).}
   \thanks{  J.~D.~McEwen is with the Department
  of Physics and Astronomy, University College London, London, U.K.~(email:~jason.mcewen@ucl.ac.uk).}

}

\maketitle
\vspace{-15mm}
\begin{abstract}
  We propose a transform for signals defined on the sphere that
  reveals their localized directional content in the spatio-spectral
  domain when used in conjunction with an asymmetric window
  function. We call this transform the directional spatially localized
  spherical harmonic transform (directional SLSHT) which extends the
  SLSHT from the literature whose usefulness is limited to symmetric
  windows. We present an inversion relation to synthesize the original
  signal from its directional-SLSHT distribution for an arbitrary
  window function. As an example of an asymmetric window, the most
  concentrated band-limited eigenfunction in an elliptical region on
  the sphere is proposed for directional spatio-spectral analysis and
  its effectiveness is illustrated on the synthetic and Mars topographic
  data-sets. Finally, since such typical data-sets on the sphere are of
  considerable size and the directional SLSHT is intrinsically
  computationally demanding depending on the band-limits of the signal
  and window, a fast algorithm for the efficient computation of the
  transform is developed.  The floating point precision numerical
  accuracy of the fast algorithm is demonstrated and a full numerical
  complexity analysis is presented.
\end{abstract}
\begin{IEEEkeywords}
\ifCLASSOPTIONonecolumn
\vspace{-5mm}
\fi
Signal analysis, spherical harmonics, 2-sphere.
\end{IEEEkeywords}

\section{Introduction}

\IEEEPARstart{S}{ignals} that are inherently defined on the sphere
appear in various fields of science and engineering, such as medical
image analysis~\cite{Chung:2010}, geodesy~\cite{Simons:2006},
computer graphics~\cite{Han:2007}, planetary
science~\cite{Audet:2011}, electromagnetic inverse
problems~\cite{Colton:1998}, cosmology~\cite{Spergel:2007}, 3D
beamforming~\cite{Ward:1995} and wireless channel
modeling~\cite{Pollock:2003}. In order to analyze and process
signals on the sphere, many signal processing techniques have been
extended from the Euclidean domain to the spherical
domain~\cite{Antoine:1999,Khalid:2012,Khalid_icassp:2012,Khalid2:2012,Marinucci:2008,McEwen:2006,McEwen:2007,McEwen:2008,Sadeghi:2012,Simons:1997,Simons:2006,Starck:2006,Wiaux:2005,Wiaux:2006,Wiaux:2008,Yeo:2008}.

Due to the ability of wavelets to resolve localized signal content in
both space and scale, wavelets have been extensively investigated for
analyzing signals on the
sphere~\cite{Antoine:1999,Marinucci:2008,McEwen:2006,McEwen:2007,Starck:2006,Wiaux:2005,Wiaux:2006,Wiaux:2008,Yeo:2008}
and have been utilized in various applications~(e.g., in
astrophysics~\cite{Barreiro:2000,McEwen:2005:WMAP,McEwen:2007:cosmology,Pietrobon:2008,Schmitt:2010,Vielva:2004} and
geophysics~\cite{Audet:2011,Simons:2011_1,Simons:2011_2}).  Some of
the wavelet techniques on the sphere also incorporate directional
phenomena in the spatial-scale decomposition of a signal~(e.g.,
\cite{Wiaux:2006,Wiaux:2008,Yeo:2008}). As an alternative to
spatial-scale decomposition, spatio-spectral~(spatial-spectral)
techniques have also been developed and applied for localized spectral
analysis, spectral estimation and spatially varying spectral filtering
of
signals~\cite{Khalid:2012,Khalid2:2012,Simons:1997,Wieczorek:2005,Wieczorek:2007}.
The spectral domain is formed through the spherical harmonic transform
which serves as a counterpart of the Fourier transform for signals on
the sphere~\cite{Colton:1998,Driscoll:1994,McEwen:2011,Sakurai:1994}.

The localized spherical harmonic transform, composed of spatial
windowing followed by spherical harmonic transform, was first devised
in \cite{Simons:1997} for localized spectral analysis.  We note that
the localized spherical harmonic transform was defined in
\cite{Simons:1997} for azimuthally asymmetric (i.e., directional)
window functions, however, it was applied and investigated for
azimuthally symmetric functions only.  Furthermore, a spectrally
truncated azimuthally symmetric window function was used for spatial
localization \cite{Simons:1997}. Due to spectral truncation, the
window used for spatial localization may not be concentrated in the
region of interest. This issue was resolved in \cite{Wieczorek:2005},
where azimuthally symmetric eigenfunctions obtained from the Slepian
concentration problem on the sphere were used as window functions (the
Slepian concentration problem is studied for arbitrary regions on the
sphere in \cite{Simons:2006}).  Following \cite{Simons:1997}, the
spatially localized spherical harmonic transform~(SLSHT) for signals
on the sphere has been devised in \cite{Khalid:2012} to obtain the
spatio-spectral representation of signals for azimuthally symmetric
window functions, where the effect of different window functions on
the SLSHT distribution is studied.  Subsequently, the SLSHT has been
used to perform spatially varying spectral filtering
\cite{Khalid2:2012}, again with azimuthally symmetric window
functions.

In obtaining the SLSHT distribution for spatio-spectral
representation of a signal, the use of an azimuthally symmetric
window function provides mathematical simplifications. However, such
an approach cannot discriminate localized directional features in
the spatio-spectral domain. This motivates the use of asymmetric
window functions in the spatio-spectral transformation of a signal
using the SLSHT. In order to serve this objective, we employ the
definition of the localized spherical harmonic transform in
\cite{Simons:1997} and define the SLSHT and the SLSHT distributions
using azimuthally asymmetric window functions for spatial
localization. Since the use of an asymmetric window function enables
the transform to reveal directional features in the spatio-spectral
domain, we call the proposed transform the directional SLSHT.  We
also provide a harmonic analysis of the proposed transform and
present an inversion relation to recover the signal from its
directional SLSHT distribution.

Since the directional SLSHT distribution of a signal is required to be
computed for each spatial position and for each spectral component, and
data-sets on the sphere are of considerable size~(e.g., three million
samples on the sphere for current data-sets~\cite{Jarosik:2010} and
fifty million samples for forthcoming
data-sets~\cite{planck:bluebook}), the evaluation of the directional
SLSHT distribution is computationally challenging. We develop fast
algorithms for this purpose.  Through experimental results we show the
numerical accuracy and efficient computation of the proposed
directional SLSHT transform. Furthermore, due to the fact that the
proposed directional SLSHT distribution depends on the window function
used for spatial localization, we analyze the asymmetric band-limited
window function with nominal concentration in an elliptical region
around the north pole, which is obtained from the Slepian
concentration problem on the sphere. We also illustrate, through an
example, the capability of the proposed directional SLSHT to reveal
directional features in the spatio-spectral domain.

The remainder of the paper is structured as follows. In
Section~\ref{sec:models}, we review mathematical preliminaries
related to the signals on the sphere, which are required in the
sequel. We present the formulation of the directional SLSHT, its
harmonic analysis and signal reconstruction from the SLSHT
distribution in Section~\ref{sec:SLSHT}. Different algorithms for
the evaluation of the SLSHT distribution are provided in
Section~\ref{sec:Algos}. In Section~\ref{sec:results}, we show
timing and accuracy results of our algorithms and an illustration of
the transform. Concluding remarks are presented in
Section~\ref{sec:conclusions}.


\section{Mathematical Background}\label{sec:models}
In order to clarify the adopted notation, we review some
mathematical background for signals defined on the sphere and the
rotation group.

\subsection{Signals on the Sphere}
In this work, we consider the square integrable complex functions
$f(\unit{x})$ defined on unit sphere $\mathbb{S}^2 \triangleq
\{\mathbf{u} \in \mathbb{R}^3 : |\mathbf{u}| = 1  \}$, where
$|\cdot|$ denotes Euclidean norm, $\unit{x}\equiv
\unit{x}(\theta,\phi) \triangleq (\sin\theta\cos\phi,\;
\sin\theta\sin\phi,\; \cos \theta)^{T}\;\in \mathbb{R}^3$ is a unit
vector and parameterizes a point on the unit sphere with $\theta \in
[0, \pi]$ denoting the co-latitude and $\phi\in [0, 2\pi)$ denoting
the longitude. The inner product of two functions $f$ and $h$ on
$\mathbb{S}^2$ is defined as~\cite{Kennedy:2011}
\begin{align}\label{eqn:innprd}
\langle f, h \rangle \triangleq  \int_{\mathbb{S}^2}
f(\hat{\boldsymbol{x}}) \overline {h(\hat{\boldsymbol{x}})}
\,ds(\hat{\boldsymbol{x}}) ,
\end{align}
where $\overline{(\cdot)}$ denotes the complex conjugate,
$ds(\hat{\boldsymbol{x}}) = \sin \theta d \theta d \phi$ and the
integration is carried out over the unit sphere. With the inner
product in \eqref{eqn:innprd}, the space of square integrable complex
valued functions on the sphere forms a complete Hilbert space
$\lsph$. Also, the inner product in \eqref{eqn:innprd} induces a norm
$\|f\| \triangleq\langle f,f \rangle^{1/2}$. We refer the functions
with finite induced norm as signals on the sphere.

The Hilbert space $\lsph$ is
separable and the spherical harmonics form the archetype complete
orthonormal set of basis functions. The spherical harmonics,
$Y_{\ell}^m(\unit{x}) = Y_{\ell}^m(\theta, \phi)$, for degree
${\ell} \geq 0$ and order $ |m| \leq {\ell}$ are defined
as~\cite{Colton:1998,Sakurai:1994}
\begin{align}\label{Eq:Sph_harmonics}
Y_{\ell}^m(\theta, \phi) &= N_\ell^m
\,P_{\ell}^{m}(\cos\theta)e^{im\phi} ,
\end{align}
where $N_\ell^m =
\sqrt{\frac{2{\ell}+1}{4\pi}\frac{({\ell}-m)!}{({\ell}+m)!}}$
denotes the normalization constant and $P_{\ell}^{m}$ are the
associated Legendre polynomials~\cite{Sakurai:1994}. With the above
definitions, the spherical harmonics form an orthonormal set of
basis functions, i.e., they satisfy $ \langle Y_{\ell}^m,
Y_{\ell'}^{m'} \rangle = \delta_{\ell \ell'} \delta_{mm'}$, where
$\delta_{\ell \ell'}$ is the Kronecker delta.

By completeness and orthonormality of the spherical harmonics, we can
expand any signal $f\in \lsph$ as
\begin{align}\label{Eq:f_expansion}
f(\hat{\boldsymbol{x}})&=\sum_{{\ell}=0}^{\infty}\sum_{m=-{\ell}}^{\ell} \shc{f}{\ell}{m}  Y_{\ell}^m(\hat{\boldsymbol{x}}),%
\end{align}
where
\begin{align}\label{Eq:fcoeff}
 \shc{f}{\ell}{m}\triangleq \langle f,Y_{{\ell}}^m \rangle &= \intsph f(\hat{\boldsymbol{x}})\overline {Y_{\ell}^m(\hat{\boldsymbol{x}})}\,ds(\hat{\boldsymbol{x}})%
\end{align}
denotes the spherical harmonic coefficient of degree $\ell$ and
order $m$. The signal $f$ is said to be band-limited with maximum
spherical harmonic degree $L_f$ if \mbox{$\shc{f}{\ell}{m}=0,\,\forall
\ell>L_f$}.

\subsection{Rotations on the Sphere and Wigner-$D$ Functions}
Rotations on the sphere are often parameterized using Euler angles
$(\alpha,\beta,\gamma)\in \textrm{SO(3)}$, where $\alpha\in
[0,\,2\pi)$, $\beta\in [0,\,\pi]$ and $\gamma \in
[0,\,2\pi)$~\cite{Sakurai:1994}. Using the `$zyz$' Euler
convention, we define the rotation operator
$\mathcal{D}_\rho$, for $\rho = (\alpha,\beta,\gamma) \in \textrm{SO(3)}$,
which rotates a function on a sphere in the sequence of $\gamma$
rotation around $z$-axis, then $\beta$ rotation about $y$-axis
followed by a $\alpha$ rotation around $z$-axis. The spherical
harmonic coefficient of a rotated signal $\mathcal{D}_\rho f$ is
related to the coefficients of the original signal by
\begin{align}
\label{eqn:rotation_non_symm} \big(\mathcal{D}_\rho f \big)_\ell^m
\,=\, \sum_{m'=-{\ell}}^{{\ell}}
D_{m,m'}^{\ell}(\rho)\shc{f}{\ell}{m'}, \: \rho =
(\alpha,\beta,\gamma),
\end{align}
where $D_{m,m'}^{\ell}(\rho)$ denotes the Wigner-$D$
function~\cite{Sakurai:1994} of degree $\ell$ and orders $m$ and
$m'$ and is given by
\ifCLASSOPTIONonecolumn
\begin{align}\label{Eq:Dlm}
D_{m,m'}^{\ell}(\rho)\,=\, D^{\ell}_{m,m'}(\alpha,\beta,\gamma) \,
&=\, \, e^{-im\alpha}d_{m,m'}^{\ell}(\beta)\, e^{-im'\gamma},\:
\rho = (\alpha,\beta,\gamma),
\end{align}
\else
\begin{align}\label{Eq:Dlm}
D_{m,m'}^{\ell}(\rho)\,&=\, D^{\ell}_{m,m'}(\alpha,\beta,\gamma)
\\ \nonumber   &=\, \, e^{-im\alpha}d_{m,m'}^{\ell}(\beta)\,
e^{-im'\gamma},\: \rho = (\alpha,\beta,\gamma),
\end{align}
\fi
where $d_{m,m'}^{\ell}(\beta)$ is the Wigner-$d$
function~\cite{Sakurai:1994}.

\subsection{Signals on the Rotation Group $\emph{\textrm{SO(3)}}$ } For $\ell \geq0$ and $m,m' \in \mathbb{Z}$
such that $|m|,|m'|\leq\ell$, the Wigner-$D$ functions in
\eqref{Eq:Dlm} form a complete set of orthogonal functions for the
space $\lsoo$ of functions defined on the rotation group SO($3$) and follow the
orthogonality relation
\begin{align}\label{Eq:WignerD_ortho}
\intso D^{\ell}_{m,m'}(\rho) \overline{D^{p}_{q,q'}(\rho)}\, d\rho =
 \frac{8 \pi^2}{2\ell +1}\,\delta_{\ell p} \delta_{mq}
\delta_{m'q'},
\end{align}
where $d\rho = d\alpha \sin \beta d\beta  d\gamma$ and the integral
is a triple integral over all rotations $(\alpha,\beta,\gamma)\in
\textrm{SO(3)}$~\cite{Sakurai:1994}. Thus, any function $f\in\lsoo$ may be expressed as
\begin{align}\label{Eq:g_expansion_Wigner}
f(\rho)\,=\, \sum_{\ell=0}^\infty
\sum_{m=-\ell}^{\ell}\sum_{m=-\ell'}^{\ell'}\sohc{f}{\ell}{m}{m'}
D^{\ell}_{m,m'}(\rho) ,
\end{align}
where
\begin{align}
\sohc{f}{\ell}{m}{m'}\,=\, \frac{2\ell +1}{8 \pi^2} \intso
f(\rho)\overline{D^{\ell}_{m,m'}(\rho)}\,d\rho.
\end{align}
The signal $f$ is said to be band-limited with maximum degree $L_f$
if $\sohc{f}{\ell}{m}{m'}=0,\,\forall \ell>L_f$.

\subsection{Discretization of $\mathbb{S}^2$ and $\emph{\textrm{SO(3)}}$}
\label{sec:models_samples}
In order to represent functions on $\mathbb{S}^2$ and
$\textrm{SO(3)}$, it is necessary to adopt appropriate tessellation
schemes to discretize both the unit sphere domain and the Euler angle
domain of SO(3). We consider tessellation schemes that support a
sampling theorem for band-limited functions, which is equivalent to
supporting an exact quadrature.

For the unit sphere domain, we adopt the equiangular tessellation
scheme~\cite{McEwen:2011} defined as $\mathfrak{S}_{L}
=\{\theta_{n_\theta} = \pi (2n_\theta+1)/(2L+1),\,\phi_{n_\phi} = 2\pi
n_\phi/(2L+1):\,0 \leq n_\theta \leq L,\, 0 \leq \,n_\phi \leq 2L\}$,
which is a grid of $(L+1)\times (2L+1)$ sample points on the sphere
(including repeated samples of the south pole) that keeps the sampling
in $\theta$ and $\phi$ independent. For a band-limited function on the
sphere $f \in \lsph$ with maximum spherical harmonic degree $L_f$, the
sampling on the grid $\mathfrak{S}_{L_f}$ ensures that all information
of the function is captured in the finite set of samples and,
moreover, that exact quadrature can be performed \cite{McEwen:2011}.
Note that this sampling theorem was developed only recently
\cite{McEwen:2011} and requires approximately half as many samples on
the sphere as required by alternative equiangular sampling theorems on
the sphere \cite{Driscoll:1994}.

For the Euler angle representation of the rotation group SO(3), we
consider the equiangular tessellation scheme $\mathfrak{E}_{L}
=\{ \alpha_{n_\alpha} = 2\pi n_\alpha/(2L+1),\, \beta_{n_\beta} = 2\pi
n_\beta/(2L+1),\,\gamma_{n_\gamma} = 2\pi n_\gamma/(2L+1):\, 0 \leq
n_\alpha,n_\gamma \leq {2L},\, 0 \leq n_\beta\leq L \}$. Again for a
function $f\in \lsoo$ with maximum spectral degree $L_f$, the sampling
of a function $f$ on $\mathfrak{E}_{L_f}$ ensures that all information
of the function is captured and also permits exact quadrature (which
follows from the results developed on the sphere \cite{McEwen:2011}).
%

\section{Directional SLSHT}\label{sec:SLSHT}

We describe in this section the directional SLSHT, which is capable of
revealing directional features of signals in the spatio-spectral\footnote{When we refer to spatio-spectral, we consider the SO(3) spatial domain, instead of $\mathbb{S}^2$. This is due to the reason that we are considering all possible rotations, parameterized using Euler angles which form the SO(3) domain.}
domain.
For spatial localization, we consider the band-limited azimuthally
asymmetric window function which is spatially concentrated in some
asymmetric region around the north pole. Since the rotation around the
\mbox{$z$-axis} does not have any affect on an azimuthally symmetric
function, the localized spherical harmonic transform using an
azimuthally symmetric window function can be parameterized on the
sphere.  However, if an azimuthally asymmetric window is used to
obtain localization in the spatial domain, the rotation of the window
function is fully parameterized with the consideration of all three
Euler angles $(\alpha,\beta,\gamma) \in\,$SO($3$). We refer to the
spatially localized transform using an asymmetric window as the
directional SLSHT. Here, we first define the directional SLSHT
distribution which presents the signal in the spatio-spectral
domain. Later in this section, we present the harmonic analysis of
SLSHT distribution and provide an inversion relation to obtain the
signal from its given directional SLSHT distribution.

\subsection{Forward Directional SLSHT }
\begin{definition}[{Directional SLSHT}] For a
  signal \mbox{$f\in\lsph$}, define the directional SLSHT distribution
  component ${g}(\rho;\ell,m) \in \lsoo$ of degree $\ell$ and order
  $m$ as the spherical harmonic transform of a localized signal where
  localization is provided by the rotation operator $\mathcal{D}_\rho$ acting on window function
  $h\in\lsph$, i.e.,
\ifCLASSOPTIONonecolumn
\begin{align}\label{Eq:stft_spatial}
g(\rho;\ell,m) 
&\triangleq \intsph f(\unit{x})\,\big(\mathcal{D}_\rho
h\big)(\unit{x})\,\overline{Y_\ell^m(\unit{x})}\,ds(\unit{x})
\end{align}
\else
\begin{align}\label{Eq:stft_spatial}
g(\rho;\ell,m) 
&\triangleq \intsph f(\unit{x})\,\big(\mathcal{D}_\rho
h\big)(\unit{x})\,\overline{Y_\ell^m(\unit{x})}\,ds(\unit{x})
\end{align}
\fi
for $0\leq \ell \leq L_g,\,|m|\leq \ell$, where $L_g=L_f+L_h$
denotes the maximum spherical harmonic degree for which the
distribution components $g(\rho;\ell,m)$ are non-zero,
and $L_f$ and $L_h$ denote the band-limits of the signal
$f$ and the window function $h$, respectively. Also, each
distribution component $g(\rho;\ell,m)$ is band-limited in
$\rho=(\alpha,\beta,\gamma)\in {\rm SO(3)}$ with maximum degree
$L_h$, i.e., when expressed in terms of Wigner-$D$ functions. We
elaborate on this shortly. Furthermore, we consider unit energy
normalized window functions such that $\langle h,h \rangle=1$.
\end{definition}

\begin{remark}
The directional SLSHT distribution component in \eqref{Eq:stft_spatial} can be
interpreted as the spherical harmonic transform of the localized
signal where the window function $h$ provides asymmetric
localization at spatial position $\unit{x} =
\unit{x}(\beta,\alpha)\in\mathbb{S}^2$ and the first rotation, through $\gamma$, determines the orientation of the window function at $\unit{x}$. If
the window function is azimuthally symmetric, this orientation of
the window function by $\gamma$ becomes invariant and the SLSHT
distribution components are defined on $\lsph$~\cite{Khalid:2012}.
\end{remark}

Since the maximum spectral degree for which the SLSHT distribution is
defined is $L_g = L_f + L_h$, we consider the band-limited window function such
that $L_h\leq L_f$ to avoid extending $L_g$ significantly above
$L_f$. We discuss the localization of the window function in spatial
and spectral domains later in the paper.

\subsection{Harmonic Analysis}

We now present the formulation of the directional SLSHT distribution
if the signal $f$ and the window function $h$ are represented in the
spectral domain. Using the expression of the spherical harmonics of a
rotated function in \eqref{eqn:rotation_non_symm}, we can write the
SLSHT distribution component $g(\rho;\ell,m)$ in
\eqref{Eq:stft_spatial} as
\ifCLASSOPTIONonecolumn
\begin{align}\label{Eq:harmonic_formulation}
g(\rho;\ell,m)\, &=\,\sum_{\ell'=0}^{L_f}\sum_{m'=-\ell'}^{\ell'}
\shc{f}{\ell'}{m'} \sum_{p=0}^{ L_h}\sum_{q=-p}^p\sum_{q'=-p}^p
\shc{h}{p}{q'}\, D^p_{q,q'}(\rho)\,T(\ell',m';p,q;\ell,m),
\end{align}
\else
\begin{align}\label{Eq:harmonic_formulation}
&g(\rho;\ell,m)\, =\,\sum_{\ell'=0}^{L_f}\sum_{m'=-\ell'}^{\ell'}
\shc{f}{\ell'}{m'} \\ \nonumber  &\,\quad \times \sum_{p=0}^{
L_h}\sum_{q=-p}^p\sum_{q'=-p}^p \shc{h}{p}{q'}\,
D^p_{q,q'}(\rho)\,T(\ell',m';p,q;\ell,m),
\end{align}
\fi
where
\begin{align}
T(\ell',m';p,q;\ell,m) = \intsph
Y_{\ell'}^{m'}(\unit{x})\,Y_p^q(\unit{x})\,\overline{Y_\ell^m(\unit{x})}\,
ds(\unit{x}) \nonumber
\end{align}
denotes the spherical harmonic triple product, which can be evaluated
using Wigner-$3j$ symbols or Clebsch-Gordan
coefficients~\cite{Sakurai:1994,Varsha:1988}.
\begin{remark}
  By comparing $g(\rho;\ell,m)$ in \eqref{Eq:harmonic_formulation}
  with \eqref{Eq:g_expansion_Wigner}, we note that the band-limit
  of $g(\rho;\ell,m)$ in $\rho$ is given by $L_h$. Since $\ell'\leq L_f$ and $p
  \leq L_h$ in \eqref{Eq:harmonic_formulation}, our statement that the
  distribution component $g(\rho;\ell,m)$ is non-zero for $\ell \leq
  L_g=L_f+L_h$ follows since the triple product $T(\ell',m';p,q;\ell,m)$ is non-zero for $\ell\leq L_f + L_h$ only.
\end{remark}

\subsection{Inverse Directional SLSHT}

Here, we define the inverse directional SLSHT to reconstruct a signal from its SLSHT distribution.
The original signal can be reconstructed from its directional SLSHT
distribution through the spectral domain marginal, that is, by
integrating the SLSHT distribution components over the spatial domain
SO(3)~\cite{Simons:1997}. Using our harmonic formulation in
\eqref{Eq:harmonic_formulation}, define $\shc{\hat{f}}{\ell}{m}$ as
the integral of the SLSHT distribution component $g(\rho;\ell,m)$ over
SO(3) giving
\ifCLASSOPTIONonecolumn
\begin{align}\label{Eq:inversion_1}
\shc{\hat{f}}{\ell}{m} &=   \intso  g(\rho;\ell,m)  d\rho ,\quad 0
\leq \ell \leq L_f \nonumber
\\ &=
\sum_{\ell'=0}^{L_h}\sum_{m'=-\ell'}^{\ell'} \shc{f}{\ell'}{m'}
\sum_{p=0}^{ L_h}\sum_{q=-p}^p\sum_{q'=-p}^p \shc{h}{p}{q'}\,
T(\ell',m';p,q;\ell,m) \intso D^p_{q,q'}(\rho)\,d\rho
\nonumber \\
&= 8\pi^2 \sum_{\ell'=0}^{L_h}\sum_{m'=-\ell'}^{\ell'}
\shc{f}{\ell'}{m'}
 \shc{h}{0}{0}\,
T(\ell',m';0,0;\ell,m)
\nonumber \\
&=\sqrt{16\pi^3} \,\shc{h}{0}{0}\,\shc{f}{\ell}{m} ,
\end{align}
\else
\begin{align}\label{Eq:inversion_1}
\shc{\hat{f}}{\ell}{m} &=   \intso  g(\rho;\ell,m)  d\rho ,\quad 0
\leq \ell \leq L_f \nonumber
\\ &=
\sum_{\ell'=0}^{L_h}\sum_{m'=-\ell'}^{\ell'} \shc{f}{\ell'}{m'}
\sum_{p=0}^{ L_h}\sum_{q=-p}^p\sum_{q'=-p}^p \shc{h}{p}{q'}\,\nonumber \\
& \quad \times T(\ell',m';p,q;\ell,m) \intso D^p_{q,q'}(\rho)\,d\rho
\nonumber \\
%
%
&=\sqrt{16\pi^3} \,\shc{h}{0}{0}\,\shc{f}{\ell}{m} ,
\end{align}
\fi
where we have used the orthogonality relation of Wigner-$D$
functions (see \eqref{Eq:WignerD_ortho}).
Using the expression in \eqref{Eq:inversion_1}, we can find the
spherical harmonic coefficient $\shc{f}{\ell}{m}$ of the signal $f$
as
\begin{align}\label{Eq:inversion_2}
\shc{f}{\ell}{m} = \frac{\shc{\hat{f}}{\ell}{m}}{\sqrt{16\pi^3}
\,\shc{h}{0}{0}} ,
\end{align}
which indicates that we only need to know the DC component of the
window function $\shc{h}{0}{0}$ in order to obtain the signal from
its directional SLSHT distribution. It further imposes the condition
that the DC component of the window function must be non-zero.
Although the distribution components in
\eqref{Eq:harmonic_formulation} are defined up to degree
$L_g=L_f+L_h$, we only require the components up to $L_f$ for signal
reconstruction.

\begin{remark}
The signal can also be reconstructed from its SLSHT distribution by
evaluating
\begin{align*}
\intso g(\rho;\ell,m)\overline{D^p_{q,q'}(\rho)} d\rho &=
8\pi^2\sum_{\ell'=0}^{L_h}\sum_{m'=-\ell'}^{\ell'}
\shc{f}{\ell'}{m'} \sum_{p=0}^{ L_h}\sum_{q=-p}^p
\nonumber \\
& \times  \sum_{q'=-p}^p \frac{\shc{h}{p}{q'}}{2p+1}
T(\ell',m';p,q;\ell,m)
\end{align*}
for all $p\leq L_h,\,|q|,\,q'|\leq p$ and for all $\ell\leq
L_g,\,|m|\leq \ell$ and then employing the orthogonality relations
of Wigner-$3j$ symbols to decouple the spherical harmonic
coefficients of the window function $h$ and the signal $f$. The
similar approach has been employed in \cite{Khalid2:2012} to invert
the signal from its modified SLSHT distribution, where the SLSHT
distribution is obtained using azimuthally symmetric window
function. This approach does not impose restriction on the DC
component of the window function to be non-zero, instead, it
requires the knowledge of the energy of the window function. In this
work, we consider the inversion of a signal presented in
\eqref{Eq:inversion_1} and \eqref{Eq:inversion_2}, as this is the
most efficient formulation.
\end{remark}

Computing the forward and inverse directional SLSHT is
computationally demanding.  Since the directional SLSHT distribution
components $g(\rho;\ell,m)$ in \eqref{Eq:stft_spatial} are defined
for $\ell\leq L_g$, the number of distribution components are of the
order $L_g^2$, while the sampling of $\rho$ is of the order $L_h^3$;
thus, the direct evaluation of the directional SLSHT distribution is
prohibitively computationally expensive.  Therefore efficient
algorithms need to be developed which reduce the computational
complexity. We address this problem in the next section.

\subsection{Window localization in Spatial and Spectral Domains}\label{sec:Window}

The directional SLSHT distribution is the spherical harmonic
transform of the product of two functions, the signal $f$ and the
rotated window function $h$ and we must be careful in interpreting
the directional SLSHT distribution in the sense that we do not
mistake using the signal to study the window because there is no
distinction mathematically. The window function should be chosen so
that it provides spatial localization in some spatial region around
the north pole~(origin). Since we have considered a band-limited
window function, the window function cannot be perfectly localized
in the spatial domain due to the uncertainty principle on the
sphere~\cite{Freeden:1999}. However, it can be optimally localized
by maximizing the energy concentration of the window function in the
desired directional region~\cite{Simons:2006}.

The interpretation and the effectiveness of the directional SLSHT
distribution depends on the chosen window function. The window
function with maximum localization in some defined asymmetric region
provides directional localization and thus reveals directional
features in the spatio-spectral domain. The more directional the
window function, the more directional features it can reveal in the
spatio-spectral domain but this tends to increase the maximum
spherical harmonic degree $L_h$. Recall that the maximum degree of
the directional SLSHT distribution components is given by $L_g =
L_f+L_h$. Thus, when the signal is expressed in the spatio-spectral
domain its spectral domain is extended by $L_h$, which results in
spectral leakage. Therefore, we want the window function to be
simultaneously maximally localized in some spatial region
$\mathcal{R}\subset\mathbb{S}^2$ and have the minimum possible
band-limit which achieves the desired level of energy concentration
in the spatial region $\mathcal{R}$.

With the consideration that there exists localization trade-off for
a window function in spatial and spectral
domain~\cite{Freeden:1999}, the choice of window function affects
the resulting SLSHT distribution. We highlight the future research
problem that there is a need to investigate the use of different
window function at different spatial positions, such that the
localization of the window function adapts to the characteristics of
the signal being analyzed. An analogous problem is well known in
time-frequency analysis~\cite{Jones:1992}, where it has been shown
that, according to several different measures of performance, the
optimal window function for short-time Fourier transform~(STFT)
depends on the signal being analyzed.

Here, we propose using a band-limited eigenfunction obtained from
the solution of the Slepian concentration problem~\cite{Simons:2006}
as a window function, concentrated in a spatially localized
elliptical region around the north pole. The elliptical region can
be parameterized using the focus colatitude $\thetac$ of the ellipse
along the positive $x$-axis and the arc length $a$ of the semi-major
axis:
\ifCLASSOPTIONonecolumn
\begin{align}\label{Eq:region_def}
\mathcal{R}_{(\thetac,a)} \triangleq \big\{ (\theta,\phi):\,
\triangle_s\big((\theta,\phi ), (\thetac,0)\big) +
\triangle_s\big((\theta,\phi ), (\thetac,\pi)\big) \leq 2a \big\}
\end{align}
\else
\begin{align}\label{Eq:region_def}
\mathcal{R}_{(\thetac,a)} \triangleq &\big\{ (\theta,\phi):
\triangle_s\big((\theta,\phi ), (\thetac,0)\big) \nonumber \\ \, & +
\triangle_s\big((\theta,\phi ), (\thetac,\pi)\big) \leq 2a \big\},
\end{align}
\fi
where $0 \leq \thetac \leq a \leq \pi/2$. Here
$\triangle_s\big((\theta,\phi ), (\theta',\phi')\big) = \arccos
\big( \sin\theta\sin\theta'\cos(\phi-\phi') +
  \cos\theta\cos\theta'\big)$ denotes the angular
distance between two points $(\theta,\phi)$ and $(\theta',\phi')$ on
the sphere.  Since the major axis is along $x$-axis, the elliptical
region is orientated along the $x$-axis.
\begin{remark}
  For a given focus $\thetac$, the region becomes more directional as the arc length $a$ approaches $\theta_c$ from $\pi/2$. For $a=\pi/2$, the region becomes
  azimuthally symmetric, i.e., we recover the polar cap of central
  angle $\pi/2$. Also, when $\theta_c=0$, the region becomes azimuthally symmetric~(polar cap) of central angle $a$.
\end{remark}

As a result of the Slepian concentration
problem~\cite{Khalid_icspcs:2011,Simons:2006} to find the
band-limited function with bandwidth $L_h$ and maximal spatial
concentration in an elliptical region $\mathcal{R}_{(\thetac,a)}$,
we obtain $(L_h+1)^2$ eigenfunctions. Due to the symmetry of the
elliptical region about $x$-$y$ plane, the eigenfunctions are real
valued~\cite{Khalid_icspcs:2011}. Here we consider the use of the
band-limited eigenfunction with maximum energy concentration in the
elliptical region for given band-limit $L_h$ and refer to such an
eigenfunction as the eigenfunction window.

\section{Efficient Computation of Directional SLSHT Distribution}\label{sec:Algos}

Here, we present efficient algorithms for the computation of the
directional SLSHT distribution of a signal and the signal
reconstruction from its directional SLSHT distribution. First, we
discuss the computational complexities if the SLSHT distribution
components are computed using direct quadrature as given in
\eqref{Eq:stft_spatial} or using the harmonic formulation in
\eqref{Eq:harmonic_formulation}. Later, we develop an alternative
harmonic formulation which reduces the computational burden.
Finally, we present an efficient algorithm that incorporates a factoring of
rotations~\cite{Risbo:1996} and exploits the FFT.

First we need to parameterize the required tessellation schemes for
$\mathbb{S}^2$ for the representation of the signal $f$ and the window
$h$ and for SO(3) which forms the spatial domain of the directional
SLSHT distribution. Since the maximum spectral degree of the signal
$f$ is $L_f$, we therefore consider the equiangular tessellation
$\mathfrak{S}_{L_f}$ to represent $f$. Since the
maximum degree for all SLSHT distribution components $g(\rho;\ell,m)$
in $\rho$ is $L_h$, we therefore consider the tessellation
$\mathfrak{E}_{L_h}$ to represent the SLSHT distribution components on
$\lsoo$.

\subsection{Direct Quadrature and Harmonic Formulation}
We define the forward spatio-spectral transform as evaluation of each
SLSHT distribution component $g(\rho;\ell,m)$. Evaluation of the
forward spatio-spectral transform using exact quadrature in
\eqref{Eq:stft_spatial} requires the computation of two dimensional
summation over the tessellation of $\mathbb{S}^2$ for each $3$-tuple $(\alpha,\beta,\gamma)$. Since there are $O(L_h^3)$ such $3$-tuples
in the tessellation scheme $\mathfrak{E}_{L_h}$ and the SLSHT
distribution components are of the order $O(L_f^2)$, the computational
complexity to compute all distribution components using direct
quadrature is $O(L_f^4 L_h^3)$. Using the harmonic formulation in
\eqref{Eq:harmonic_formulation}, the complexity to compute each SLSHT
distribution component is $O(L_f^2 L_h^6)$ and to compute all SLSHT
distribution components is $O(L_f^4 L_h^6)$.  Although the harmonic
formulation in \eqref{Eq:harmonic_formulation} is useful to establish that the
signal can be reconstructed from the directional SLSHT distribution, it is
much more computationally demanding than direct quadrature. We develop
efficient algorithms in the next subsection which improve the
computational complexity of the harmonic formulation and make it more
efficient than direct quadrature.

For the inverse directional SLSHT distribution, we only need
to integrate over SO(3) to obtain the signal in the
spherical harmonic domain as proposed in \eqref{Eq:inversion_1}.
Since the integral can be evaluated by a summation over all Euler
angles using quadrature weights, an efficient way to recover
the signal from its SLSHT distribution is through direct quadrature,
with complexity of $O(L_h^3)$ for each distribution component and
$O(L_f^2 L_h^3)$ for all components.

In order to evaluate the integral in \eqref{Eq:inversion_1} exactly,
we need to define quadrature weights along Euler angle $\beta$ in the
tessellation $\mathfrak{E}_{L_h}$. We evaluate the integral in
\eqref{Eq:inversion_1} by the following summation\footnote{In the
  evaluation of \eqref{Eq:sig_reconstruction_discrete} we have
  computed the summation over $2L_h+1$ sample points in both $\alpha$
  and $\gamma$. This is due to the tessellation $\mathfrak{E}_{L_h}$
  required to capture all information content of
  $g(\alpha,\beta,\gamma;\ell,m)$. However,
  if one were considered in recovering $f$ only, then
  given the quadrature rule in \cite{McEwen:2011}
  $\shc{\hat{f}}{\ell}{m}$ in \eqref{Eq:sig_reconstruction_discrete}
  could be computed exactly with only $L_h+1$ sample points in
  $\alpha$ and $\gamma$. }
\ifCLASSOPTIONonecolumn
\begin{align}\label{Eq:sig_reconstruction_discrete}
\shc{\hat{f}}{\ell}{m} \,=\,
\frac{1}{(2L_h+1)^3}\sum_{n_\alpha=0}^{2L_h}
 \sum_{n_\beta=0}^{L_h} \sum_{n_\gamma=0}^{2L_h}\,
 g(\alpha_{n_\alpha},\beta_{n_\beta},\gamma_{n_\gamma};\ell,m)\,q(\beta_{n_\beta}) ,
\end{align}
\else
\begin{align}\label{Eq:sig_reconstruction_discrete}
\shc{\hat{f}}{\ell}{m} \,=&\,
\frac{1}{(2L_h+1)^3}\sum_{n_\alpha=0}^{2L_h}
 \sum_{n_\beta=0}^{L_h} \sum_{n_\gamma=0}^{2L_h}\, \nonumber \\&
 \times g(\alpha_{n_\alpha},\beta_{n_\beta},\gamma_{n_\gamma};\ell,m)\,q(\beta_{n_\beta}) ,
\end{align}
\fi
where the quadrature weights $q(\beta_{n_\beta})$ follow from \cite{McEwen:2011}, with
\begin{align}
q(\beta_{n_\beta}) = \begin{cases}
                        4\pi^2 \big(\lfloor \frac{L_h}{2} \rfloor + \frac{1}{2}\big)^{-1},&  \beta_{n_\beta}=0 \\
                        8\pi^2\sum\limits_{m=-L_h}^{L_h}w(-m)\,\cos{m\beta_{n_\beta}},&  \textrm{otherwise} \\
                     \end{cases}
\end{align}
where $w(m)$ is defined as~\cite{McEwen:2011}
\begin{align}
w(m) = \begin{cases}
        \frac{\pm i\pi}{2},\quad&m=\pm 1, \\
        0, & m\, \textrm{odd},\,m\neq1,\\
        \frac{2}{1-m^2}, & m\,\textrm{even}.
        \end{cases}
\end{align}

\subsection{Fast Algorithm for Forward Directional SLSHT}
\label{sec:Algos:fast}
Here, we develop a fast algorithm to reduce the computational
complexity of the forward SLSHT.  We first consider an alternative
harmonic formulation of the forward SLSHT and then employ the
factoring of rotations approach which was first proposed in
\cite{Risbo:1996} and has been used in the implementations of the fast
spherical convolution~\cite{Wandelt:2001} and the directional
spherical wavelet transform~\cite{McEwen:2007}.

We may write the
directional SLSHT distribution component $g(\rho;\ell,m)$ in
\eqref{Eq:stft_spatial} as a spherical convolution~\cite{McEwen:2007} of $h$ and
the spherical harmonic modulated signal $\overline{f}\,Y_\ell^m$, giving
\begin{align}\label{Eq:alternative_harmonic_formulation}
g(\rho;\ell,m)\, &=\, \sum_{p=0}^{ L_h}\sum_{q=-p}^p\sum_{q'=-p}^p
\overline{\shc{\overline{f}Y_\ell^m}{p}{q}} \shc{h}{p}{q'}\,
D^p_{q,q'}(\alpha,\beta,\gamma),
\end{align}
which can be expressed, using the definition of the Wigner-$D$ function in
\eqref{Eq:Dlm}, as
\ifCLASSOPTIONonecolumn
\begin{align}\label{Eq:alternative_harmonic_formulation_wigner}
g(\rho;\ell,m)\, &=\, \sum_{p=0}^{ L_h}\bigg(\sum_{q=-p}^p
\big(\sum_{q'=-p}^p \overline{\shc{\overline{f}Y_\ell^m}{p}{q}}
\shc{h}{p}{q'} \, d^p_{q,q'}(\beta) e^{-iq'\gamma}\big)
e^{-iq\alpha}\bigg).
\end{align}
\else
\begin{align}\label{Eq:alternative_harmonic_formulation_wigner}
g(\rho;\ell,m)\, &=\, \sum_{p=0}^{ L_h}\sum_{q=-p}^p \sum_{q'=-p}^p
\overline{\shc{\overline{f}Y_\ell^m}{p}{q}} \nonumber \\ & \quad
\times \shc{h}{p}{q'} \, d^p_{q,q'}(\beta) e^{-iq'\gamma}
e^{-iq\alpha}.
\end{align}
\fi

The band-limit of the spherical harmonic modulated signal
$\overline{f}\,Y_\ell^m$ is $L_f+\ell$.  Since the maximum $\ell$ for
which $g$ is non-zero is $L_f+L_h$, we must compute up to
$\overline{f}\,Y_{L_f+L_h}^m$, which is band-limited to $2L_f+L_h$.
However, we \emph{only} need to compute the spherical harmonic
coefficients $\shc{\overline{f}Y_\ell^m}{p}{q}$ of the modulated signal up
to degree $p\leq L_h$.
Therefore, the computation of the spherical harmonic transform of
$\overline{f}\,Y_\ell^m$ is an interesting sub-problem.  We show in
\appref{App:SSHT} that the spherical harmonic coefficients
$\overline{\shc{\overline{f}Y_\ell^m}{p}{q}}$ for $0 \leq p \leq
L_h,\, |q| \leq p$ of the signal $\overline{f}\,Y_\ell^m$ can be
computed in $ O(L_f^3 L_h^2)$ time for all $\ell$ and $m$.

By factoring the single rotation by $(\alpha,\beta, \gamma)$ into
two rotations~\cite{McEwen:2007,Risbo:1996,Wandelt:2001}
\ifCLASSOPTIONonecolumn
\begin{align}
\mathcal{D}_\rho = \mathcal{D}_{\rho_1}\,\mathcal{D}_{\rho_2},\quad
\rho = (\alpha,\beta,\gamma),\,\rho_1 =
(\alpha-pi/2,-\pi/2,\beta),\,\rho_2 = (0,\pi/2,\gamma+\pi/2),
\end{align}
\else
\begin{align}
\mathcal{D}_\rho = \mathcal{D}_{\rho_1}\,\mathcal{D}_{\rho_2}, \quad
&\rho = (\alpha,\beta,\gamma),\, \rho_1 =
(\alpha-\pi/2,-\pi/2,\beta),\nonumber \\  & \,\rho_2 =
(0,\pi/2,\gamma+\pi/2),
\end{align}
\fi
and noting the effect of rotation on spherical harmonic coefficients
in \eqref{eqn:rotation_non_symm}, we can write the Wigner-$D$
function in \eqref{Eq:Dlm} as
\begin{align}\label{Eq:WignerD_expansion_new}
D^p_{q,q'}(\alpha,\beta,\gamma)
&= \, i^{q-q'}\,\sum_{q''=-p}^p \Delta^{p}_{q''q}\,\Delta^p_{q''q'}
\,e^{-iq\alpha- iq''\beta - iq'\gamma},
\end{align}
where $\Delta^p_{q q'} = d^p_{q,q'}(\pi/2)$ and we have used the
following symmetry properties of Wigner-$d$
functions~\cite{Varsha:1988}
\ifCLASSOPTIONonecolumn
\begin{align}
d_{q,q'}^{p}(\beta)= (-1)^{q-q'}d_{q,q'}^{p}(-\beta) =  (-1)^{q-q'}
d_{-q,-q'}^{p}(\beta) = (-1)^{q-q'} d_{q',q}^{p}(\beta) =
d_{-q',-q}^{p}(\beta)
\end{align}
\else
\begin{align}
d_{q,q'}^{p}(\beta) &= (-1)^{q-q'}d_{q,q'}^{p}(-\beta) = (-1)^{q-q'}
d_{-q,-q'}^{p}(\beta) \nonumber \\ &
= (-1)^{q-q'}
d_{q',q}^{p}(\beta) = d_{-q',-q}^{p}(\beta) .
\end{align}
\fi
Using the Wigner-$D$ expansion given in
\eqref{Eq:WignerD_expansion_new}, we can write the alternative
harmonic formulation of the SLSHT distribution component
$g(\rho;\ell,m)$ in
\eqref{Eq:alternative_harmonic_formulation} as
\ifCLASSOPTIONonecolumn
\begin{align}\label{Eq:alternative_harmonic_formulation_2}
g(\rho;\ell,m)\, &=\, \sum_{p=0}^{ L_h}\sum_{q=-p}^p \sum_{q'=-p}^p
\overline{\shc{\overline{f}Y_\ell^m}{p}{q}} \shc{h}{p}{q'}\,
i^{q-q'}\,\sum_{q''=-p}^p \Delta^{p}_{q''q}\,\Delta^p_{q''q'}
\,e^{-iq\alpha- iq''\beta - q'\gamma},
\end{align}
\else
\begin{align}\label{Eq:alternative_harmonic_formulation_2}
g(\rho;\ell,m)\, &=\, \sum_{p=0}^{ L_h}\sum_{q=-p}^p \sum_{q'=-p}^p
\overline{\shc{\overline{f}Y_\ell^m}{p}{q}} \shc{h}{p}{q'}\,
i^{q-q'}\,\nonumber \\ & \! \! \! \times  \sum_{q''=-p}^p
\Delta^{p}_{q''q}\,\Delta^p_{q''q'} \,e^{-iq\alpha- iq''\beta -
q'\gamma},
\end{align}
\fi
where $\rho = (\alpha,\beta,\gamma)$. By reordering the summations we
can write
\ifCLASSOPTIONonecolumn
\begin{align}\label{Eq:alternative_harmonic_formulation_2_2}
g(\rho;\ell,m)\, &=\, \sum_{q=-L_h}^{L_h}\sum_{q'=-L_h}^{L_h}
\,\sum_{q''=-L_h}^{L_h} C_{q,q',q''}(\ell,m) \,e^{-iq\alpha-
iq''\beta - q'\gamma}, \quad \rho = (\alpha,\beta,\gamma).
\end{align}
\else
\begin{align}\label{Eq:alternative_harmonic_formulation_2_2}
g(\rho;\ell,m)\, &=\, \sum_{q=-L_h}^{L_h}\sum_{q'=-L_h}^{L_h}
\,\sum_{q''=-L_h}^{L_h} C_{q,q',q''}(\ell,m) \nonumber
\\ & \quad \, \times e^{-iq\alpha- iq''\beta - q'\gamma}, \quad \rho =
(\alpha,\beta,\gamma),
\end{align}
\fi
where
\ifCLASSOPTIONonecolumn
\begin{align}\label{Eq:constant_alternative_formulation}
C_{q,q',q''}(\ell,m) = i^{q-q'} \sum_{p=\max(|q|,|q'|,|q''|)}^{ L_h}
\Delta^{p}_{q''q}\,\Delta^p_{q''q'}
\overline{\shc{\overline{f}Y_\ell^m}{p}{q}} \shc{h}{p}{q'}.
\end{align}
\else
\begin{align}\label{Eq:constant_alternative_formulation}
C_{q,q',q''}(\ell,m) = i^{q-q'}\!\! \!\! \!\!\!\!\!\!\!\!
\sum\limits_{p=\max(|q|,|q'|,|q''|)}^{ L_h}\!\! \!\!\!\!\!\!
\Delta^{p}_{q''q}\,\Delta^p_{q''q'}
\overline{\shc{\overline{f}Y_\ell^m}{p}{q}} \shc{h}{p}{q'}.
\nonumber
\end{align}
\fi
Comparatively, the computation of the SLSHT distribution components
using the expression given by
\eqref{Eq:alternative_harmonic_formulation_2_2} is not more efficient than
the initial expression
\eqref{Eq:alternative_harmonic_formulation_wigner}. However, the
presence of complex exponentials can be exploited by employing FFTs
to evaluate the involved summations.

The objective of factoring the rotations is to carry out the $\beta$
rotation along the $y$-axis as a rotation along the $z$-axis. The
rotations along the $z$-axis are expressed using complex
exponentials and thus these rotations can be applied with much less
computational burden, by exploiting the power of an FFT, relative to a
rotation about the $y$-axis.
All the three rotations which characterize the spatial domain of the
SLSHT distribution components appear in complex exponentials in
\eqref{Eq:alternative_harmonic_formulation_2_2} and thus we can use
FFTs to evaluate the summation of $C_{q,q',q''}(\ell,m)$ over $q$,
$q'$ and $q''$. First we need to compute $C_{q,q',q''}(\ell,m)$ for
each $\ell$ and for each $m$ which requires the one-dimensional
summation over three dimensional grid formed by $q$, $q'$ and $q''$
and thus can be computed in $O(L_h^4)$. Using $C_{q,q',q''}(\ell,m)$,
the summation over the complex exponentials in
\eqref{Eq:alternative_harmonic_formulation_2_2} can be carried out in
$O(L_h^3\log_2 L_h)$ using FFTs.  The overall complexity of this
approach is dominated by the computation of $C_{q,q',q''}(\ell,m)$,
that is, $O(L_h^4)$ for each SLSHT distribution component and $O(L_f^2
L_h^4)$ for the complete SLSHT distribution.
We note that the evaluation of $C_{q,q',q''}(\ell,m)$ requires the
computation of $\Delta^p_{q q'}$ which can be evaluated over the
$(q,q')$ plane for each $p$ using the recursion formula
of~\cite{Trapani:2006} with a complexity of $O(L_h^2)$. The $\Delta$
matrices are independent of the signal under analysis and therefore
can be computed offline. However, we compute $\Delta$ matrices
on-the-fly to minimize storage requirements. Since $p$ is of the
order $L_h$, the $\Delta$ matrices can be evaluated in $O(L_h^3)$,
which does not change the overall complexity of our proposed
algorithm. The overall asymptotic complexity of our fast algorithm
is thus $O(L_f^3 L_h^2 + L_f^2 L_h^4)$.

\begin{remark}
Since the complexity to compute the spherical harmonic transform of
the modulated signal ${\overline{f}\,Y_\ell^m}$ up to degree $L_h$
is $O(L_f^2 \log_2 L_f + L_f L_h^2)$ for each $\ell,\,m$ as shown in
\appref{App:SSHT}, the complexity of our fast algorithm to compute
one SLSHT distribution component is $O(L_f^2 \log_2 L_f + L_f L_h^2+
L_h^4)$. The factor $L_f^2 \log_2 L_f$ in the complexity does not
change if we compute spherical harmonic transform of
${\overline{f}\,Y_\ell^m}$ up to degree $L_h$ for all $\ell,\,m$
instead of each $\ell,\,m$.
\end{remark}

\begin{remark}
  In order to evaluate
  \eqref{Eq:alternative_harmonic_formulation_wigner}, we note that the
  separation of variables approach~\cite{Wiaux:2006} can be used as an
  alternative to the factoring of rotation approach to develop a fast
  algorithm. This is due to the factorized form of \mbox{Wigner-$D$}
  function and the consideration of equiangular tessellation scheme
  for SO(3), which keeps the independence between the samples along
  different Euler angles. In terms of the computational complexity,
  the separation of variable approach has the same computational
  complexity as the factoring of rotation approach. However, the
  separation of variable approach needs to compute Wigner-$d$
  functions for all values of $\beta$ but only requires a two dimensional FFT,
  whereas the factoring of rotation approach only requires the
  evaluation of Wigner-$d$ function for $\pi/2$ but requires a three dimensional
  FFT. Since both approaches have the same complexity, we use the
  factoring of rotation in our implementation of the fast algorithm.
\end{remark}

\begin{remark}
  If we want to analyze the signal $f$ with multiple window functions,
  then we do not need to recalculate the spherical harmonic transform
  of the modulated signal $\overline{f}\,Y_\ell^m$, which accounts for
  the $O(L_f^3 L_h^2)$ factor in the overall complexity. Once it is
  computed, the SLSHT distribution can be computed in $O(L_f^2L_h^4)$
  time for each window function of the same band-limit using the proposed efficient
  implementation.
\end{remark}

Our proposed formulation and efficient implementation can be further
optimized in the case of a steerable window function.  Steerable
functions have an azimuthal harmonic band-limit in $m$ that is less
than the band-limit in $\ell$ (see \cite{Wiaux:2005,Wiaux:2006} for further
details about steerability on the sphere).  In this case, the $L_f^2
L_h^4$ factor contributing to the overall asymptotic complexity of the
fast algorithm is reduced to $L_f^2 L_h^3$.  Furthermore, we may then
compute the directional SLSHT for any continuous $\gamma\in[0,2\pi)$
from a small number of basis orientations (due to the linearity of the
SLSHT).

If the signal and window function are real, the computational time can
be further reduced by considering the conjugate symmetry relation of
the spherical harmonic coefficients. Furthermore, in this setting, the
SLSHT distribution components also satisfy the conjugate symmetry
property
\begin{align}
g(\rho;\ell,-m) = (-1)^m\,\overline{g(\rho;\ell,m)}
\end{align}
and we do not need to compute the SLSHT distribution
components of negative orders.

\ifCLASSOPTIONonecolumn
\begin{figure*}[!ht]
    \centering
    \vspace{-5mm}
    \subfloat[Computation time to evaluate $(\overline{f} Y_\ell^m)_p^q$]
        {\label{fig:time1a}\includegraphics[scale=0.45]{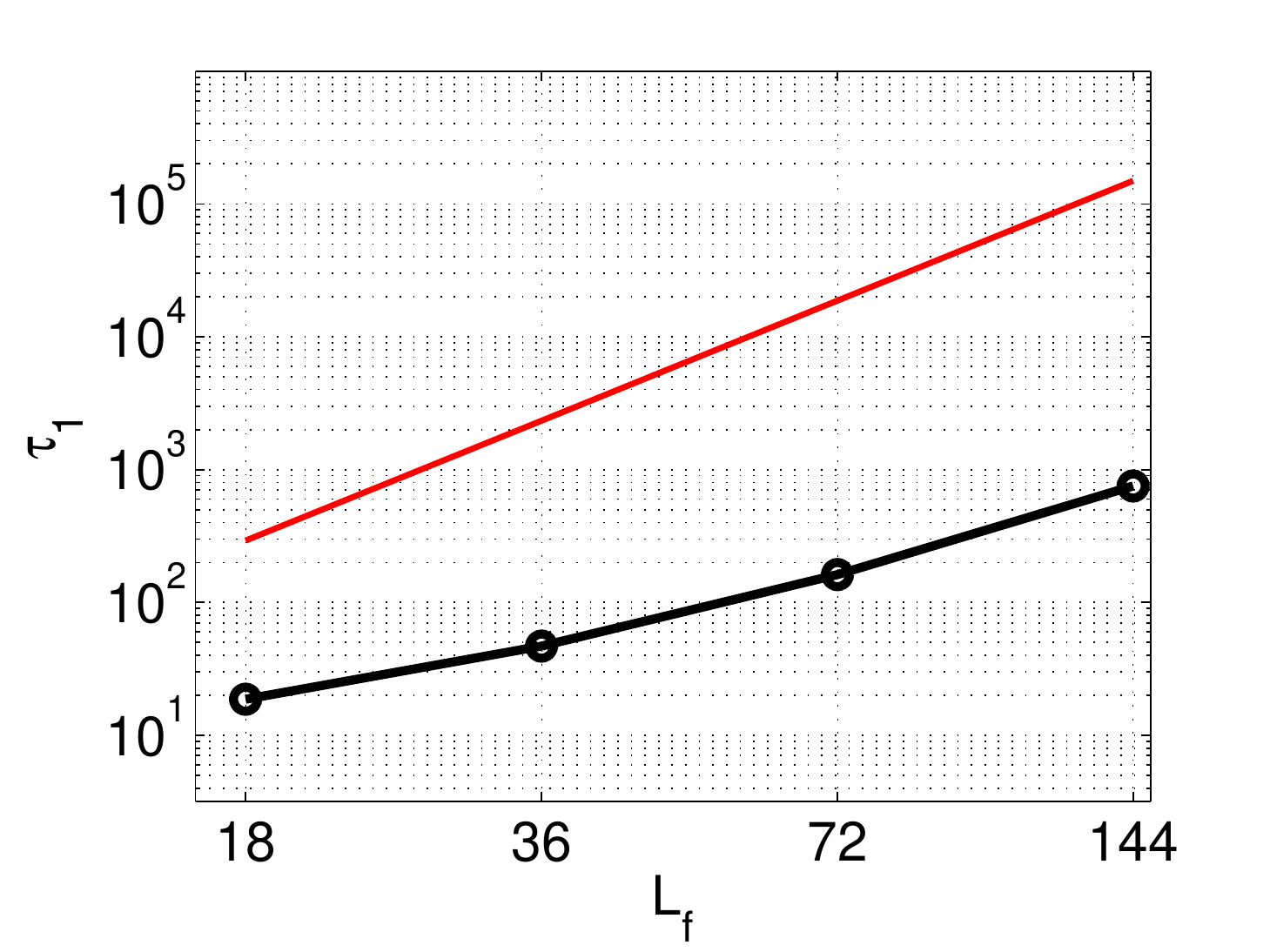}}
        \hspace{5mm}
    \subfloat[Computation time to evaluate all SLSHT distribution
    components given $(\overline{f} Y_\ell^m)_p^q$ ]
        {\label{fig:time1b}\includegraphics[scale=0.45]{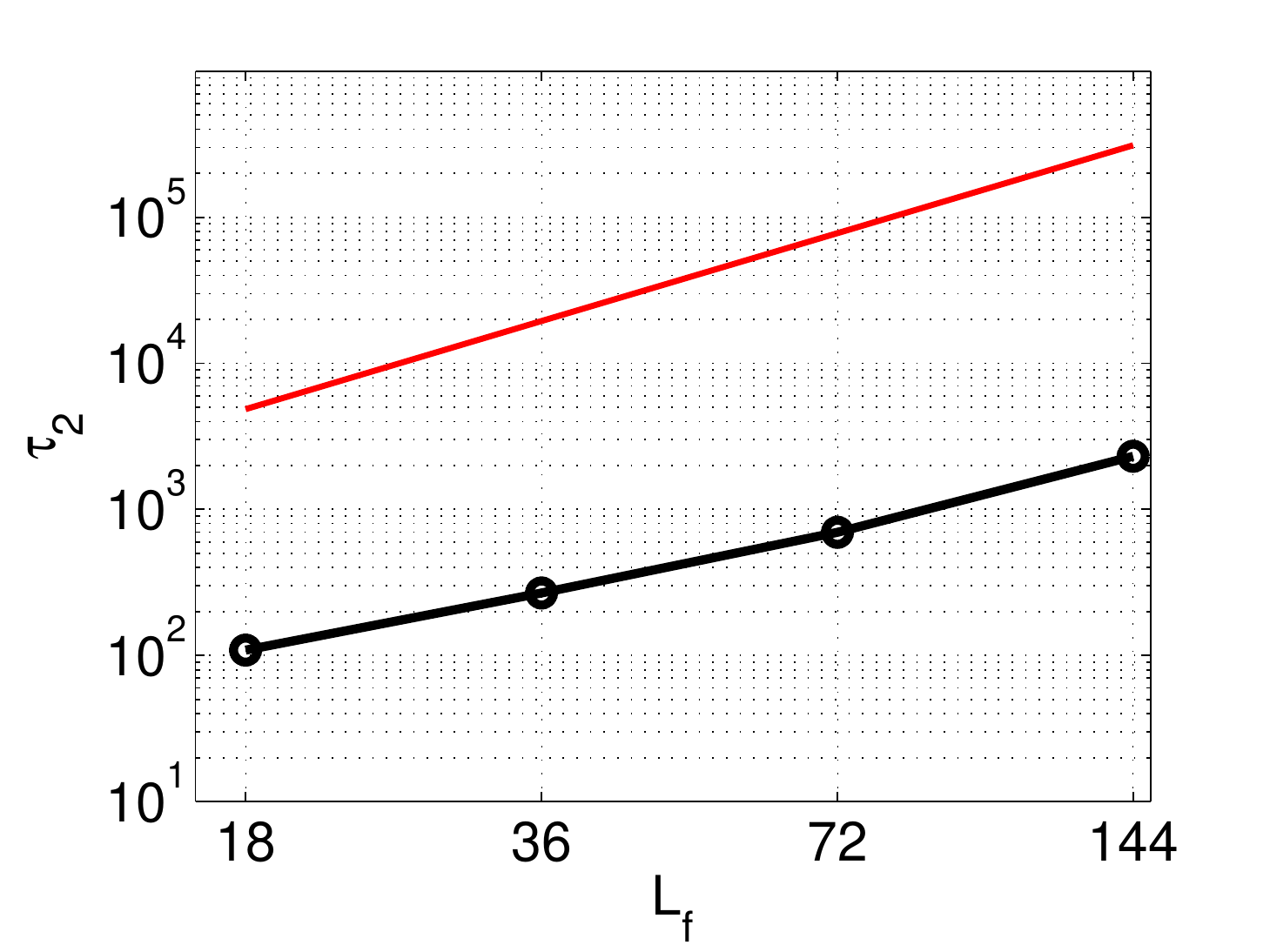}}

        \vspace{-5mm}
    \subfloat[Inverse transform computation time]
        {\label{fig:time2}\includegraphics[scale=0.45]{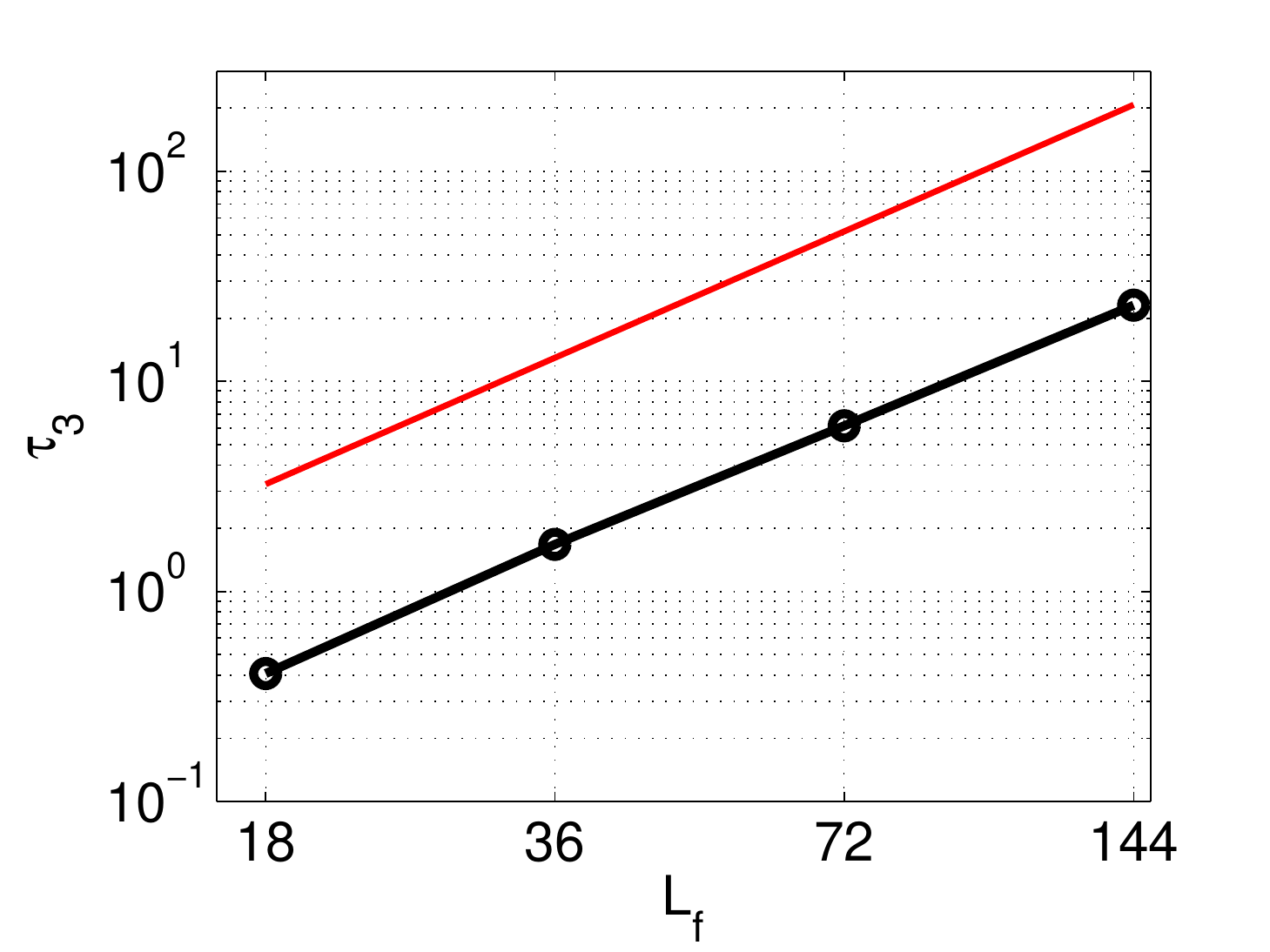}}
        \hspace{5mm}
    \subfloat[Numerical validation]
        {\label{fig:error}\includegraphics[scale=0.45]{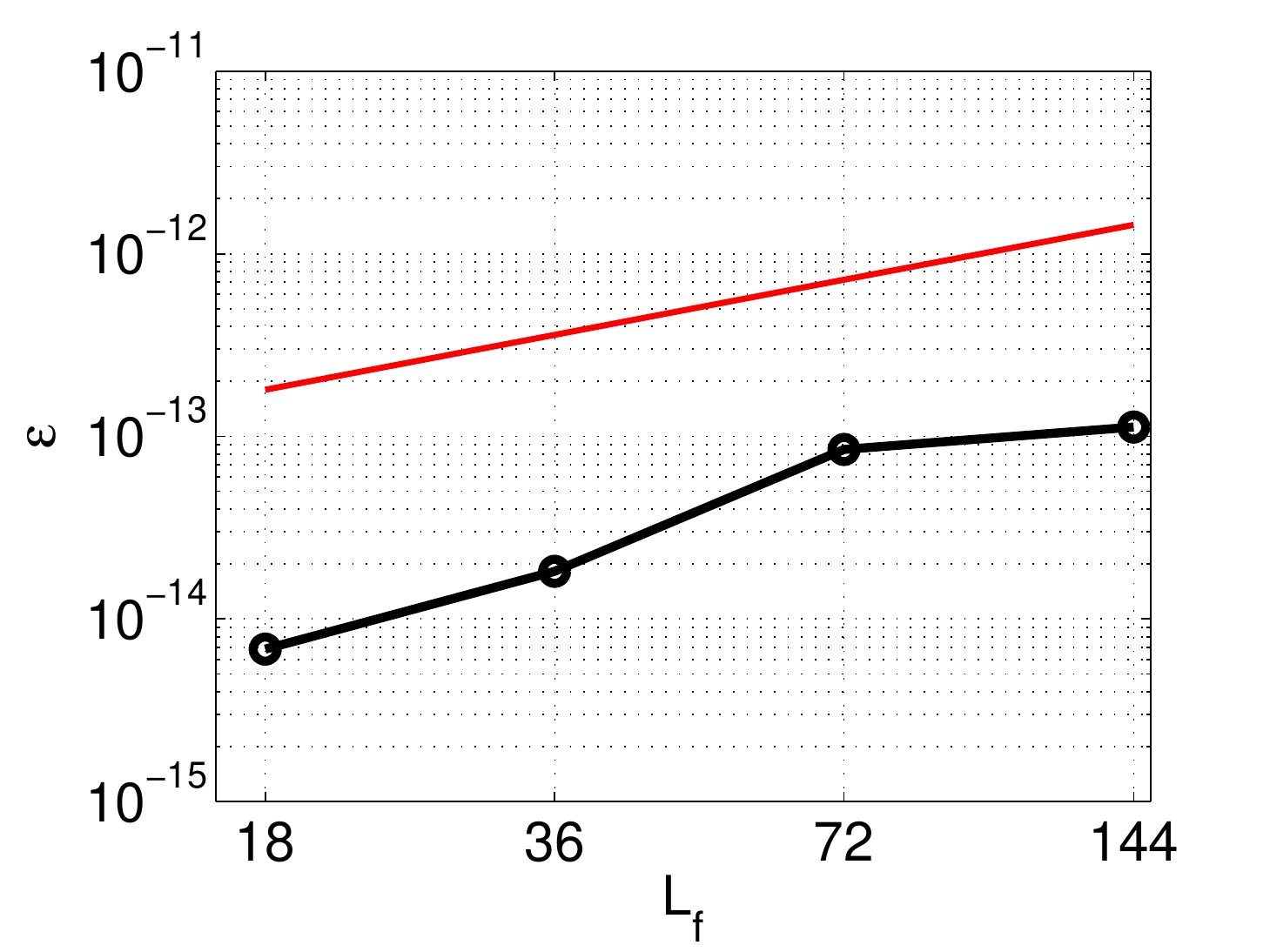}}
    \vspace{3mm}
    \caption{Numerical validation and computation time of the proposed
      algorithms. The computation time in seconds: (a) $\tau_1$ (b)
      $\tau_2$ and (c) $\tau_3$. For fixed $L_h$, $\tau_1$ evolves as $O(L_f^3)$ and
      both $\tau_2$ and $\tau_3$ scale as $O(L_f^2)$ as shown by the
      solid red lines~(without markers). (d) The maximum error
      $\epsilon$, which empirically appears to scale as $O(L)$, as
      shown by the solid red line. }
    \label{fig:simulations_1}
\end{figure*}
\else
\begin{figure*}[!ht]
    \centering
    \vspace{-5mm}
    \subfloat[Computation time to evaluate $(\overline{f} Y_\ell^m)_p^q$]
        {\label{fig:time1a}\includegraphics[scale=0.5]{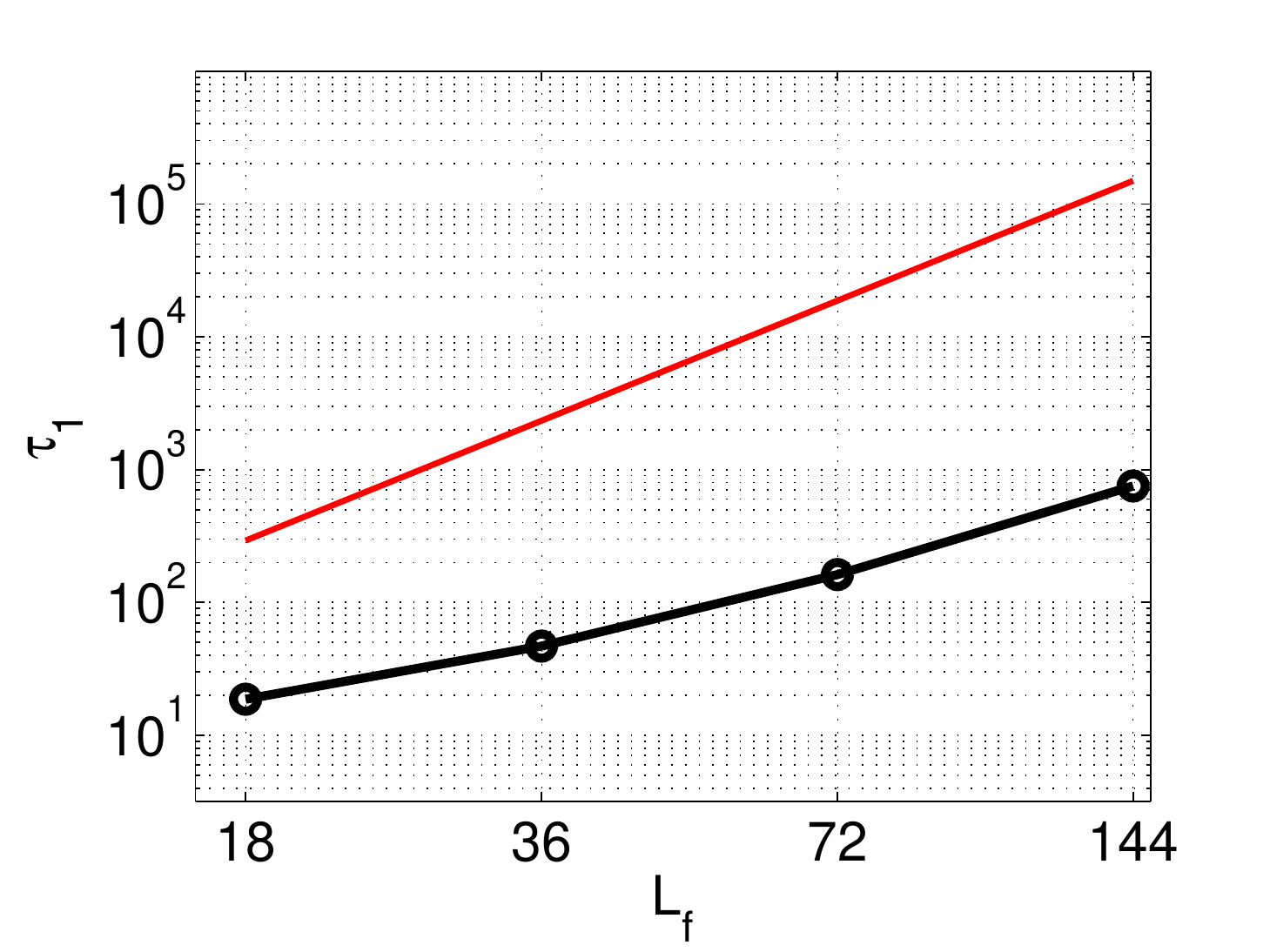}}
        \hspace{10mm}
    \subfloat[Computation time to evaluate all SLSHT distribution
    components given $(\overline{f} Y_\ell^m)_p^q$ ]
        {\label{fig:time1b}\includegraphics[scale=0.5]{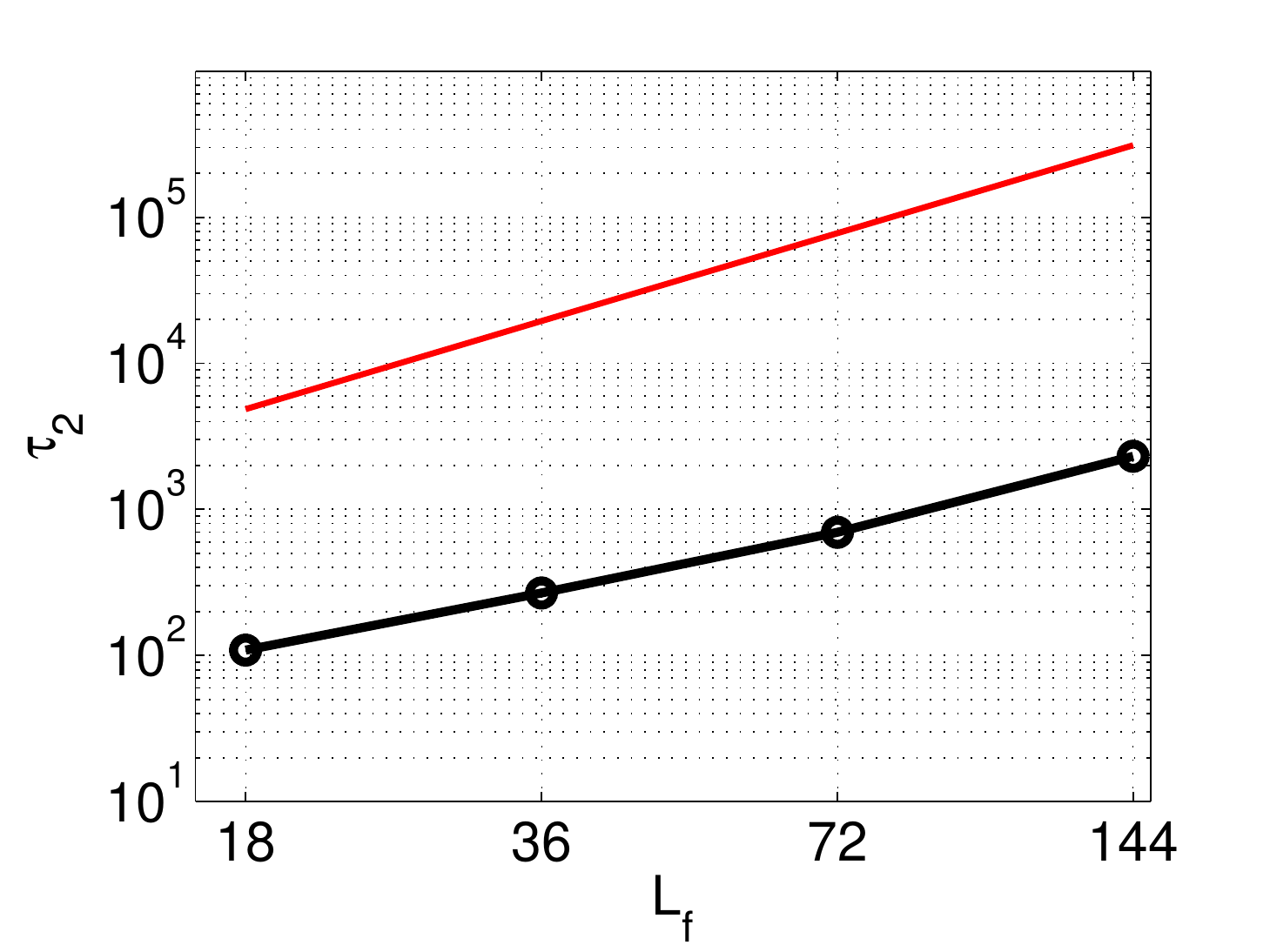}}

\vspace{-4mm}
    \subfloat[Inverse transform computation time]
        {\label{fig:time2}\includegraphics[scale=0.5]{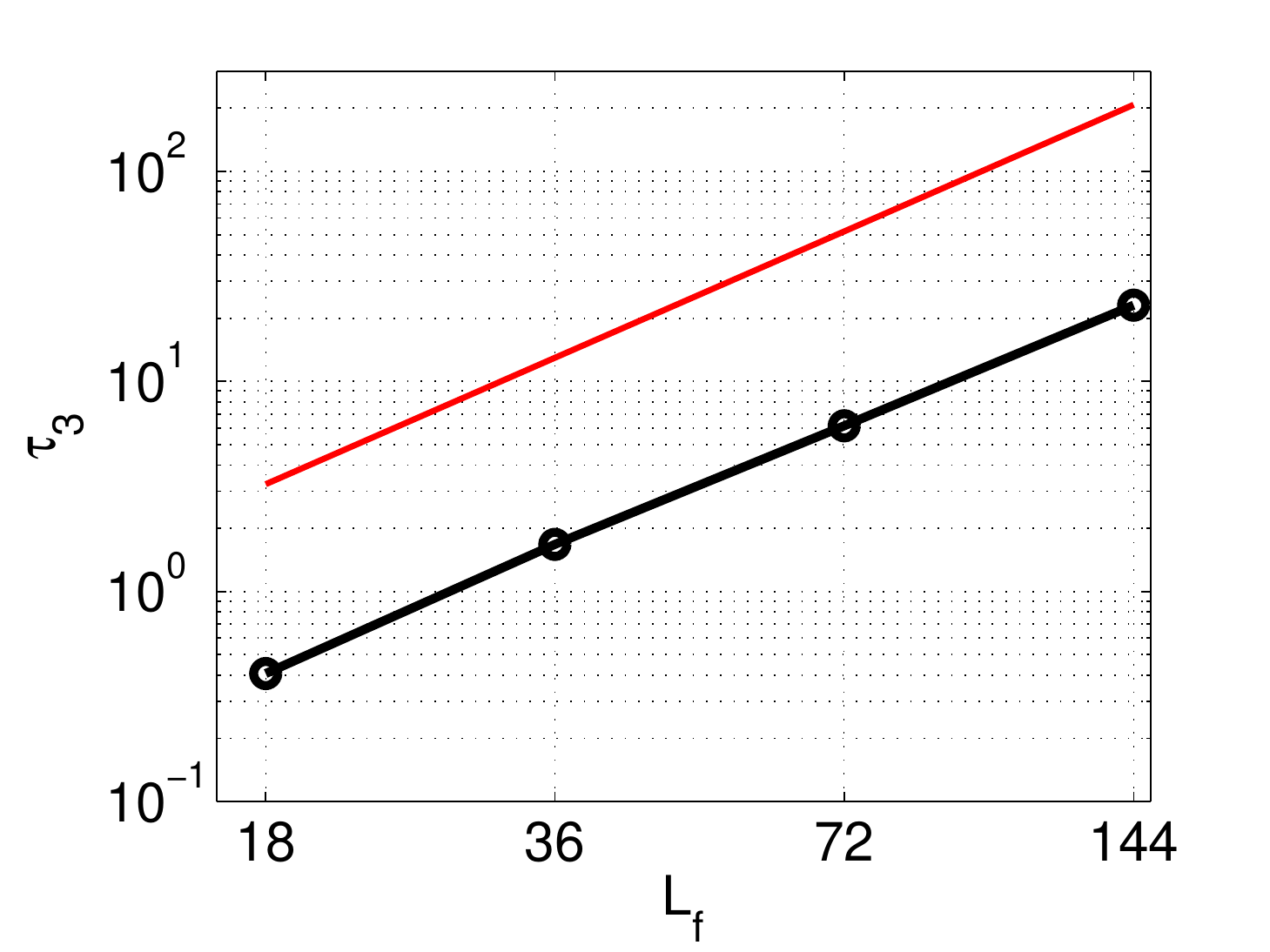}}
        \hspace{10mm}
    \subfloat[Numerical validation]
        {\label{fig:error}\includegraphics[scale=0.5]{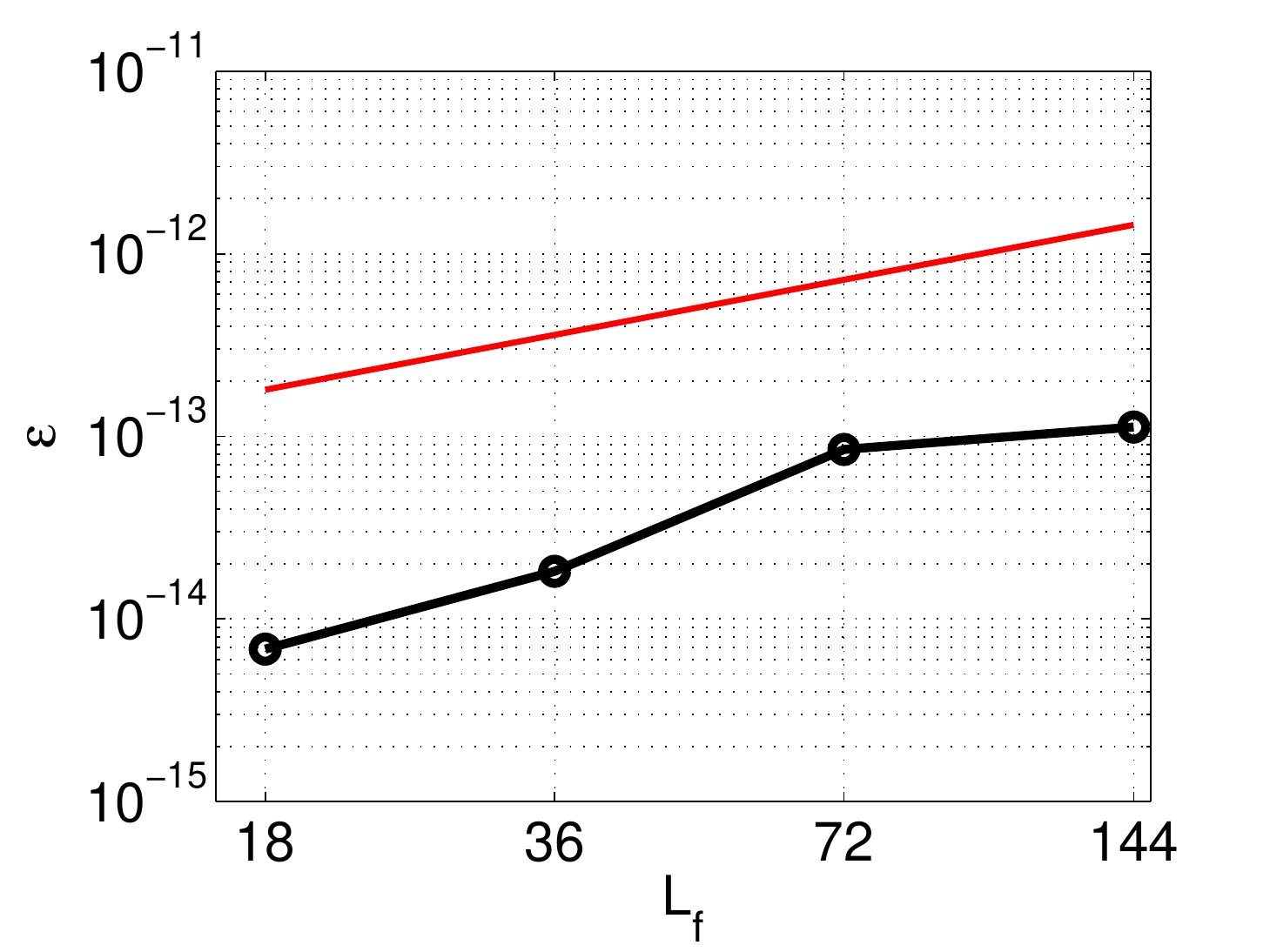}}
    \caption{Numerical validation and computation time of the proposed
      algorithms. The computation time in seconds: (a) $\tau_1$ (b)
      $\tau_2$ and (c) $\tau_3$. For fixed $L_h$, $\tau_1$ evolves as $O(L_f^3)$ and
      both $\tau_2$ and $\tau_3$ scale as $O(L_f^2)$ as shown by the
      solid red lines~(without markers). (d) The maximum error
      $\epsilon$, which empirically appears to scale as $O(L)$, as
      shown by the solid red line. }
    \label{fig:simulations_1}
\end{figure*}
\fi

\section{Results}\label{sec:results}

In this section, we first demonstrate the numerical validation and
computation time of our algorithms to evaluate the directional SLSHT
components. Later, we provide an example to illustrate the capability
of the directional SLSHT, showing that it reveals the directional
features of signals in the spatio-spectral domain. The implementation
of the our algorithms is carried out in \texttt{MATLAB}, using the
\texttt{MATLAB} interface of the
\texttt{SSHT}\footnote{\url{http://www.jasonmcewen.org/}} package
(the core algorithms of which are written in $\texttt{C}$ and which
also uses the \texttt{FFTW}\footnote{\url{http://www.fftw.org/}} package
to compute Fourier transforms) to efficiently compute forward and
inverse spherical harmonic transforms~\cite{McEwen:2011}.

\subsection{Numerical Validation and Computation Time}
In order to evaluate the numerical accuracy and the computation time,
we carry out the following numerical experiment. We use the
band-limited function $h$ for spatial localization with band-limit
$L_h=18$ and spatial localization in the region
$\mathcal{R}_{(\pi/6,\pi/6+\pi/240)}$. We generate band-limited test
signals with band-limits $18\leq L_f\leq 130$ by generating spherical
harmonic coefficients with real and imaginary parts uniformly
distributed in the interval $[0,1]$.

For the given test signal, we measure the computation time $\tau_1$ to
evaluate spherical harmonic transform of the modulated signal, i.e., $(\overline{f}Y_\ell^m)_p^q$ for $p\leq L_h, \,q\leq|p|$ and for all $\ell\leq L_f+L_h,\, m\leq|\ell|$, using the method presented in \appref{App:SSHT}.
Given the spherical harmonic transform of the modulated signal, we then measure the computation time $\tau_2$ to compute
all directional SLSHT distribution components $g(\rho;\ell,m)$
for $\ell\leq L_f+L_h$ and $m\leq|\ell|$ using our fast algorithm presented in Section~\ref{sec:Algos:fast}, where we compute the
Wigner-$d$ functions on-the-fly for the argument $\pi/2$ by using the
recursion of Trapani~\cite{Trapani:2006}. We also record the computation time $\tau_3$ to recover a signal from its SLSHT
distribution components. All numerical
experiments are performed using \texttt{MATLAB} running on a 2.4 GHz
Intel Xeon processor with 64 GB of RAM and the results are averaged
over ten test signals. The computation time $\tau_1$ and $\tau_2$ are plotted against the band-limit $L_f$ of the test signal in
\figref{fig:time1a} and \figref{fig:time1b}, which respectively evolve as $O(L_f^3)$ and $O(L_f^2)$ for fixed $L_h$ and thus corroborate
the theoretical complexity. The computation time $\tau_3$ for the inverse directional SLSHT is
plotted in \figref{fig:time2}, which scales as $O(L_f^2)$ for fixed
$L_h$, again supporting the theoretical complexity.

We reconstruct the original signal from its SLSHT distribution components
using \eqref{Eq:sig_reconstruction_discrete} and
\eqref{Eq:inversion_1}, in order to assess the numerical
accuracy of our algorithms by measuring the maximum absolute
error between the original spherical harmonic coefficients of the
test signal and the reconstructed values. The maximum absolute error
is plotted in \figref{fig:error} for different band-limits $L_f$,
which illustrates that our algorithms achieve very good numerical
accuracy with numerical errors at the level of floating point precision.

\subsection{Directional SLSHT Illustration}
In this subsection, we provide examples to illustrate the capability
of the proposed transform to reveal the localized contribution of
spectral contents and probe the directional features in the
spatio-spectral domain.

\ifCLASSOPTIONonecolumn
\begin{figure*}[t]
\else
\begin{figure*}[!ht]
\fi
 \centering
    \vspace{-5mm}
    \hspace{-5mm}
    \subfloat[$f_1$]{
        \includegraphics[scale=0.28]{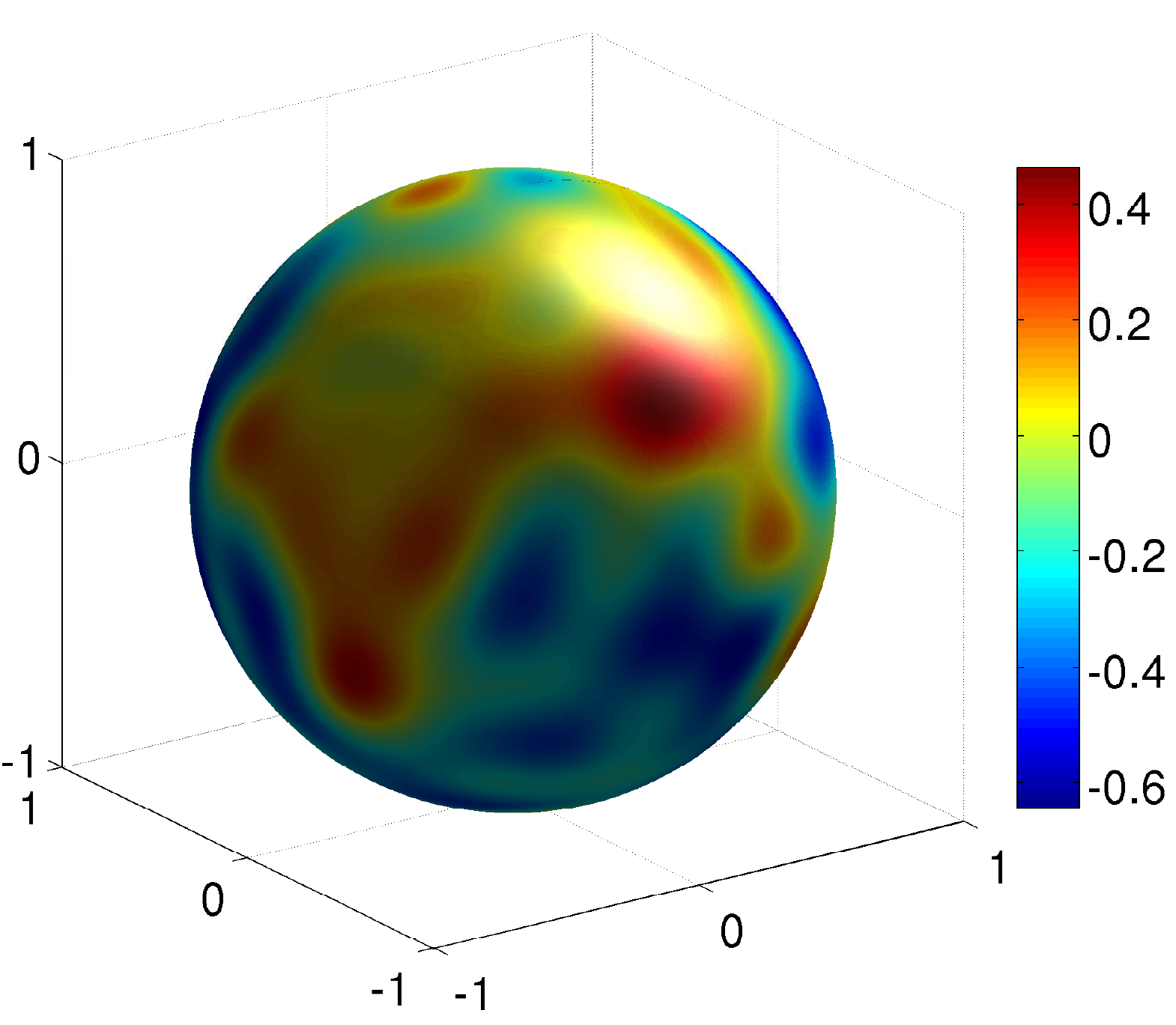}}
        \hspace{5mm}
   \subfloat[$f_2$]{
        \includegraphics[scale=0.28]{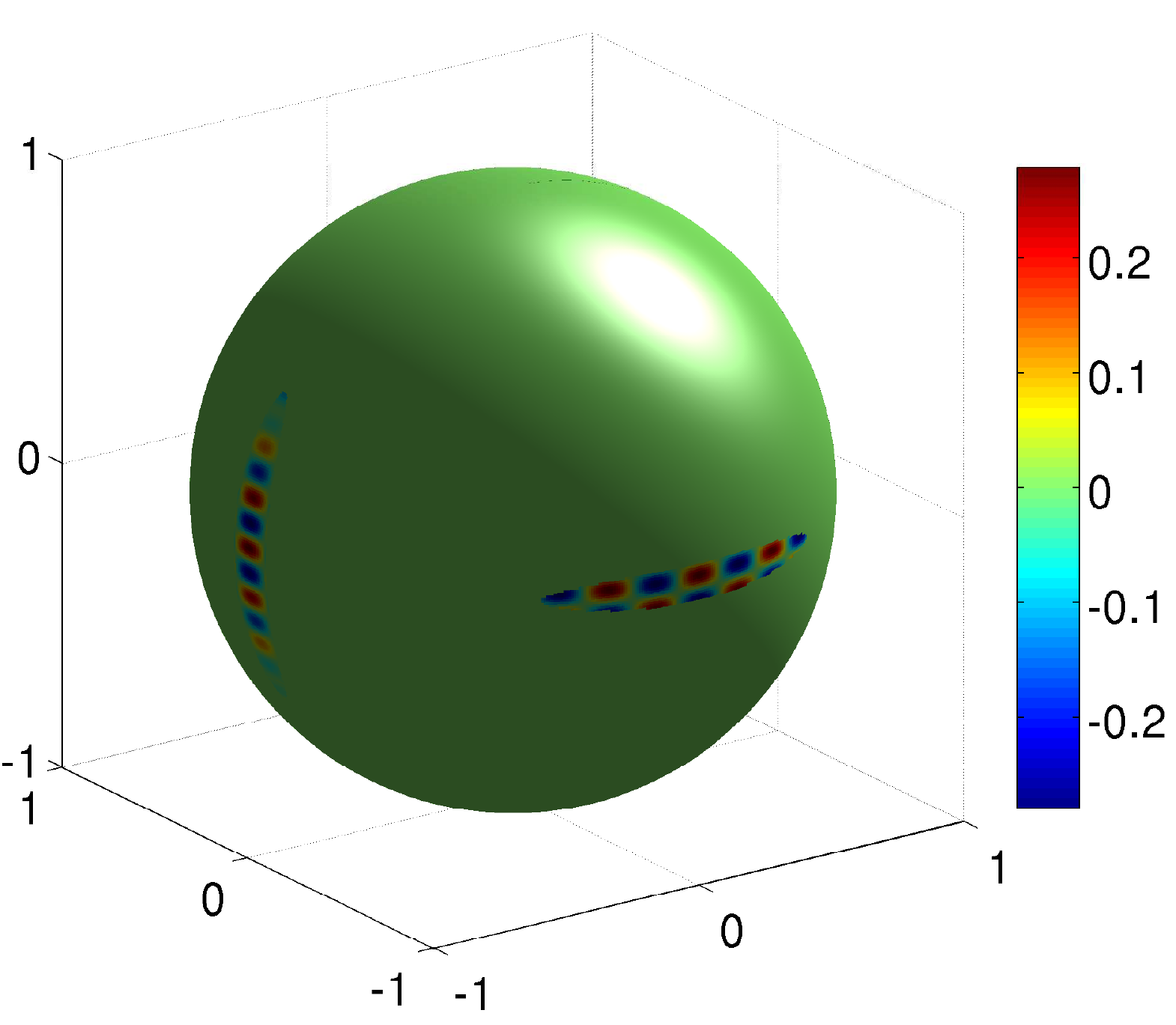}}
        \hspace{5mm}
    \subfloat[$f$]{
        \includegraphics[scale=0.28]{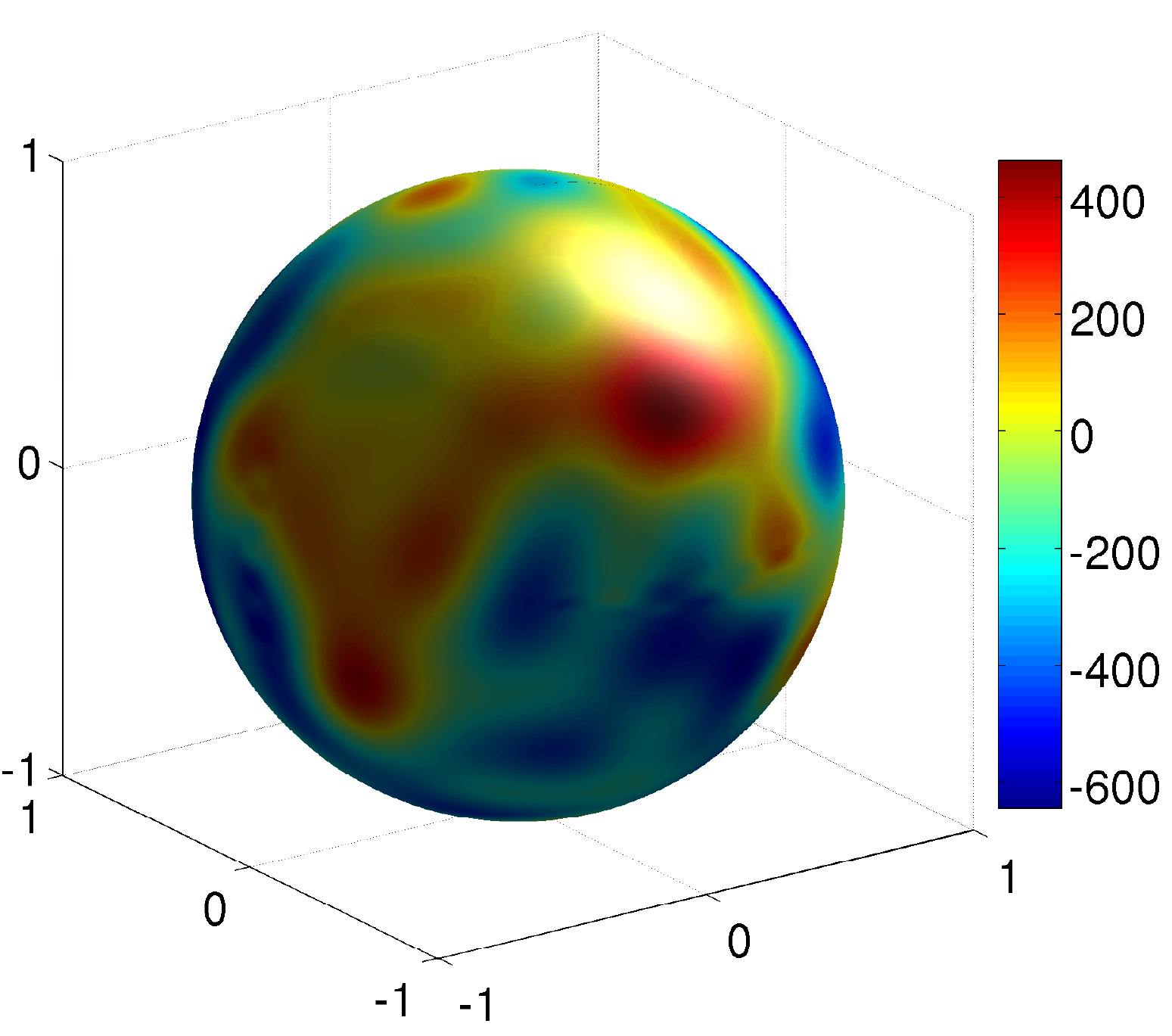}}

 \caption{(a) Spectrally truncated unit energy normalized Earth topographic map $f_1$ and (b) signal $f_2$ composed
of higher degree spherical harmonics localized in elliptical
regions. (c) Weighted sum of $f_1$ and $f_2$ as defined in
\eqref{Eq:syn_signal_sum}}. \label{fig:syn_sig}\vspace{-0.25in}
\end{figure*}
%
\subsubsection{Example 1 - Synthetic Data Set}
We first construct a signal having localized contribution of higher
degree spectral contents and then analyze the signal using proposed
directional SLSHT. Let signal $f_1$ be the spectrally truncated,
unit energy normalized Earth topographic map with band-limit
$L_{f_1}=30$, which is obtained by using spherical harmonic model of
topography of Earth and is shown in \figref{fig:syn_sig}a. Also
consider the signal $f_2$ composed of higher degree spherical
harmonics localized in two non-overlapping elliptical regions with
different orientation. We obtain such a signal $f_2$ by spectrally
truncating the following signal $\tilde{f}_2$ with in the band-limit
$L_{f_2} = 128$,
\begin{align}
\tilde{f}_2(\unit{x}) &= \begin{cases}
\sum\limits_{\ell=40}^{45}\left( Y_\ell^{20}(\unit{x})+
Y_\ell^{-20}(\unit{x})\right) &
\unit{x}\in\mathcal{R} = \mathcal{R}_1 \cup \mathcal{R}_1 \\
0 & \unit{x}\in \mathbb{S}^2 \setminus \mathcal{R},
\end{cases}
\end{align}
where $\mathcal{R}_1$ and $\mathcal{R}_2$ are the elliptical regions
of the form $\mathcal{R}_{(\pi/6,\pi/6+\pi/240)}$, respectively
rotated by $(\pi/2,\pi/2,0)\in\textrm{SO(3)}$ and
$(3\pi/2,\pi/2,\pi/2)\in\textrm{SO(3)}$. The unit energy normalized
signal $f_2$ is shown in \figref{fig:syn_sig}b. We note that the
regions $\mathcal{R}_1$ and $\mathcal{R}_2$ have orientation along
colatitude and longitude respectively.

We analyze the following synthetic signal using the proposed
transform
\begin{align}\label{Eq:syn_signal_sum}
f(\unit{x}) &= 10^3 \times \left( \frac{f_1(\unit{x})}{\|f_1\|} +
\frac{f_2(\unit{x})}{4\|f_2\|} \right),
\end{align}
which can be considered as a sum of low frequency signal and high
frequency localized signal. The signal $f$ is shown in
\figref{fig:syn_sig}c, where it can be observed that the information
cannot be obtained about the presence of higher degree spherical
harmonics localized in different directional regions. Furthermore,
the spherical harmonic coefficients provide details about the
presence of higher degree spherical harmonics in the signal, but do
not reveal any information about the localized contribution of
higher degree spherical harmonics.

If we analyze the signal by employing the SLSHT using an azimuthally
symmetric window function, the presence of localized contributions
of higher degree spectral contents can be determined in the
spatio-spectral domain~\cite{Khalid:2012}. However, the presence of
directional features cannot be extracted. Here, we illustrate that
the use of the directional SLSHT enables the identification of
directional features in the spatio-spectral domain, which is due to
the consideration of an asymmetric window function for spatial
localization.

We obtain the directional SLSHT distribution components
$g(\rho;\ell,m)$ of the signal $f$ using the band-limited
eigenfunction window $h$ with $L_h=18$ and $90\%$ concentration in
the spatial domain in an elliptical region
$\mathcal{R}_{(\pi/6,\pi/6+\pi/240)}$. The magnitude of the SLSHT
distribution components $g(\rho;\ell,m)$ for order $m=20$ and for
degrees $\ell \in \{41,43,45\}$ are shown in \figref{fig:syn_comps}
for Euler angle (a) $\gamma=0$ and (b) $\gamma=100\pi/201
\approx\pi/2$, and for degrees $\ell \in \{21,23,25\}$, the
components are shown for (c) $\gamma=0$ and (d)
$\gamma\approx\pi/2$. Since the elliptical region is oriented along
the $x$-axis, the window with orientation $\gamma=0$ provides
localization along colatitude and the window with orientation
$\gamma\approx\pi/2$ provides localization along longitude. It can
be observed that the localized contribution of higher degree
directional spectral contents is extracted in spatio-spectral
domain. The localized higher degree directional features along the
orientation $\gamma=0$ and $\gamma\approx\pi/2$ are revealed in the
spatio-spectral domain as shown in \figref{fig:syn_comps}a and
\figref{fig:syn_comps}b respectively, which are not visible in lower
degree distribution components as shown in \figref{fig:syn_comps}c
and \figref{fig:syn_comps}d.

Due to the ability of the directional SLSHT to reveal the localized
contribution of spectral contents and the directional or oriented
features in the spatio-spectral domain, it can be useful in many
applications where the signal on the sphere is localized in position
and orientation. We further illustrate the capability of our
proposed transform by analyzing the Mars topographic map in
spatio-spectral domain
%
\begin{figure}[!ht]
 \centering
    \vspace{-5mm}
    \hspace{-5mm}
    \subfloat{
        \includegraphics[scale=0.18]{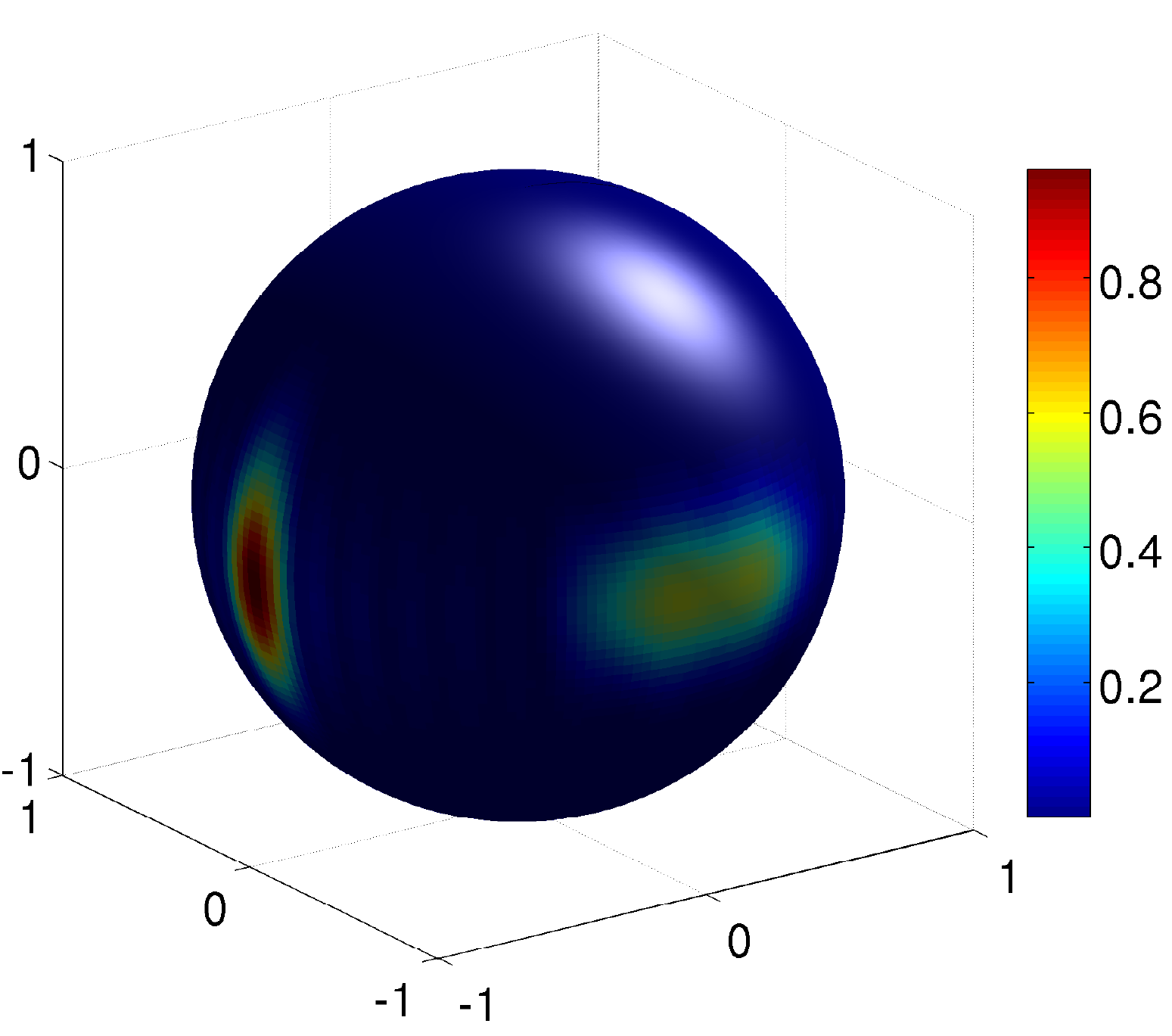}}
   \subfloat{
        \includegraphics[scale=0.18]{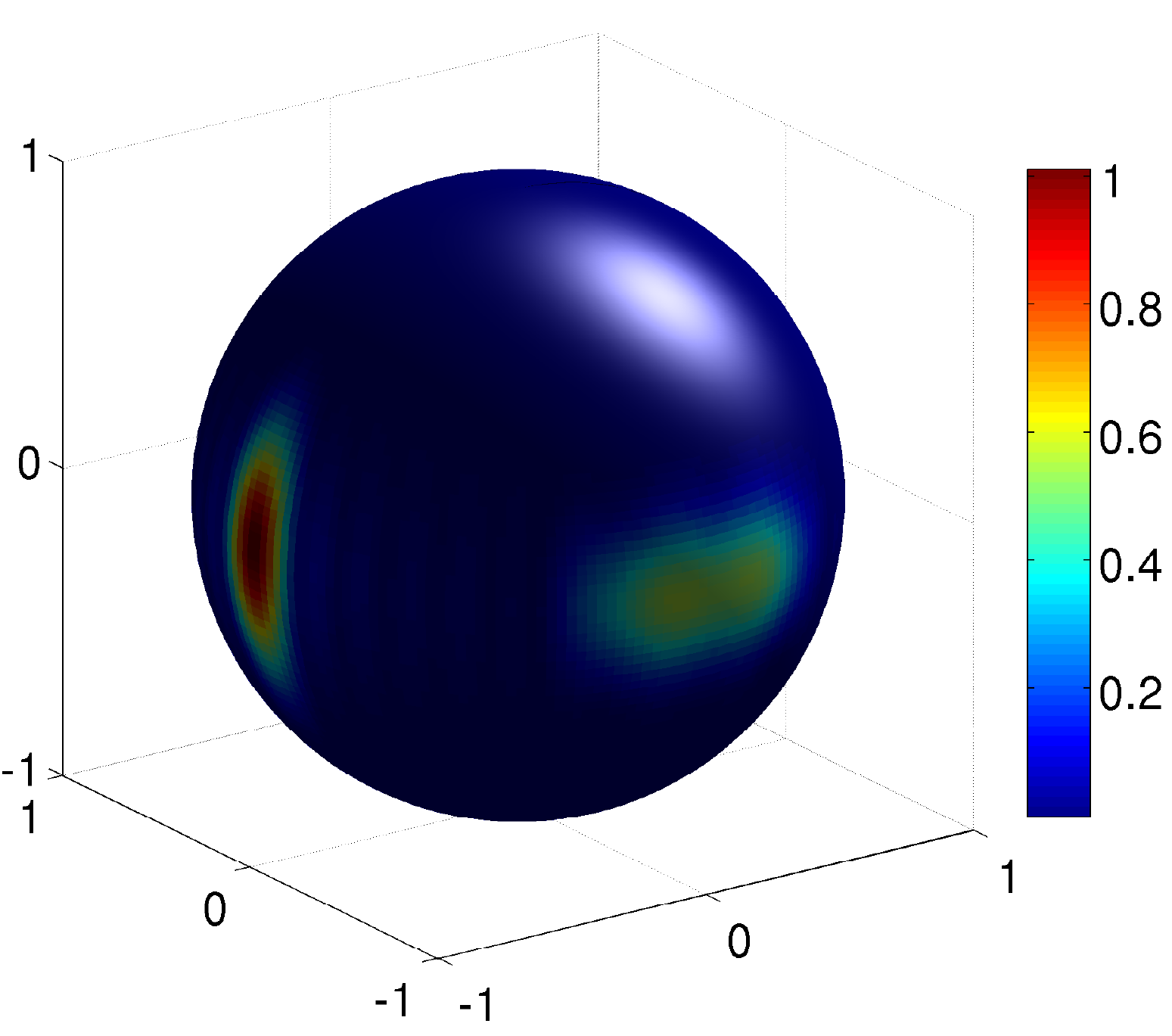}}
    \subfloat{
        \includegraphics[scale=0.18]{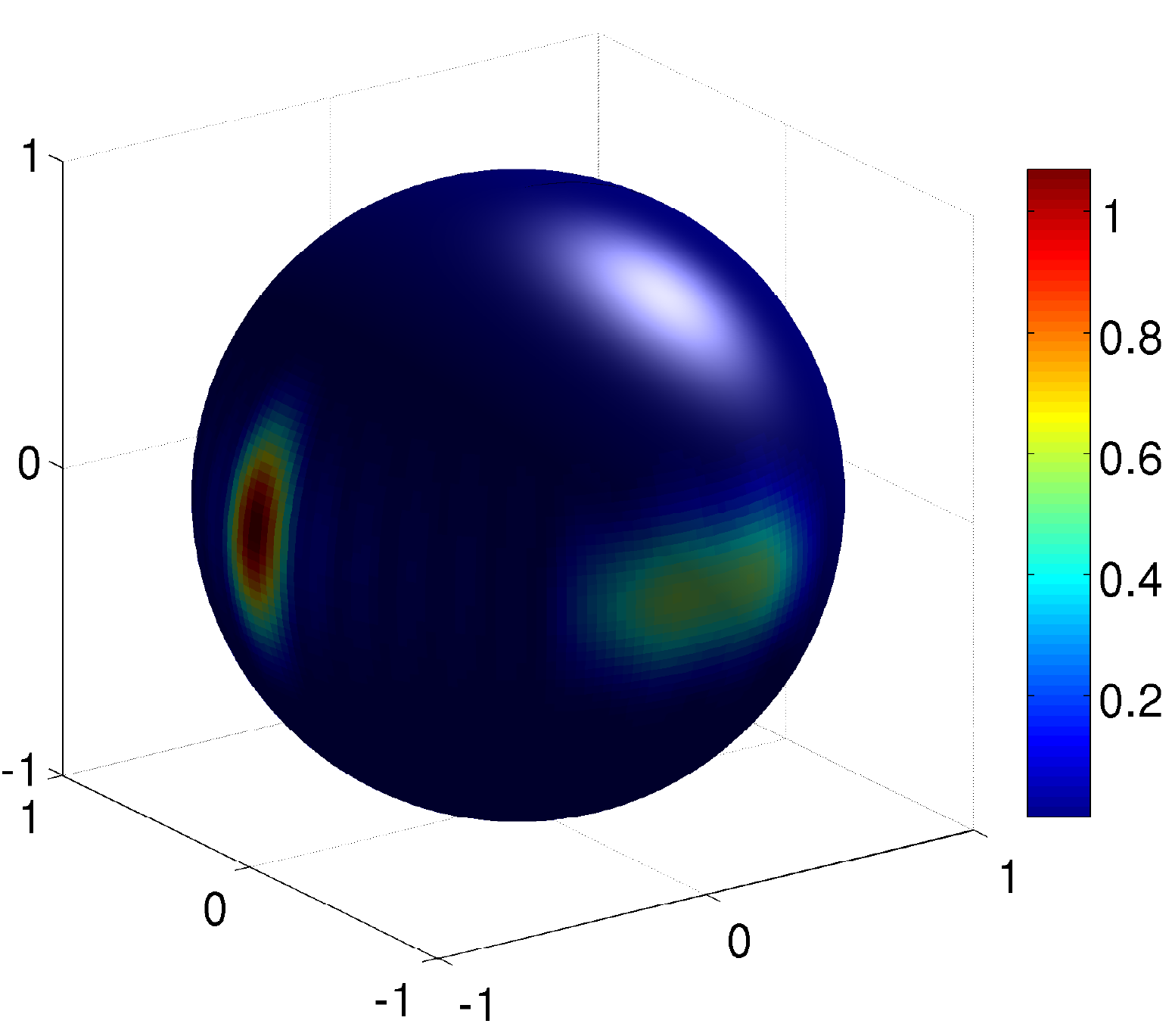}}

        \vspace{-3mm}
       \setcounter{subfigure}{0}
       \centering
        \subfloat[]{}

    \hspace{-8mm}
    \subfloat{
        \includegraphics[scale=0.18]{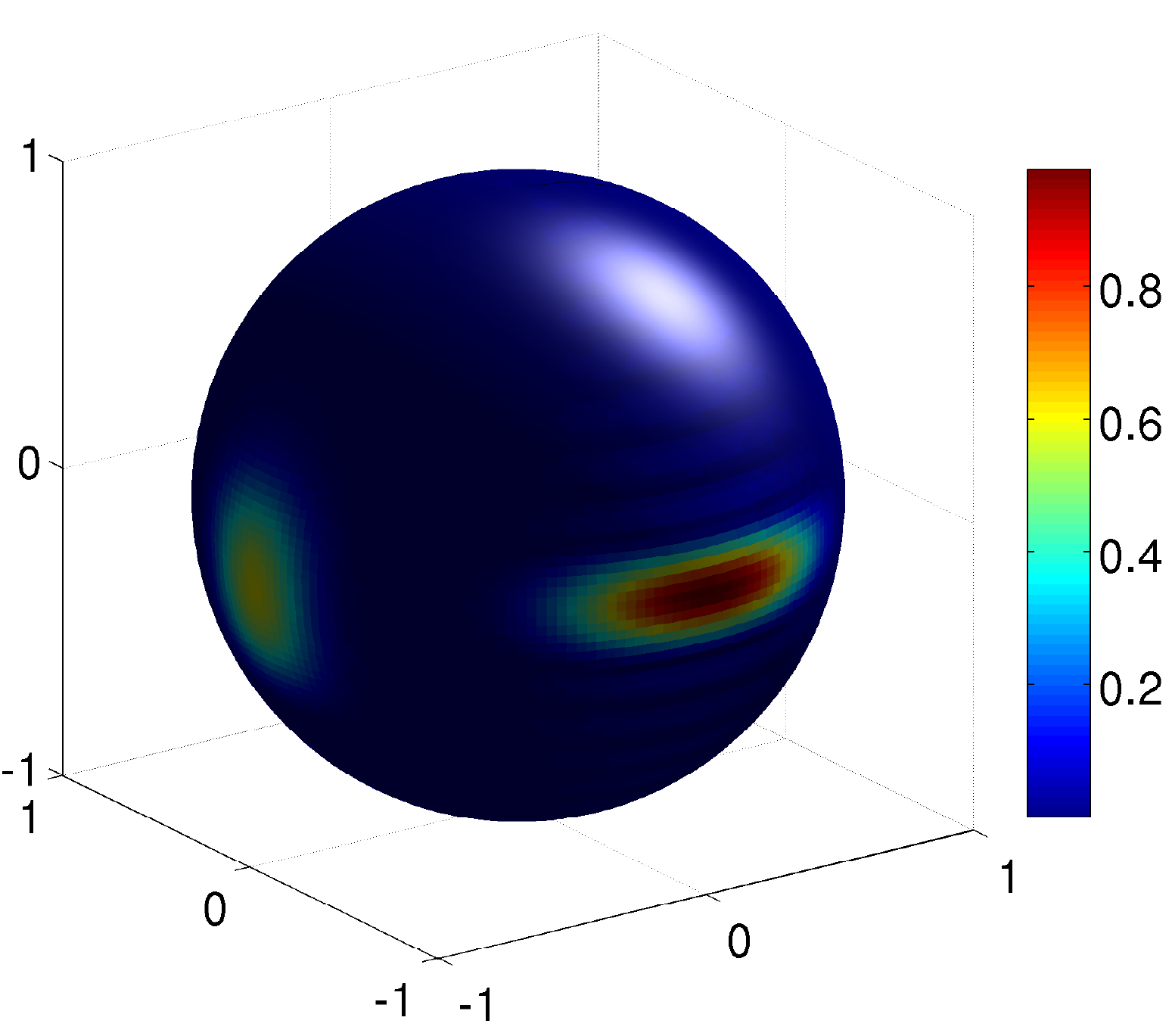}}
   \subfloat{
        \includegraphics[scale=0.18]{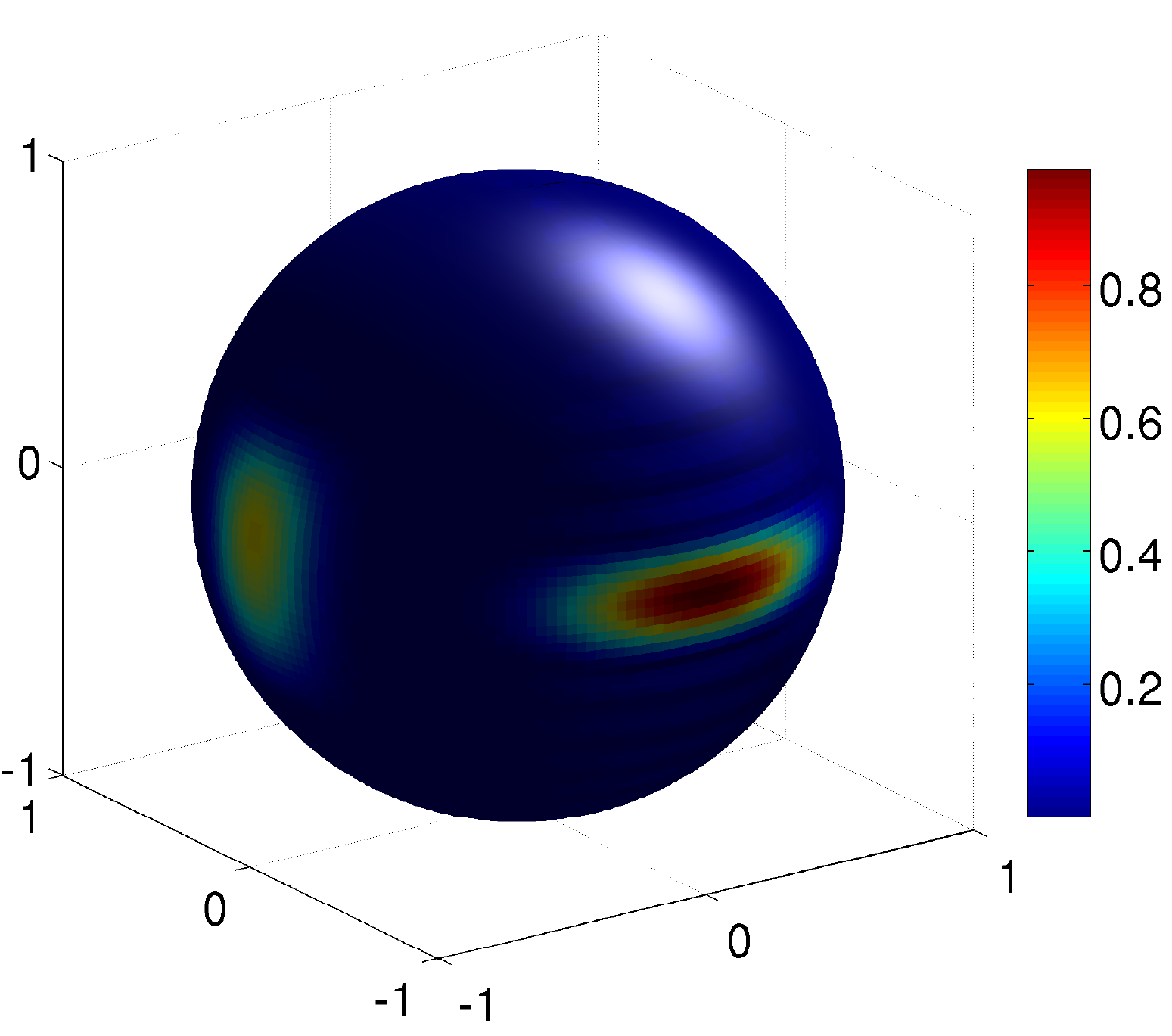}}
    \subfloat{
        \includegraphics[scale=0.18]{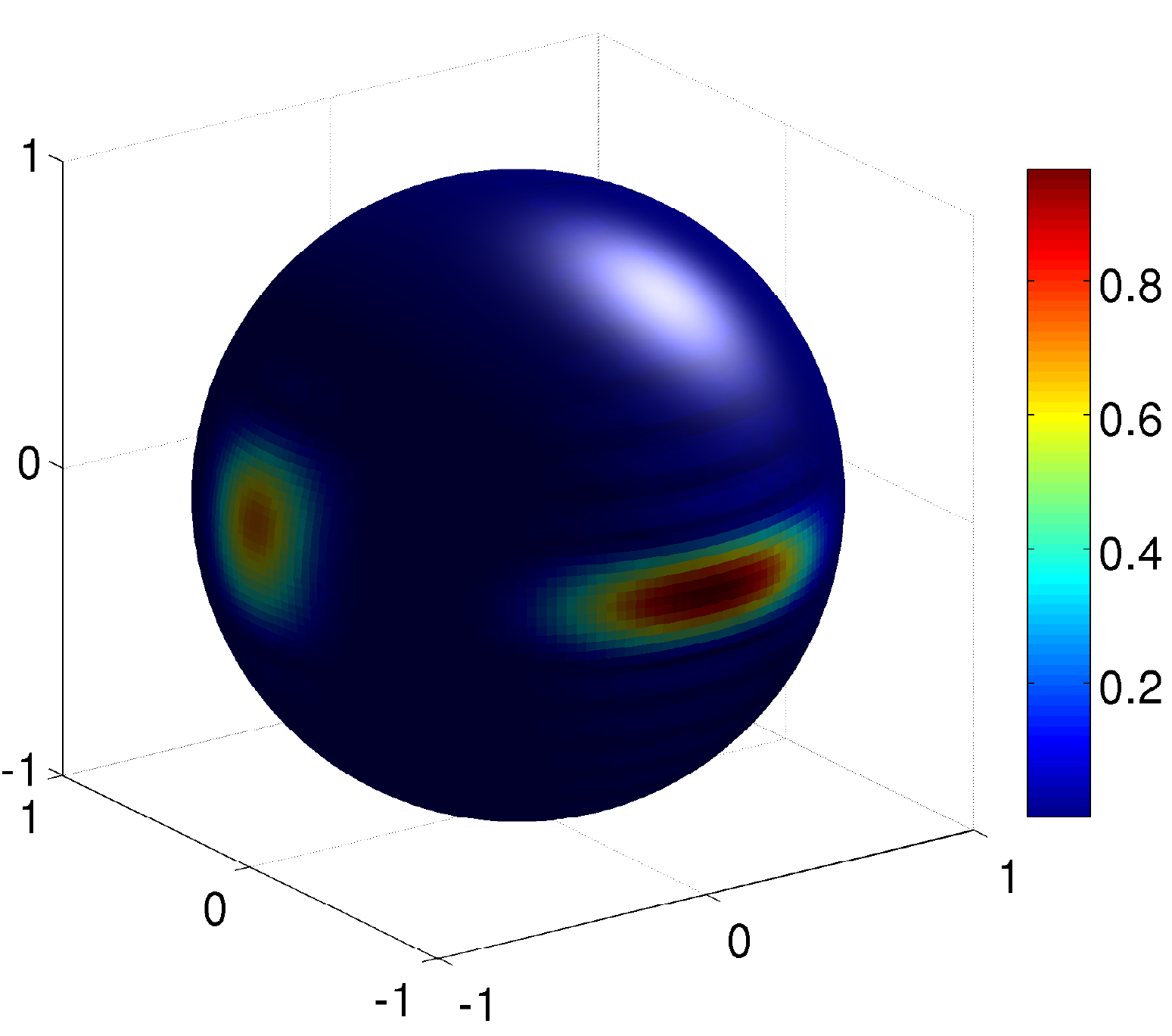}}

        \vspace{-3mm}
\setcounter{subfigure}{1}
        \subfloat[]{}

    \hspace{-8mm}
    \subfloat{
        \includegraphics[scale=0.18]{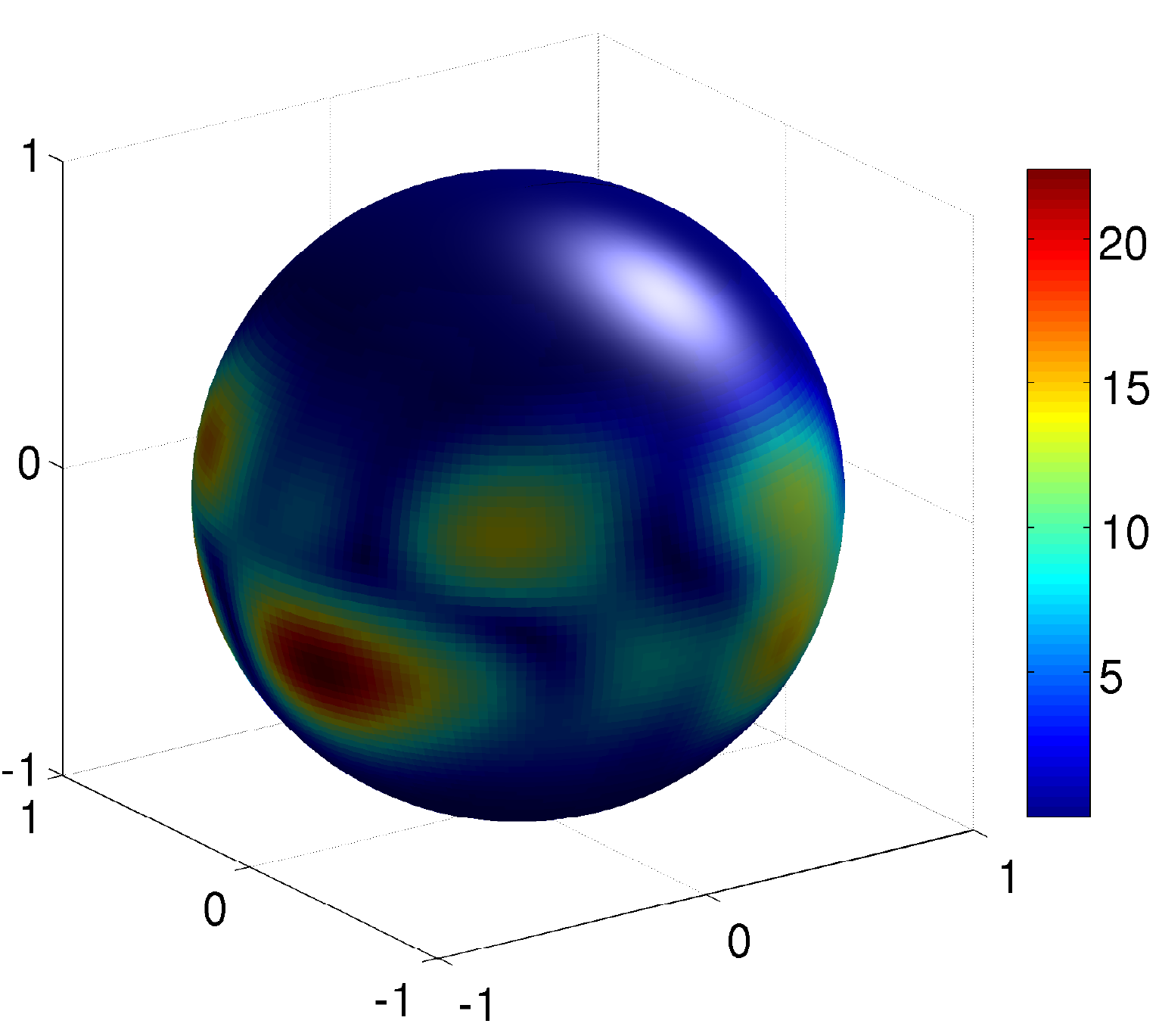}}
   \subfloat{
        \includegraphics[scale=0.18]{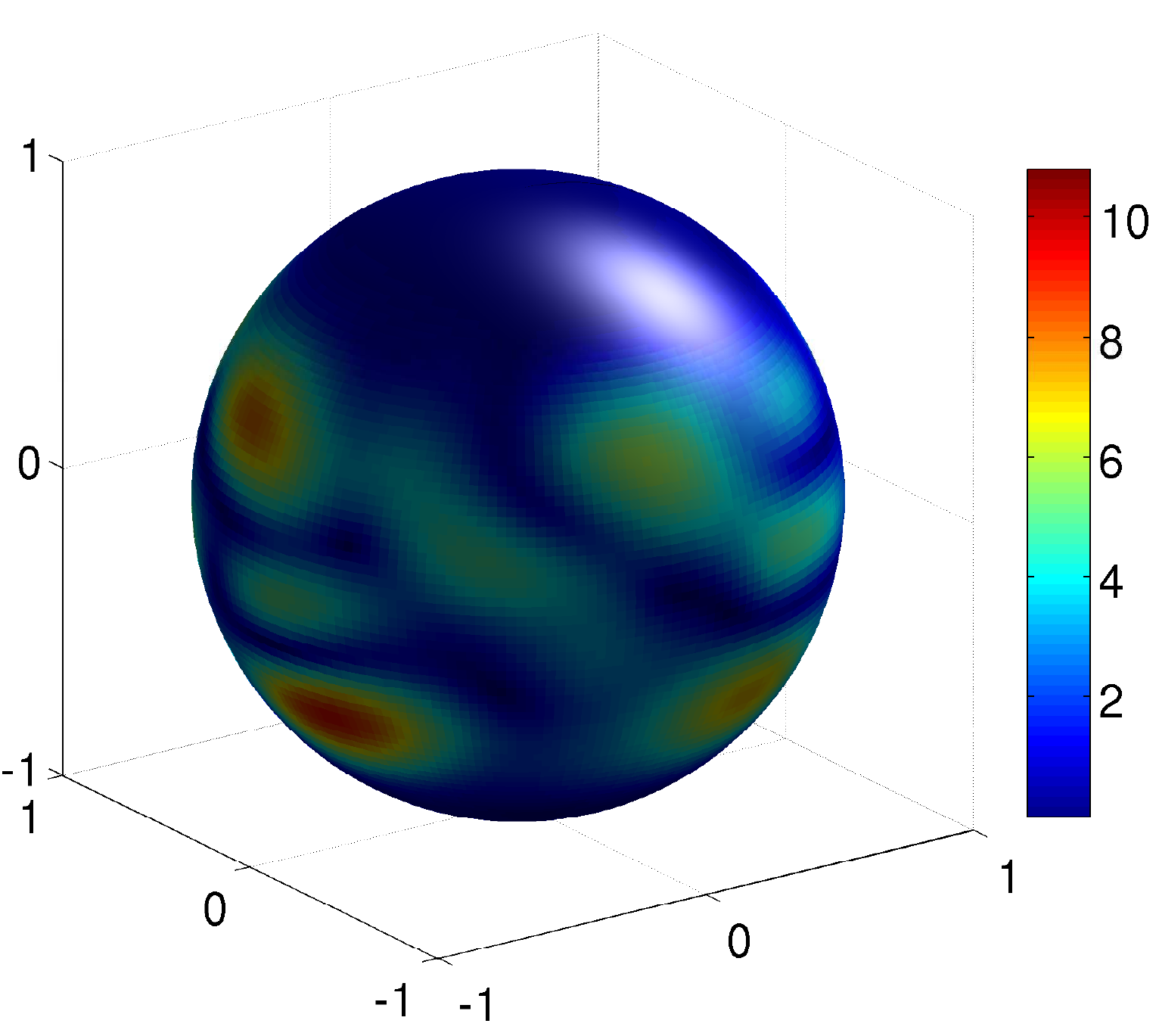}}
    \subfloat{
        \includegraphics[scale=0.18]{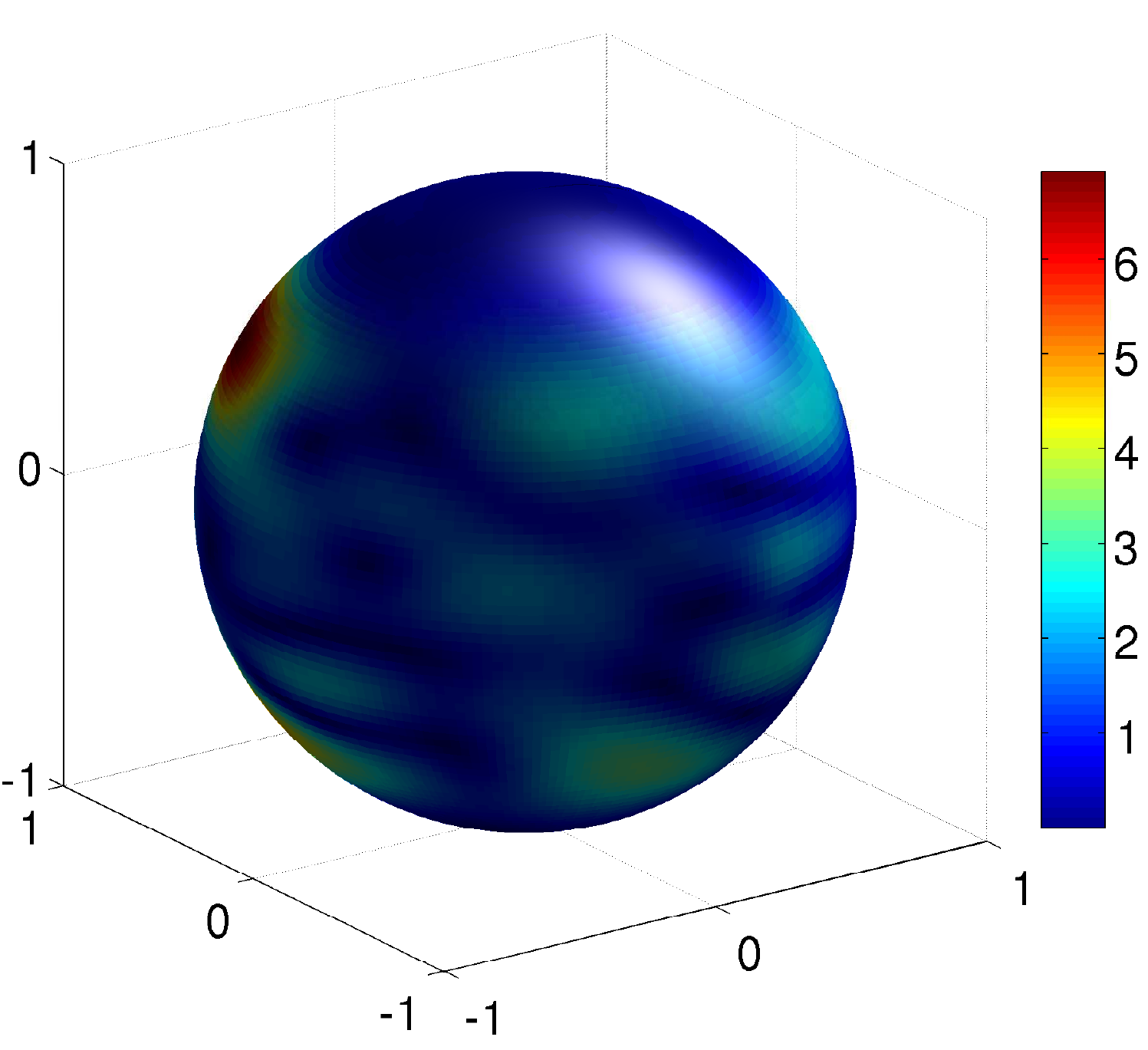}}

        \vspace{-3mm}
       \setcounter{subfigure}{2}
        \subfloat[]{}

    \hspace{-8mm}
    \subfloat{
        \includegraphics[scale=0.18]{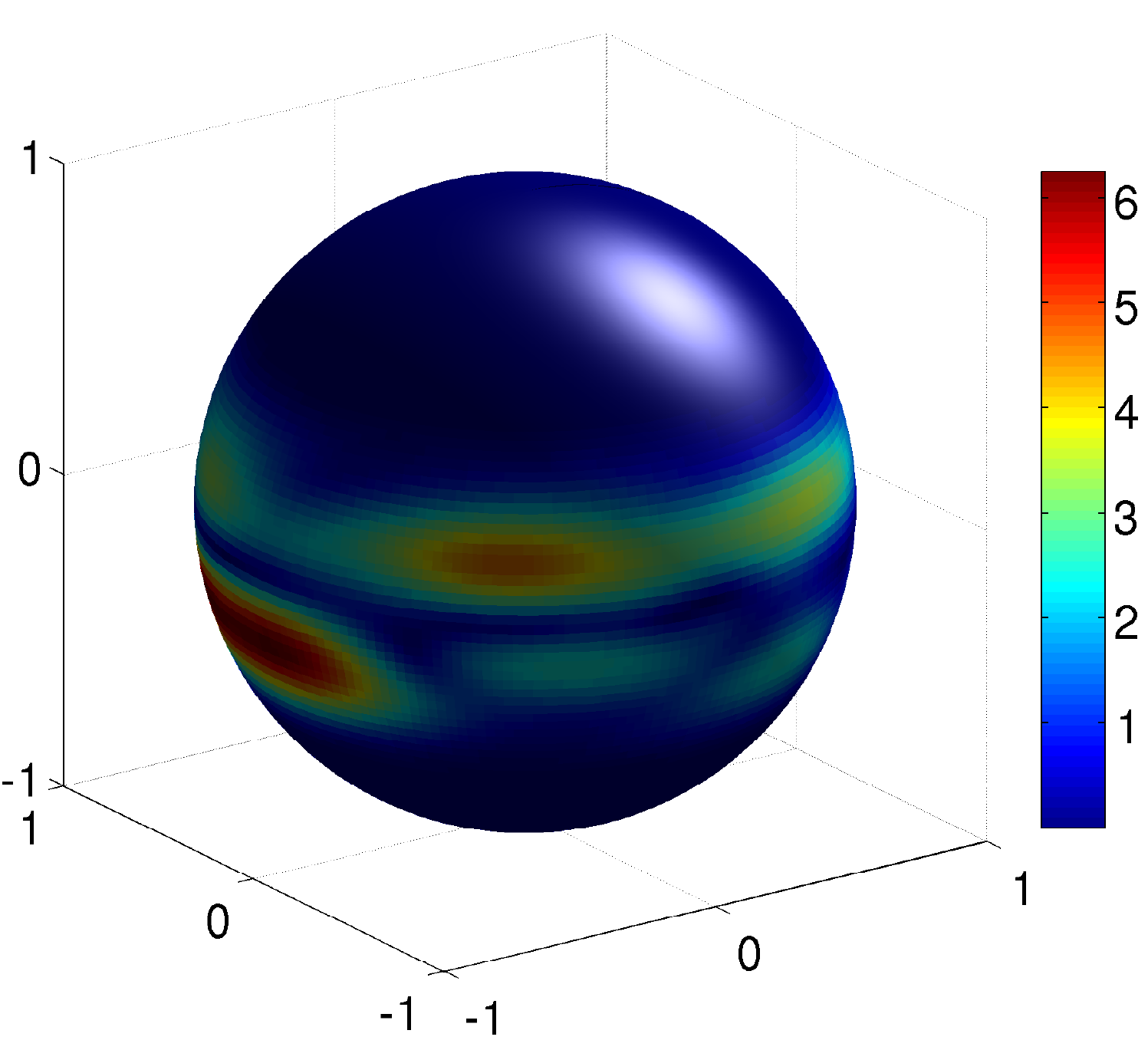}}
   \subfloat{
        \includegraphics[scale=0.18]{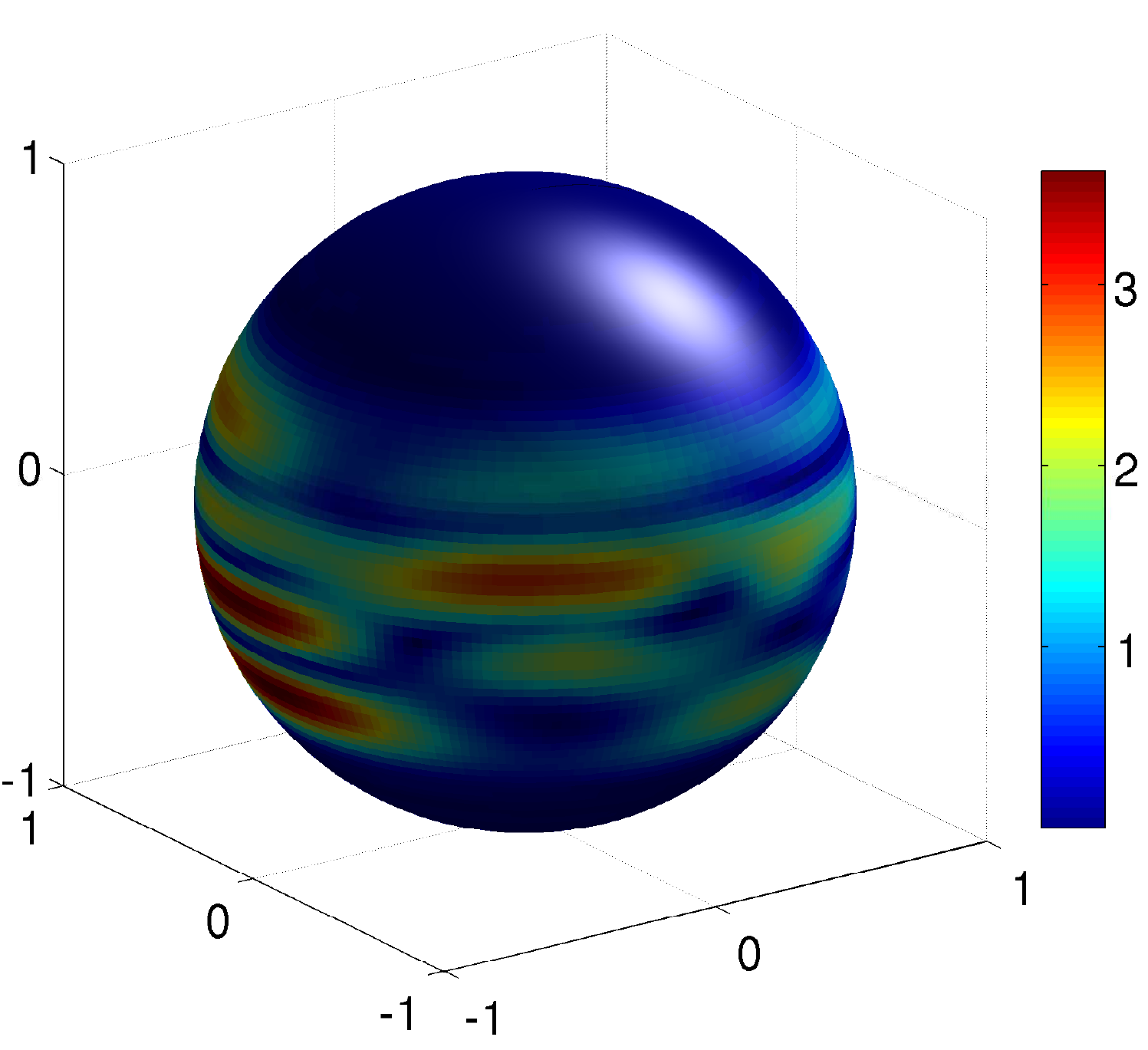}}
    \subfloat{
        \includegraphics[scale=0.18]{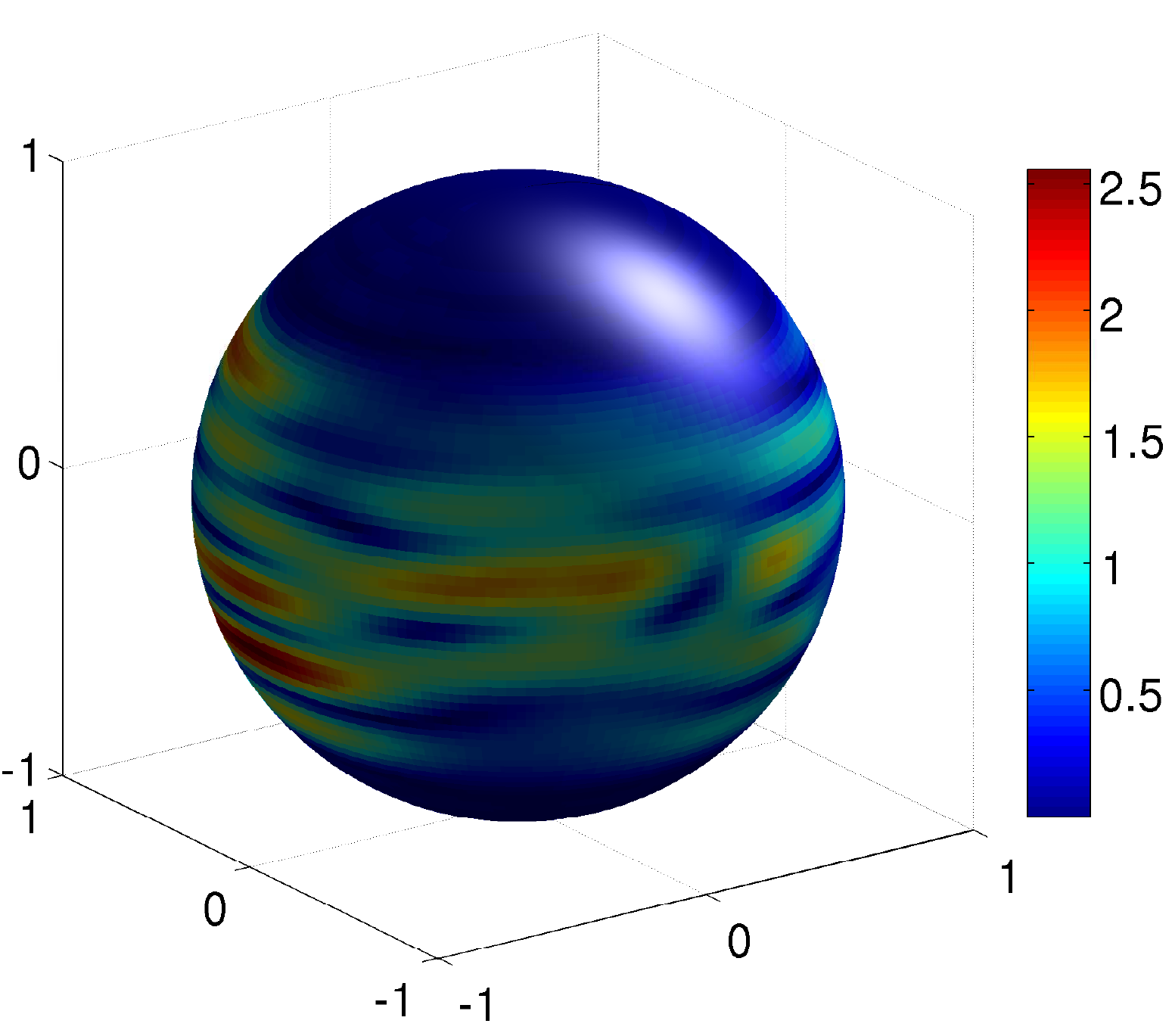}}

    \centering
        \vspace{-3mm}
    \setcounter{subfigure}{3}
    \subfloat[]{}

 \caption{Magnitude of the components of the
directional SLSHT distribution of the synthetic signal shown in
\figref{fig:syn_sig}c. For fixed orientation $\gamma$ of the window
function around $z$-axis, the distribution components
$g(\rho;\ell,m)$ are mapped on the sphere using $\rho =
(\phi,\theta,\gamma)$ for order $m=20$. The components are shown for
degrees $\ell\in \{41,43,45\}$ and for orientation (a) $\gamma=0$
and (b) $\gamma\approx\pi/2$ of the window function around $z$-axis,
and the components are shown for degrees $\ell\in \{21,23,25\}$ and
for orientation (c) $\gamma=0$ and (d) $\gamma\approx\pi/2$. Top
left: $g(\rho;41,20)$, top right: $g(\rho;45,20)$. }
\label{fig:syn_comps}\vspace{-0.25in}
\end{figure}

\subsubsection{Example 2 - Mars Data Set}
Now, we consider the Mars topographic map~(height above geoid) as a
signal on the sphere, which is obtained by using the spherical
harmonic model of the topography of
Mars\footnote{\url{http://www.ipgp.fr/~wieczor/SH/}}. The Mars
topographic map is shown in \figref{fig:mars_spatial} in the spatial
domain, where the grand canyon Valles Marineris and the mountainous
regions of Tharsis Montes and Olympus Montes are shown, leading to
the high frequency contents. We note that the mountainous regions
are non-directional features of the Mars map, whereas the grand
canyon serves as a directional feature with direction orientated
along a line of approximate constant latitude.

The directional SLSHT distribution components $g(\rho;\ell,m)$ of
the Mars map $f$ are obtained using the band-limited eigenfunction
window $h$ with $L_h=60$ and $90\%$ concentration in the spatial
domain in an elliptical region $\mathcal{R}_{(\pi/16,\pi/15)}$. The
magnitude of the SLSHT distribution components $g(\rho;\ell,m)$ for
order $m=15$ and degrees $80 \leq \ell \leq 85$ and $20 \leq \ell
\leq 25$ are shown in \figref{fig:mars_comps_1}a and
\figref{fig:mars_comps_1}b respectively for $\gamma\approx\pi/2$. It
is evident that using orientation of the window $\gamma\approx\pi/2$
probes the information about the grand canyon Valles
Marineris~(directional feature) along longitude in the
spatio-spectral domain. The localized contribution of higher degree
spherical harmonics towards the mountainous region can also be
observed in \figref{fig:mars_comps_1}b for degree $20\leq \ell \leq
25$. However, there is no significant contribution of spherical
harmonics of degree $80\leq\ell\leq 85$ towards mountainous region
as indicated in \figref{fig:mars_comps_1}a, but the localization of
the directional features along the orientation $\gamma\approx\pi/2$
is revealed in the spatio-spectral domain.

\ifCLASSOPTIONonecolumn
\begin{figure*}[t]
    \centering
    \includegraphics[scale=0.5]{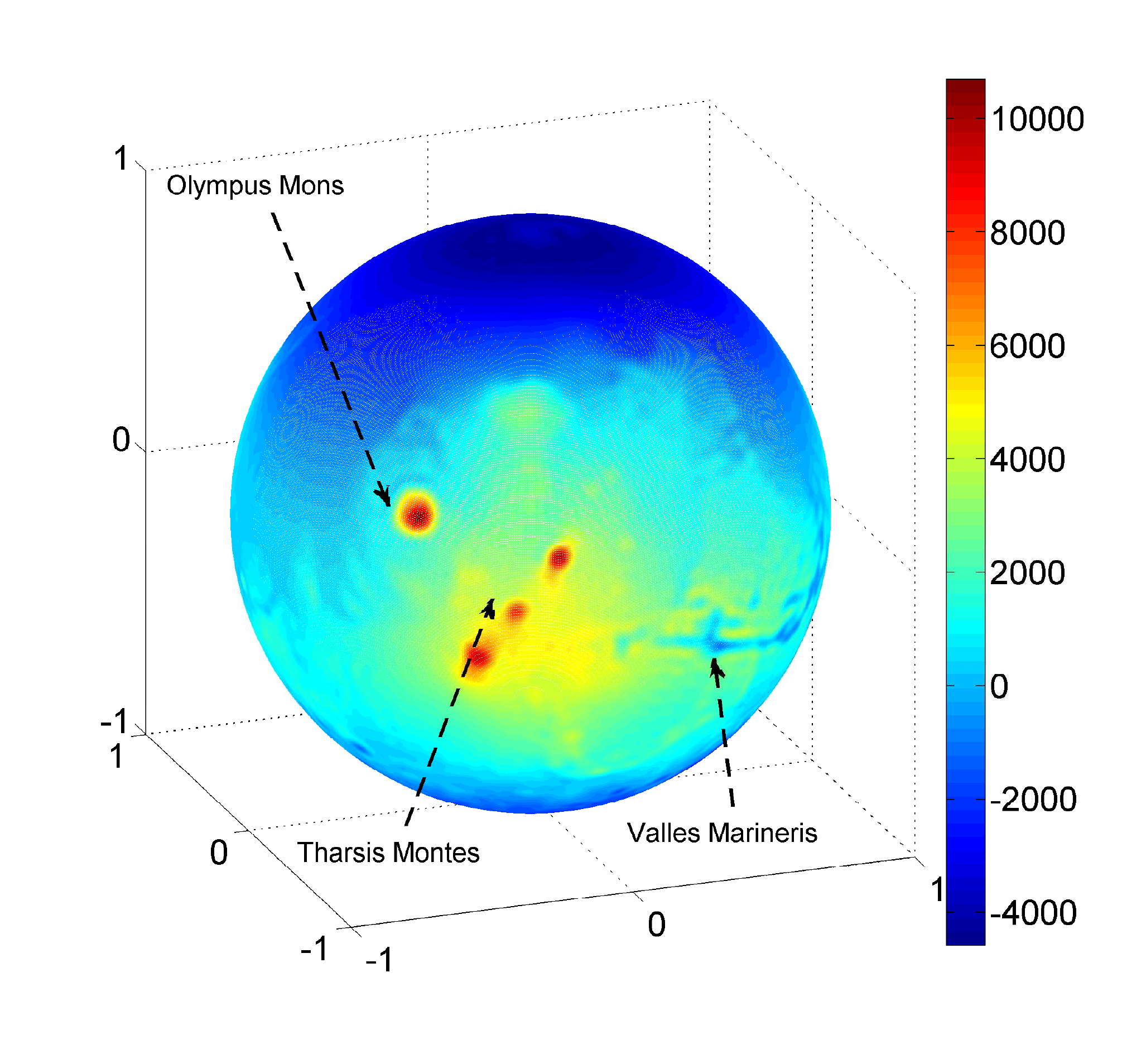}\vspace{-2mm}
    \caption{Mars signal in the spatial domain. The grand canyon Valles Marineris and the mountainous regions of Tharsis Montes and Olympus Montes are indicated. }
    \label{fig:mars_spatial}
\end{figure*}
\else
\begin{figure}[t]
    \centering
    \includegraphics[scale=0.28]{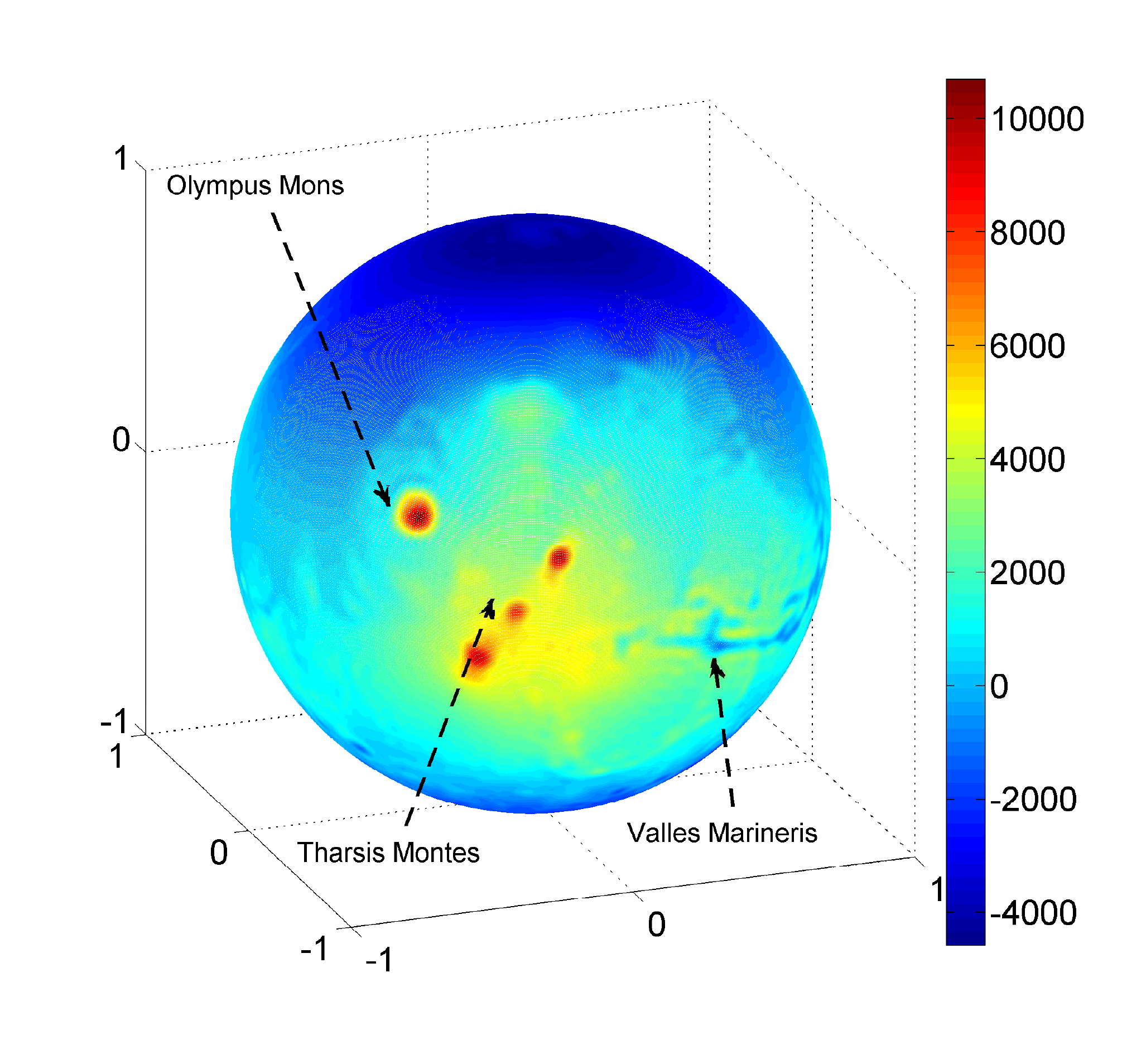}\vspace{-2mm}
    \caption{Mars signal in the spatial domain. The grand canyon Valles Marineris and the mountainous regions of Tharsis Montes and Olympus Montes are indicated. }
    \vspace{-4mm}
    \label{fig:mars_spatial}
\end{figure}
\fi
\begin{figure}[th]
 \centering
    \vspace{-5mm}
    \hspace{-2mm}
    \subfloat{
        \includegraphics[scale=0.19]{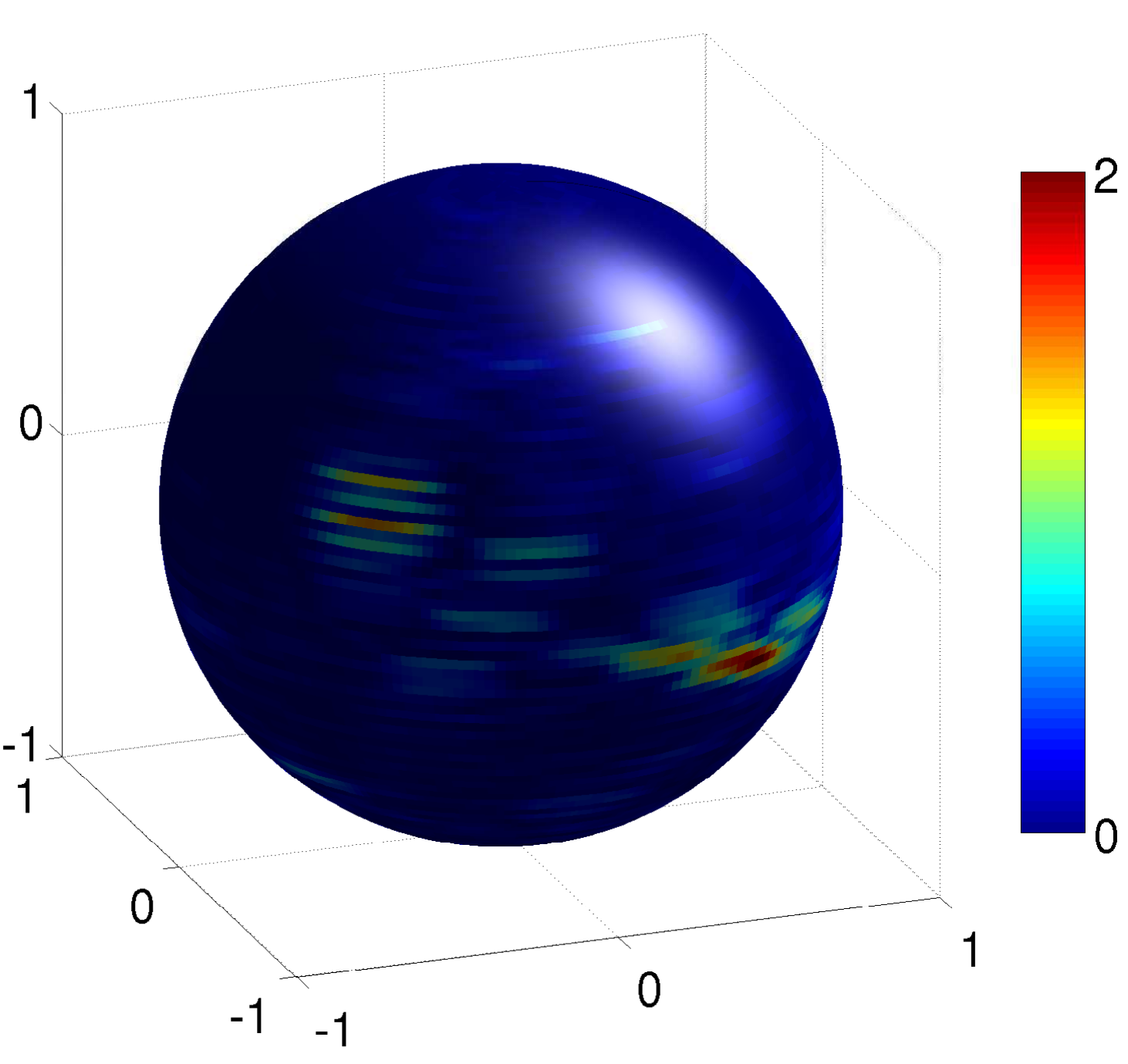}}
   \subfloat{
        \includegraphics[scale=0.19]{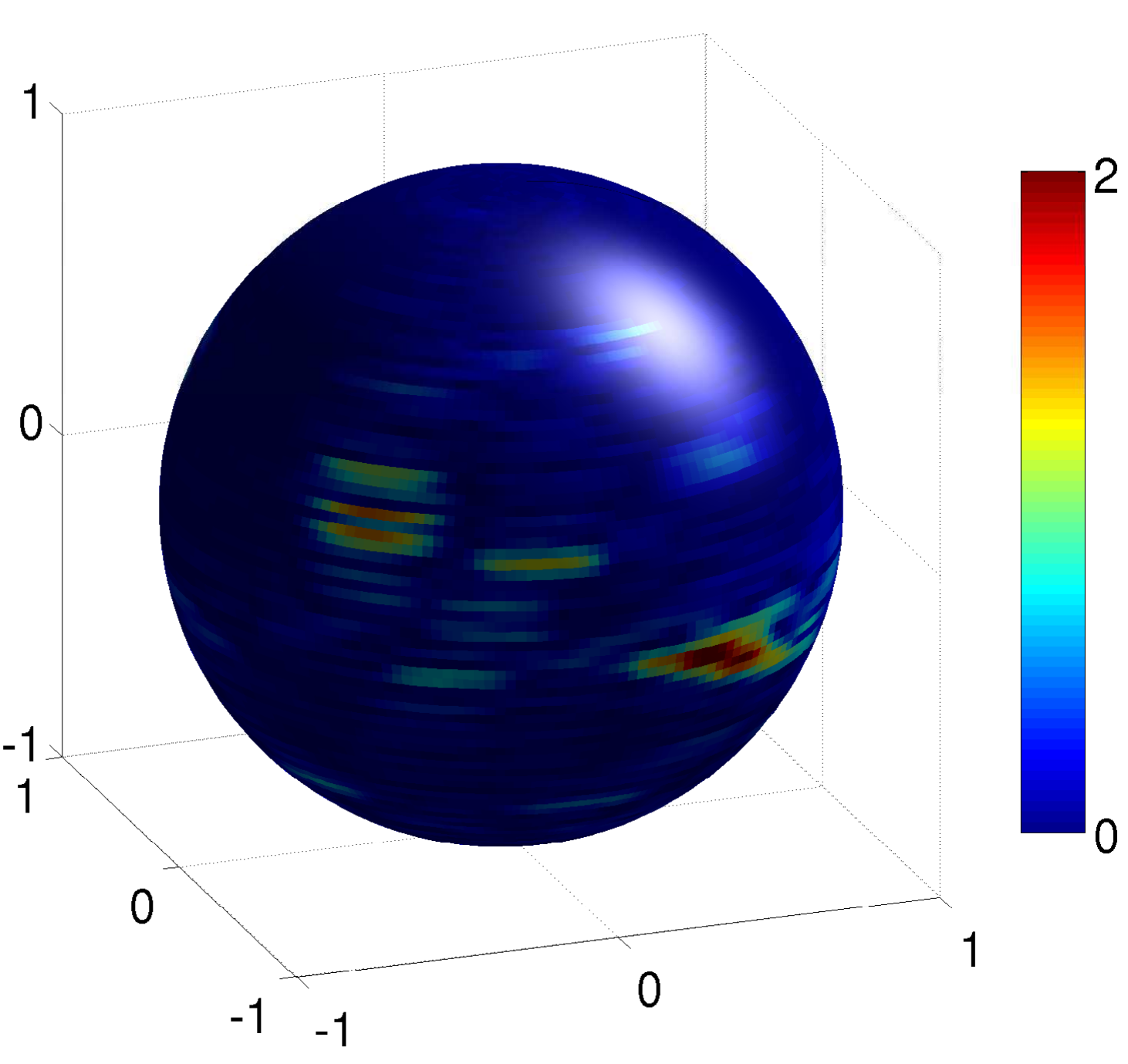}}
    \subfloat{
        \includegraphics[scale=0.19]{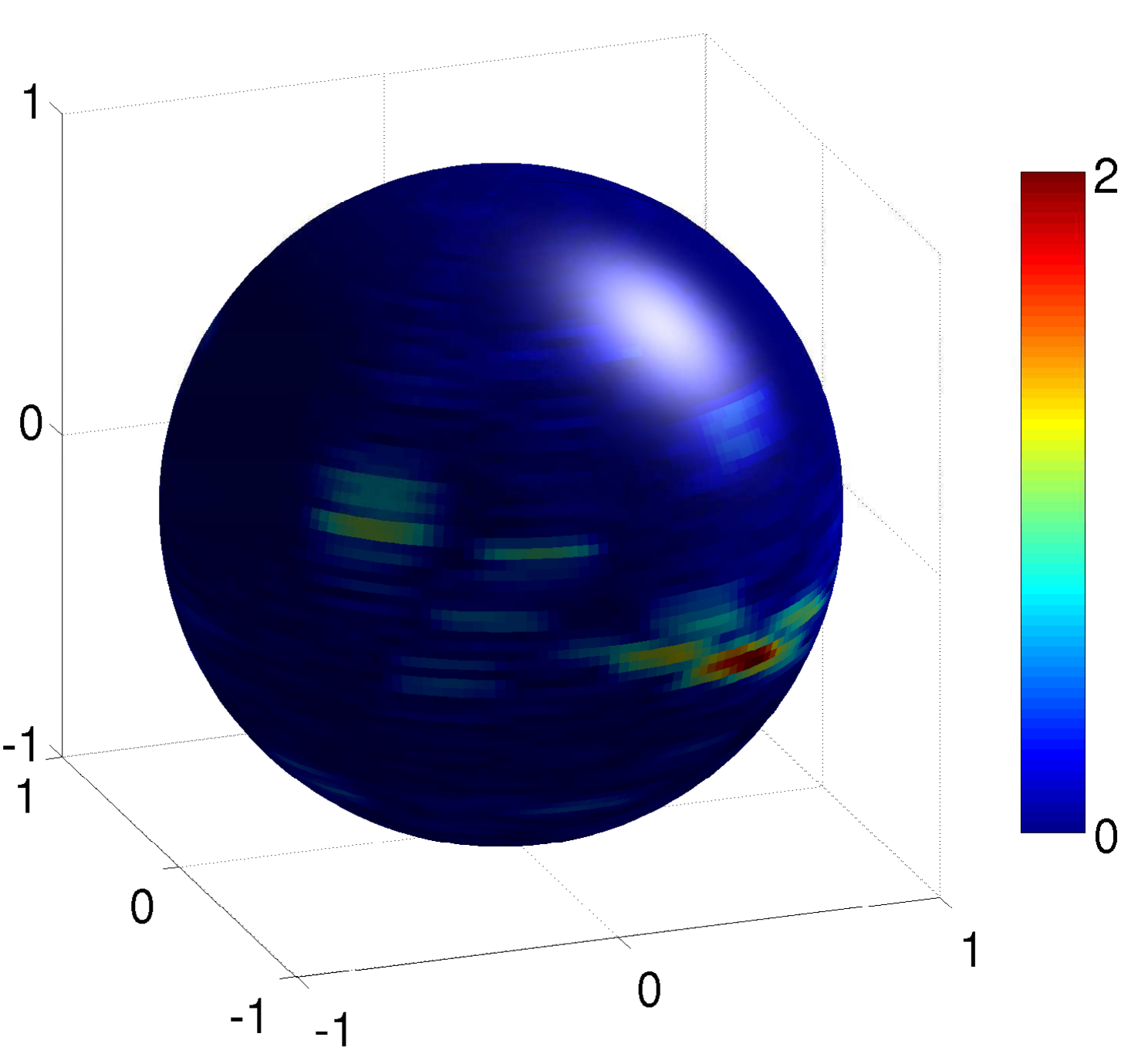}}

    \hspace{-3mm}
    \subfloat{
        \includegraphics[scale=0.19]{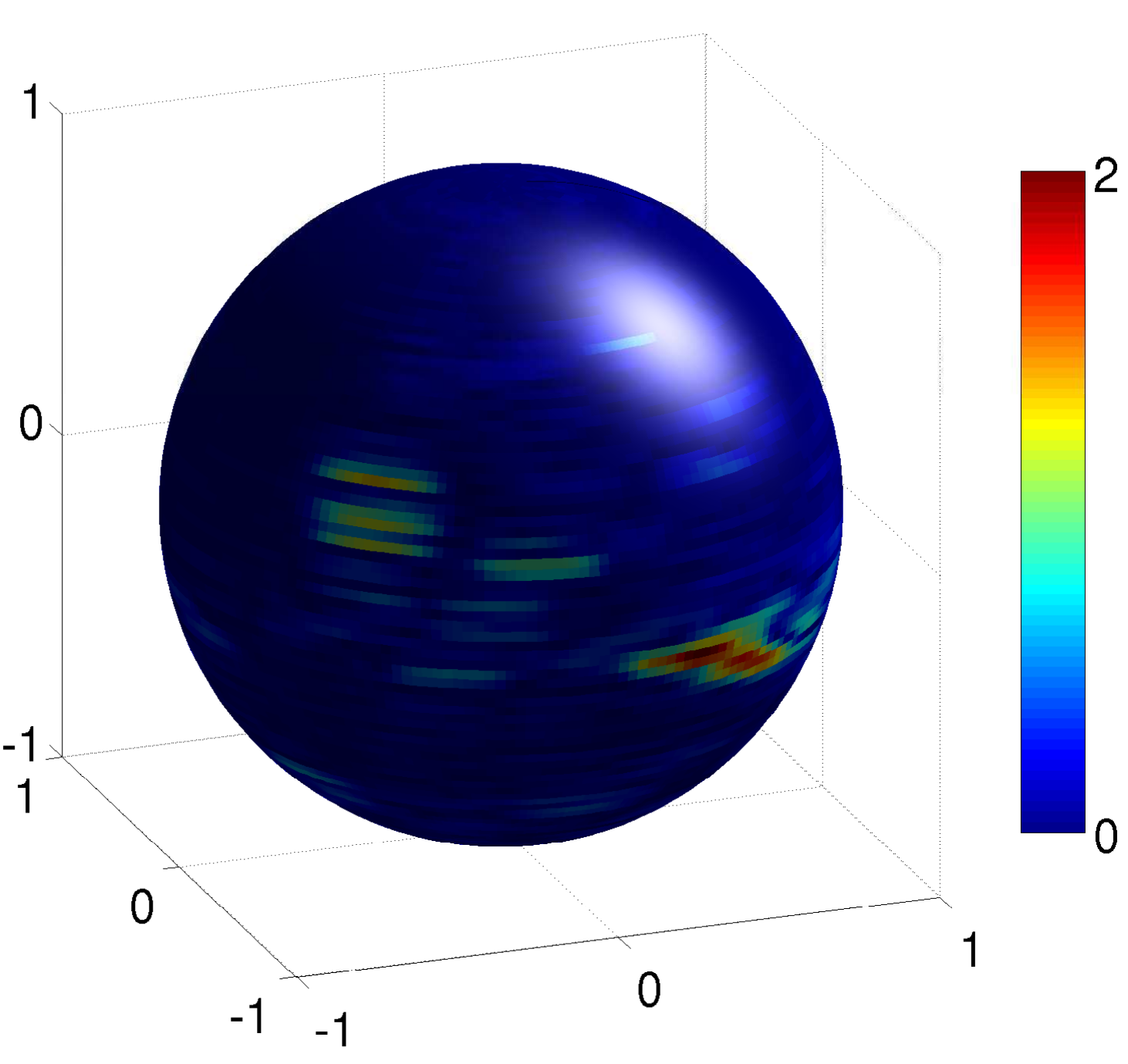}}
   \subfloat{
        \includegraphics[scale=0.19]{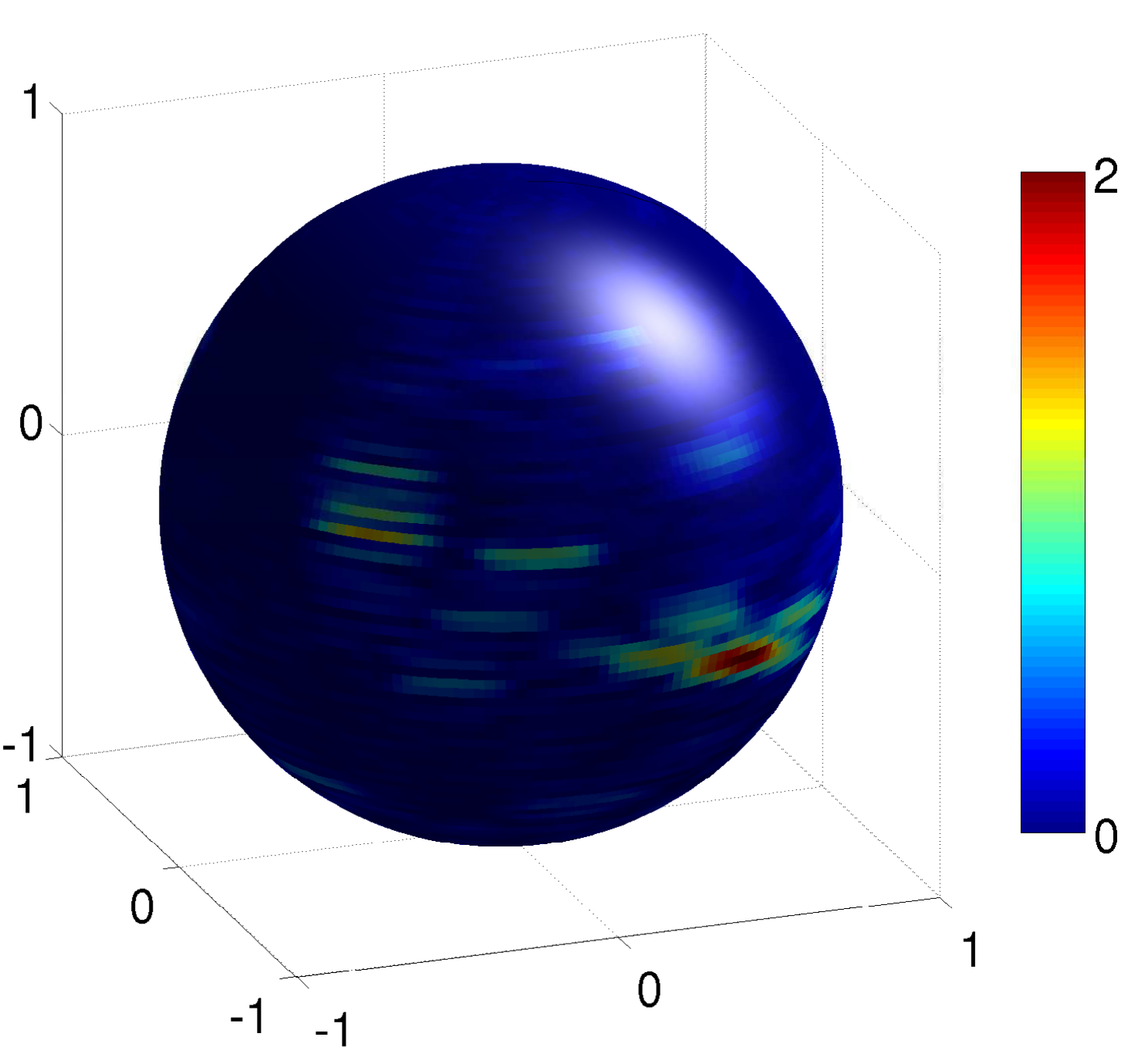}}
    \subfloat{
        \includegraphics[scale=0.19]{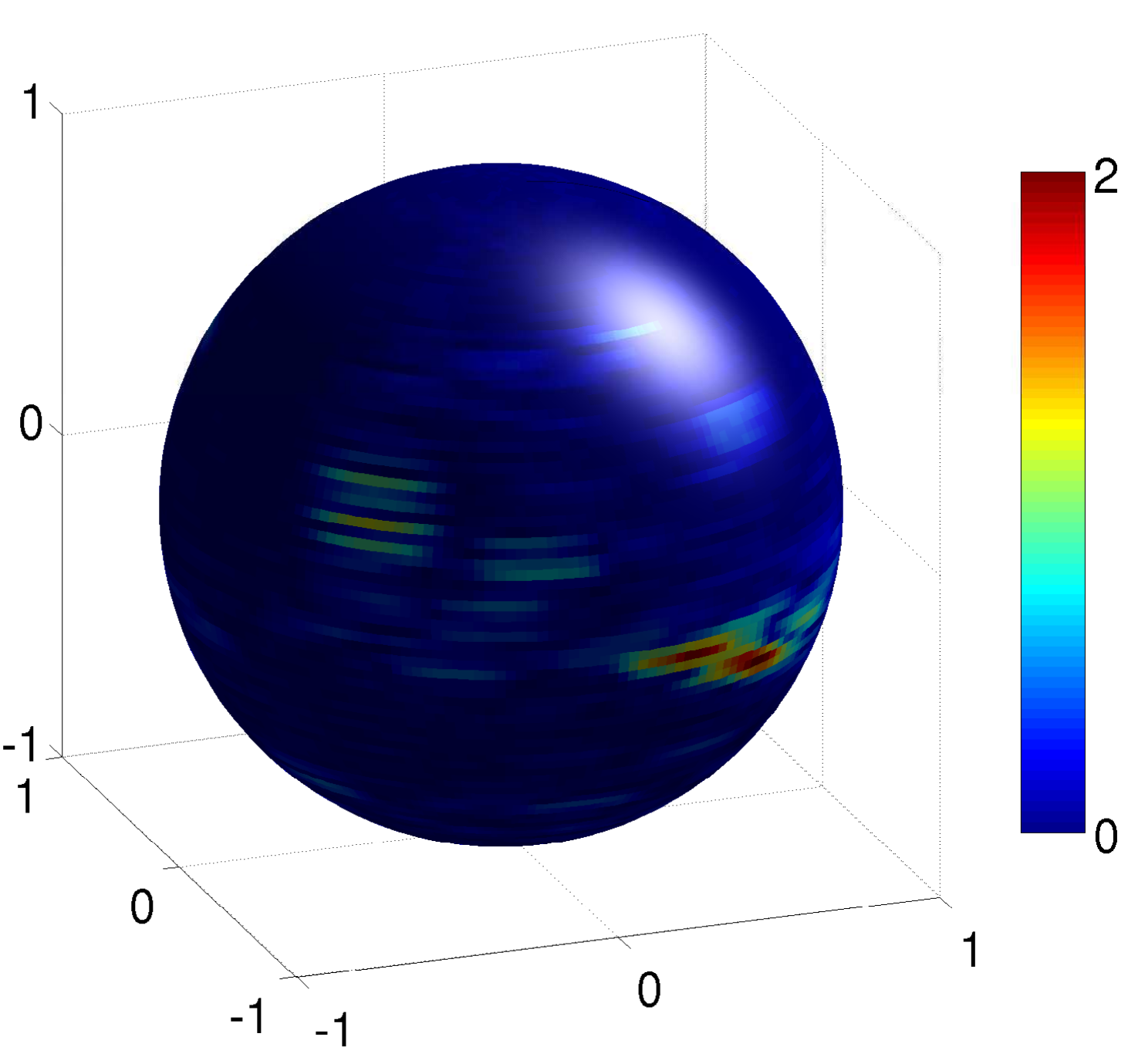}}
       \setcounter{subfigure}{0}

\ifCLASSOPTIONonecolumn
    \vspace{-10mm}
\else
    \vspace{-5mm}
\fi
   \subfloat[]{}

    \hspace{-2mm}
    \subfloat{
        \includegraphics[scale=0.19]{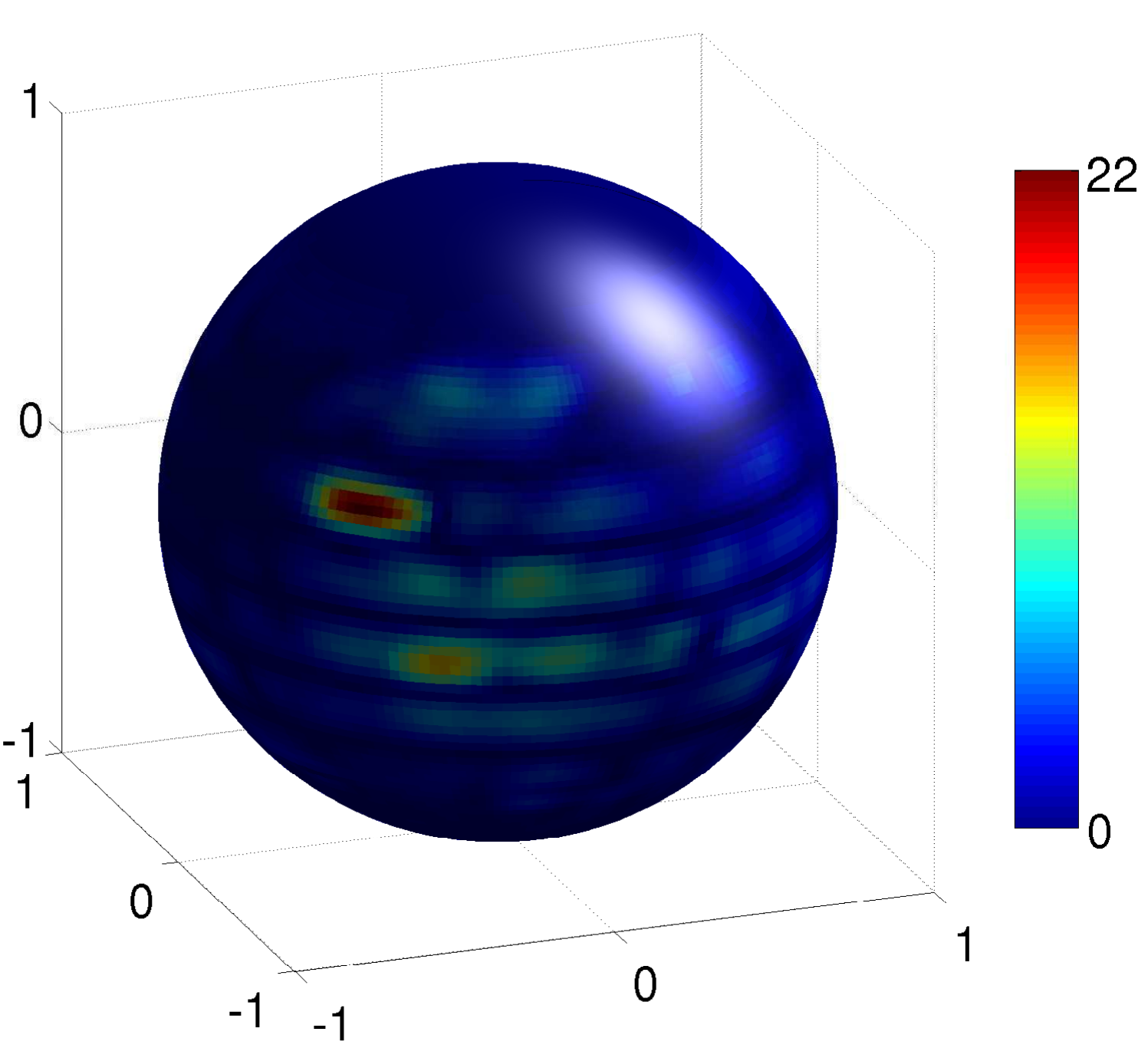}}
   \subfloat{
        \includegraphics[scale=0.19]{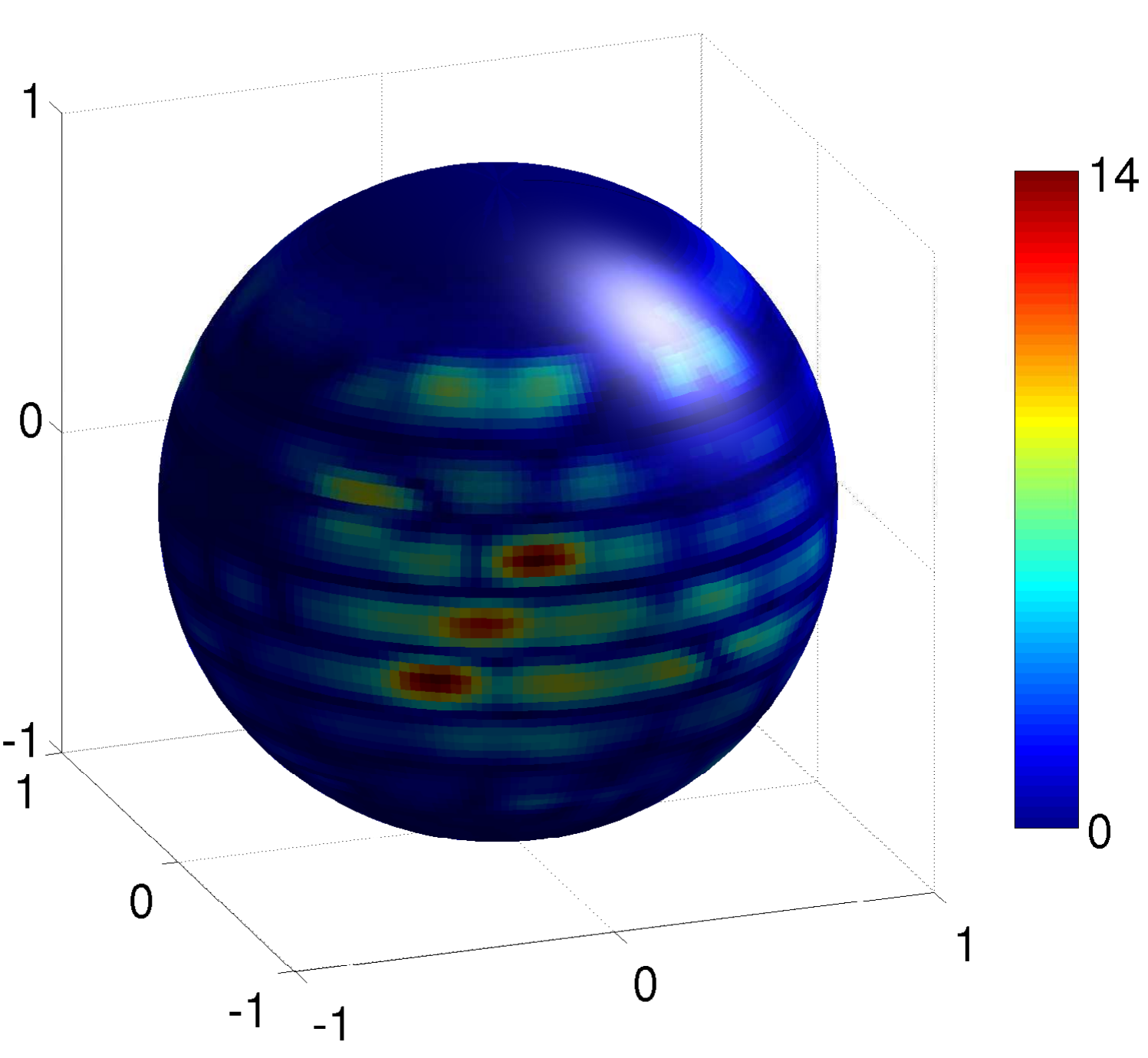}}
    \subfloat{
        \includegraphics[scale=0.19]{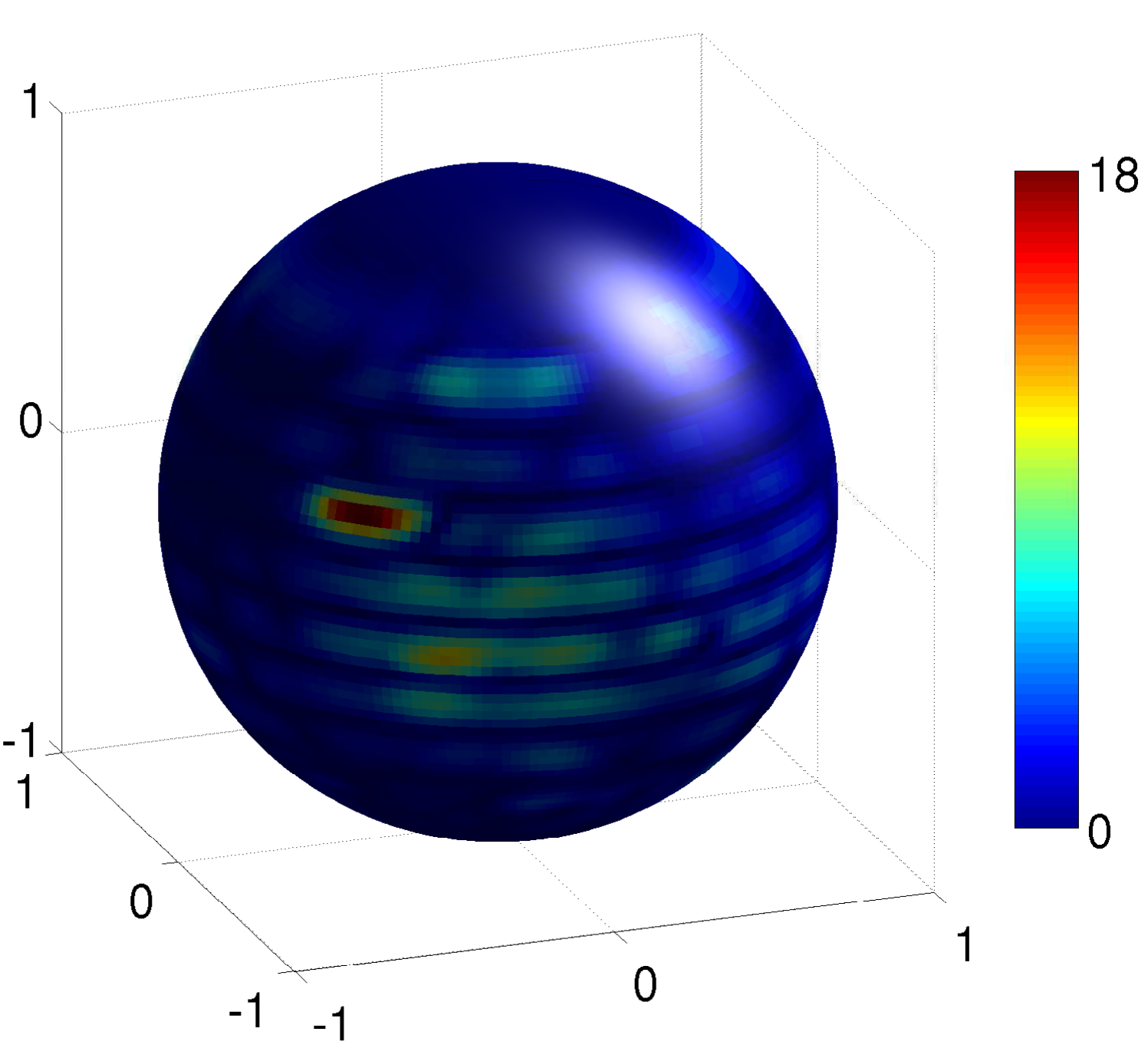}}

    \hspace{-3mm}
    \subfloat{
        \includegraphics[scale=0.19]{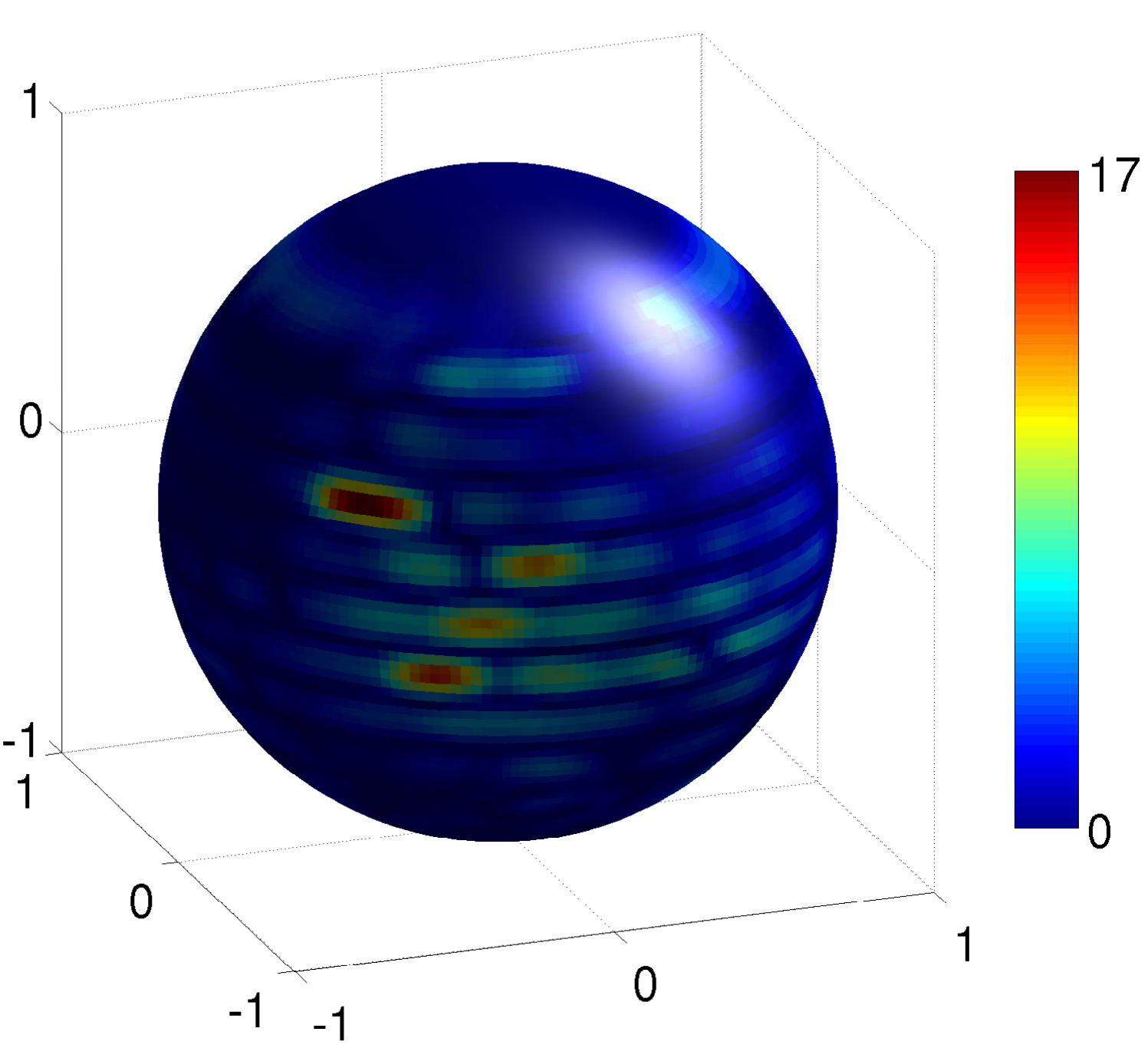}}
   \subfloat{
        \includegraphics[scale=0.19]{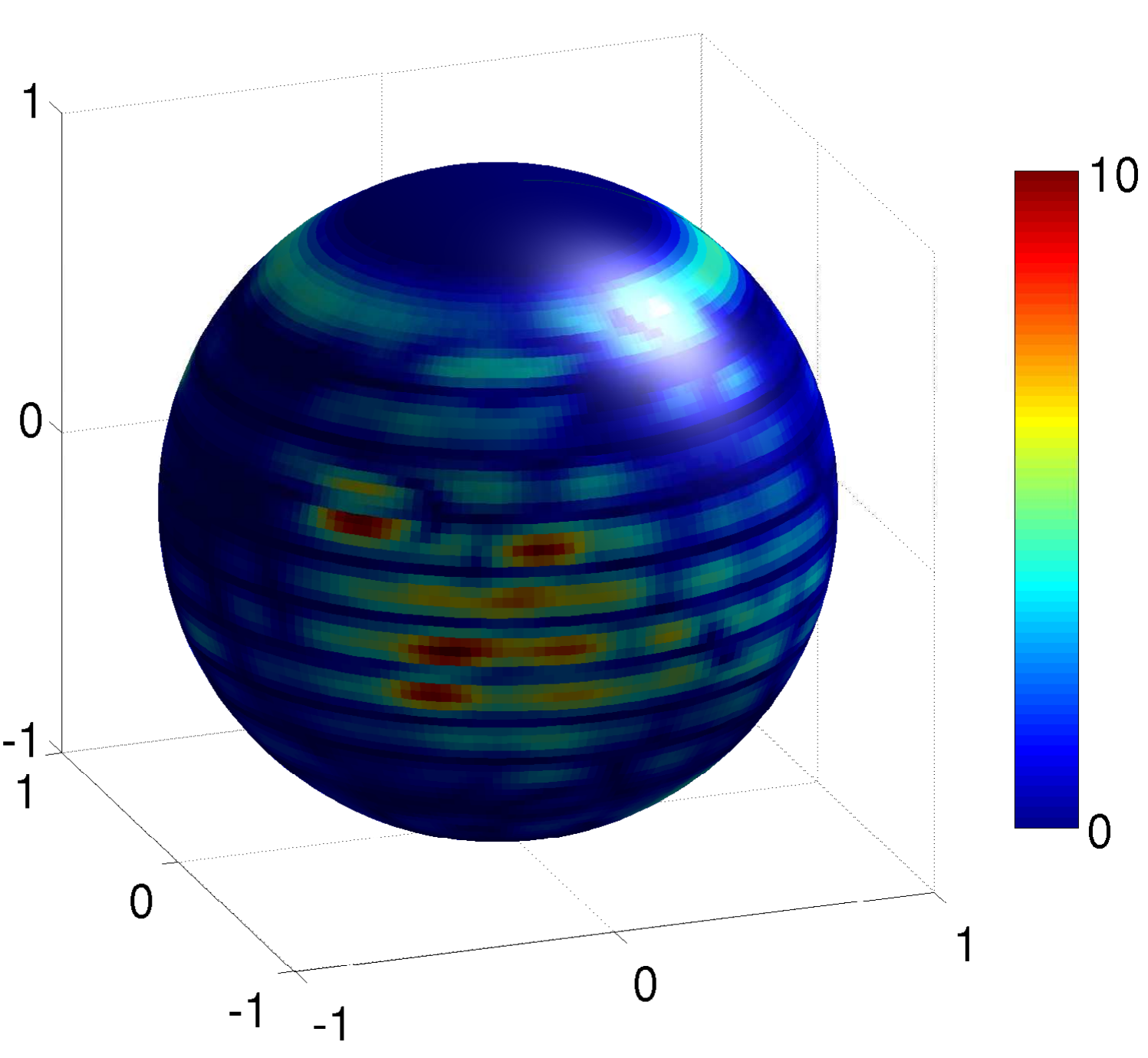}}
    \subfloat{
        \includegraphics[scale=0.19]{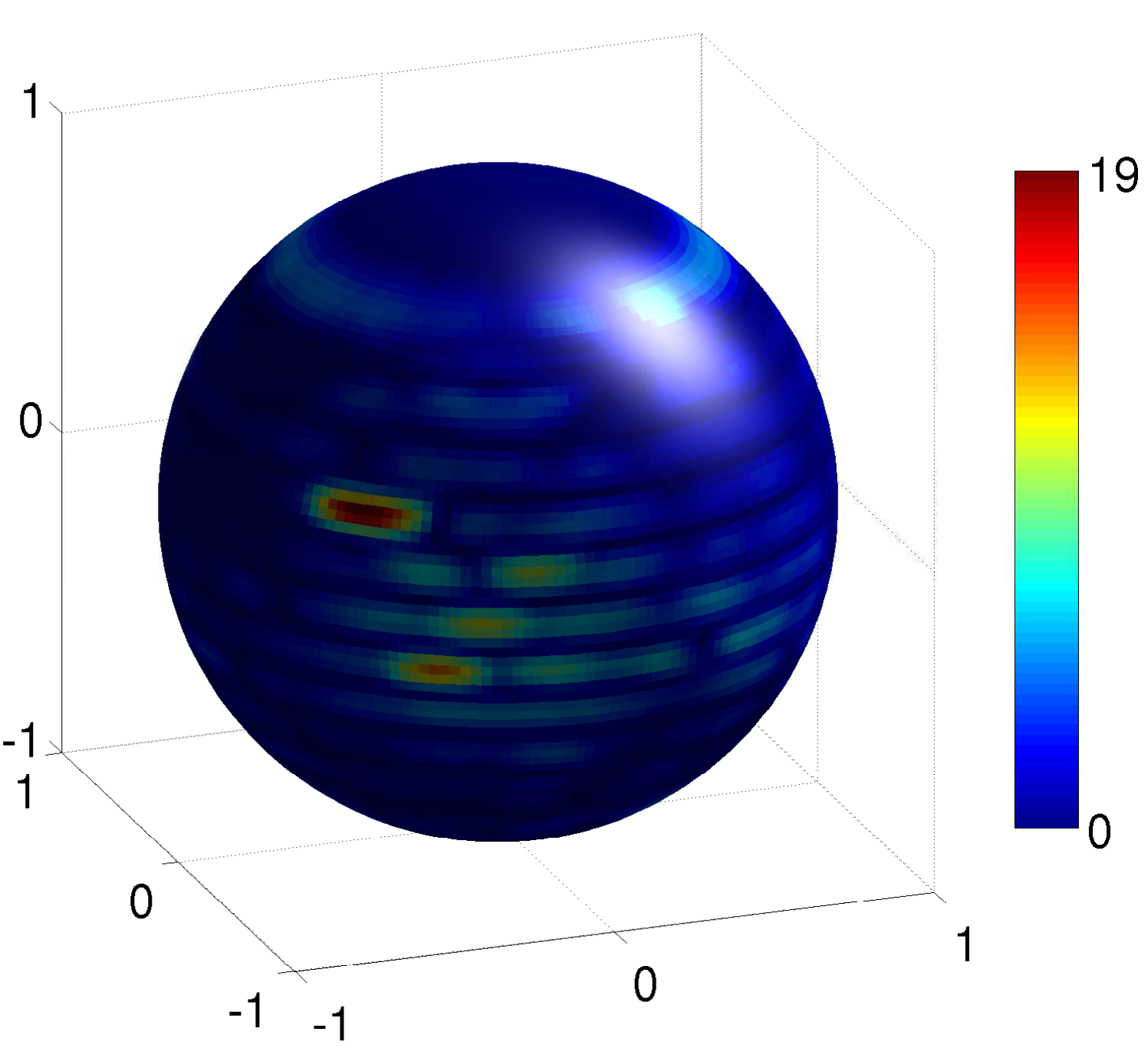}}
       \setcounter{subfigure}{1}

\ifCLASSOPTIONonecolumn
    \vspace{-10mm}
\else
    \vspace{-3mm}
\fi

   \subfloat[]{}

\caption{Magnitude of the components of the directional SLSHT
distribution of the Mars signal obtained using the eigenfunction
window concentrated in an elliptical region of focus
$\thetac=\pi/16$ and major axis $a=\pi/15$. For fixed orientation
$\gamma$, the distribution components $g(\rho;\ell,m)$ are mapped on
the sphere using $\rho = (\phi,\theta,\gamma)$ for order $m=15$ and
degrees (a) $80\leq \ell\leq85$ and (b) $20\leq \ell\leq 25$. The
components are shown for orientation $\gamma\approx\pi/2$ of the
window function around the $z$-axis. Top left: $g(\rho;20,15)$, top
right: $g(\rho;22,15)$. } \label{fig:mars_comps_1}\vspace{-0.25in}
\end{figure}

\section{Conclusions}\label{sec:conclusions}

We have presented the directional SLSHT to project a signal on the
sphere onto its joint spatio-spectral domain as a directional SLSHT
distribution. In spirit, the directional SLSHT is composed of SO(3)
spatial localization followed by the spherical harmonic transform.
Here, we have proposed the use of an azimuthally asymmetric window
function to obtain spatial localization, which enables the transform
to resolve directional features in the spatio-spectral domain. We have
also presented an inversion relation to synthesize the original signal
from its directional SLSHT distribution. Since data-sets on the sphere
are of considerable size, we have developed a fast algorithm for the
efficient computation of the directional SLSHT distribution of a
signal.  The computational complexity of computing the directional
SLSHT is reduced by providing an alternative harmonic formulation of
the transform and then exploiting the factoring of rotation
approach~\cite{Risbo:1996} and the fast Fourier transform.  The
computational complexity of the proposed fast algorithm to evaluate
SLSHT distribution of a signal with band-limit $L_f$ using window
function with band-limit $L_h$ is $O(L_f^3 L_h^2 + L_f^2 L_h^4)$ as
compared to the complexity of direct evaluation, which is
$O(L_f^4L_h^3)$. The numerical accuracy and the speed of our fast
algorithm has also been studied. The directional SLSHT distribution
relies on a window function for spatial localization; we have analyzed
the band-limited window function obtained from the Slepian
concentration problem on the sphere, with nominal concentration in an
elliptical region around the north pole. We provided an illustration
which highlighted the capability of the directional SLSHT to reveal
directional features in the spatio-spectral domain, which is likely to
be of use in many applications.

\begin{appendices}

\section{Spherical Harmonic Transform of Modulated Signal}\label{App:SSHT}
Our objective is to compute the spherical harmonic transform of the
modulated signal $\overline{f}{Y_\ell^m}$,
up to degree $L_h$, for all $\ell$ and $m$. In order to serve the purpose, we
use a separation variable technique given by
\ifCLASSOPTIONonecolumn
\begin{align}\label{Eq:sht_sep_variable}
\shc{\overline{f}\,Y_\ell^m}{p}{q}  \,= N_\ell^m N_p^q \int_0^\pi
P_\ell^m(\cos\theta) P_p^q (\cos\theta) \underbrace{ \int_0^{2\pi}
\overline{f(\theta,\phi)}e^{i(m-q)\phi} d\phi}_{I(\theta,m-q)} \sin\theta
d\theta.
\end{align}
\else
\begin{align}\label{Eq:sht_sep_variable}
\shc{\overline{f}\,Y_\ell^m}{p}{q}  \,&= N_\ell^m N_p^q \int_0^\pi
P_\ell^m(\cos\theta) P_p^q (\cos\theta) \nonumber \\ &\quad \times
\underbrace{ \int_0^{2\pi}
\overline{f(\theta,\phi)}e^{i(m-q)\phi} d\phi}_{I(\theta,m-q)} \sin\theta
d\theta.
\end{align}
\fi Since $0\leq \ell \leq L_f + L_h$ and $0\leq p \leq L_h$, we
need to consider the signal $\overline{f}\,Y_\ell^m$ sampled on the
grid $\mathfrak{S}_{2L_f+2L_h}$ for the explicit evaluation of exact
quadrature (note that sampling in $\phi$ could be optimized given
\mbox{$|m-q|\leq L_f+2L_h$} but this would require a different
tessellation of the sphere and will not alter the overall complexity
of the computation). Using \eqref{Eq:sht_sep_variable}, the integral
over $\phi$, giving $I(\theta,m-q)$, can be computed first in
$O(L_f^2 \log_2 L_f)$ for all $m-q$. Once $I(\theta,m-q)$ is
computed, the exact quadrature weights that follow from
\cite{McEwen:2011} can be used to evaluate the integral over
$\theta$ in $O(L_f)$ for each $p,\,q,\, \ell,\,m$ and in $O(L_f
L_h^2)$ for all $p,\,q,$ and each $ \ell,\,m$.  Thus the overall
complexity to compute the spherical harmonic transform of the
modulated signal ${\overline{f}\,Y_\ell^m}$ up to degree $L_h$ is
$O(L_f^2 \log_2 L_f + L_f L_h^2) $ for each $\ell,\,m$ and  $O(L_f^2
\log_2 L_f + L_f^3 L_h^2) = O(L_f^3 L_h^2)$ for all $\ell,\,m$.

\end{appendices}
\ifCLASSOPTIONonecolumn \else
 \renewcommand{\baselinestretch}{0.98}  
 \fi
\bibliography{SLSHT_new} 

\ifCLASSOPTIONonecolumn \else
 \renewcommand{\baselinestretch}{1} 
 \fi
\begin{IEEEbiography}[{\includegraphics[width=1in,height=1.25in,clip,keepaspectratio]{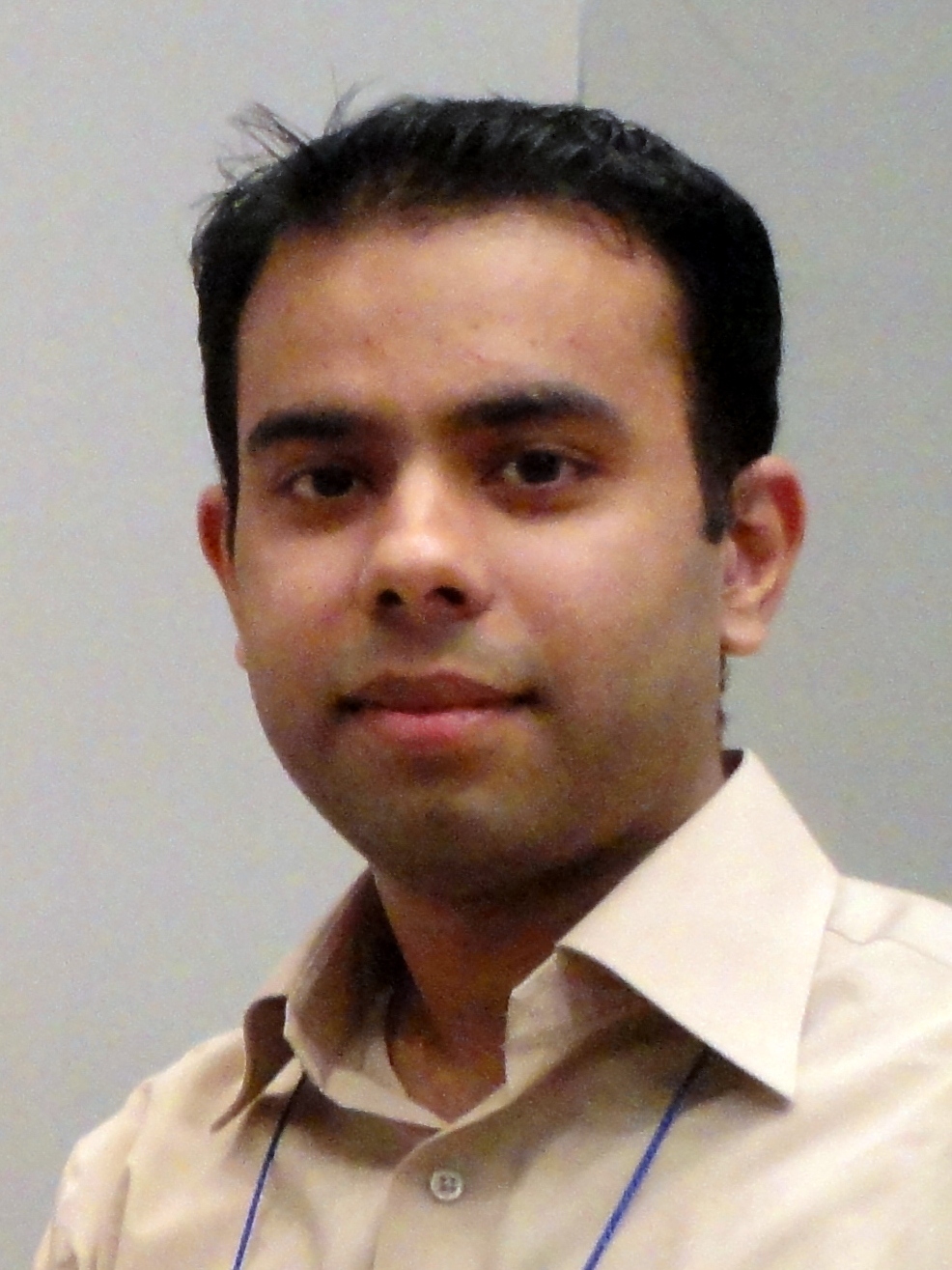}}]{Zubair Khalid}
(S'10) received his B.Sc. (first-class hons.) degree in
Electrical Engineering from the University of Engineering $\&$
Technology (UET), Lahore, Pakistan in 2008. He is currently pursuing
his PhD degree from the Research School of Engineering, the
Australian National University, Canberra, Australia. Zubair was
awarded University Gold Medal and Industry Gold Medals from Siemens
and Nespak for his overall outstanding performance in Electrical
Engineering during the his undergraduate studies. He is a recipient
of an Endeavour International Postgraduate Award for the duration of
his Ph.D. He was also awarded an ANU Vice Chancellor's Higher Degree
Research (HDR) travel grant in 2011. His research interests are in
the area of development of novel signal processing techniques for
signals on the sphere.
\end{IEEEbiography}
\begin{IEEEbiography}[{\includegraphics[width=1in,height=1.25in,clip,keepaspectratio]{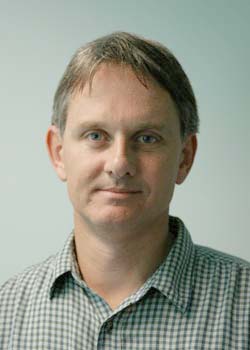}}]{Rodney A. Kennedy}
(S'86-M'88-SM'01-F'05) received the B.E. degree from the University of New South Wales, Sydney, Australia, the M.E. degree from the University of Newcastle, and the Ph.D. degree from the Australian National University, Canberra.

He is currently a Professor in the Research School of Engineering, Australian National University. He is a Fellow of the IEEE. His research interests include digital signal processing, digital and wireless communications, and acoustical signal processing.
\end{IEEEbiography}
\begin{IEEEbiography}[{\includegraphics[width=1in,height=1.25in,clip,keepaspectratio]{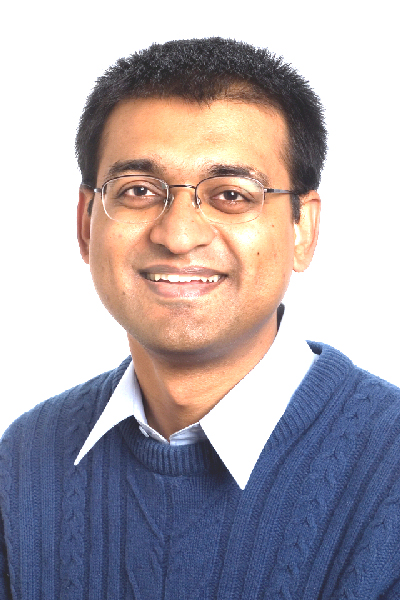}}]{Salman Durrani}
(S'00-M'05-SM'10) received the B.Sc. (1st class honours) degree in Electrical Engineering from the University of Engineering $\&$ Technology, Lahore, Pakistan in 2000. He received the PhD degree in Electrical Engineering from the University of Queensland, Brisbane, Australia in Dec. 2004. He has been with the Australian National University, Canberra, Australia, since 2005, where he is currently a Senior Lecturer in the Research School of Engineering, College of Engineering $\&$ Computer Science.

His current research interests are in wireless communications and signal processing, including synchronization in cooperative communication systems, connectivity of ad-hoc networks and vehicular networks and signal processing on the unit sphere. He serves as a Technical Program Committee Member for international conferences such as ICC '13, PIMRC '12 and AusCTW '12. He was awarded an ANU Vice-Chancellor's Award for Teaching Excellence in 2012. He was a recipient of an International Postgraduate Research Scholarship from the Australian Commonwealth during 2001-2004. He was awarded a University Gold Medal during his undergraduate studies. He has 55 publications to date in refereed international journals and conferences. He is a Member of Institution of Engineers, Australia and a Senior Member of IEEE.
\end{IEEEbiography}
\begin{IEEEbiography}[{\includegraphics[width=1in,height=1.25in,clip,keepaspectratio]{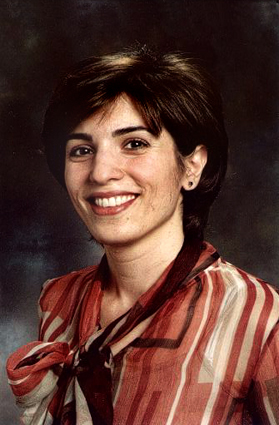}}]{Parastoo Sadeghi}
(S'02-M'06-SM'07) received the B.E. and M.E. degrees in
electrical engineering from
Sharif University of Technology, Tehran, Iran, in 1995 and 1997,
respectively, and the Ph.D. degree
in electrical engineering from The University of New South Wales,
Sydney, Australia, in 2006. From
1997 to 2002, she worked as a Research Engineer and then as a Senior
Research Engineer at Iran
Communication Industries (ICI) in Tehran, Iran and at Deqx (formerly
known as Clarity Eq) in Sydney,
Australia. She is currently a Fellow at the Research School of
Engineering, The
Australian National University, Canberra, Australia. She has visited
various
research institutes, including the Institute for Communications
Engineering,
Technical University of Munich, from April to June 2008 and MIT from
February to May 2009.

Dr. Sadeghi has co-authored more than 80 refereed journal or conference
papers and is a Chief Investigator in a number of Australian Research
Council Discovery and Linkage Projects. In 2003 and 2005, she received
two IEEE Region 10 student paper awards for her research in the
information theory of time-varying fading channels. Her research
interests are mainly in the area of wireless communications systems and
signal processing.
\end{IEEEbiography}

\begin{IEEEbiography}[{\includegraphics[width=1in,height=1.25in,clip,keepaspectratio]
 {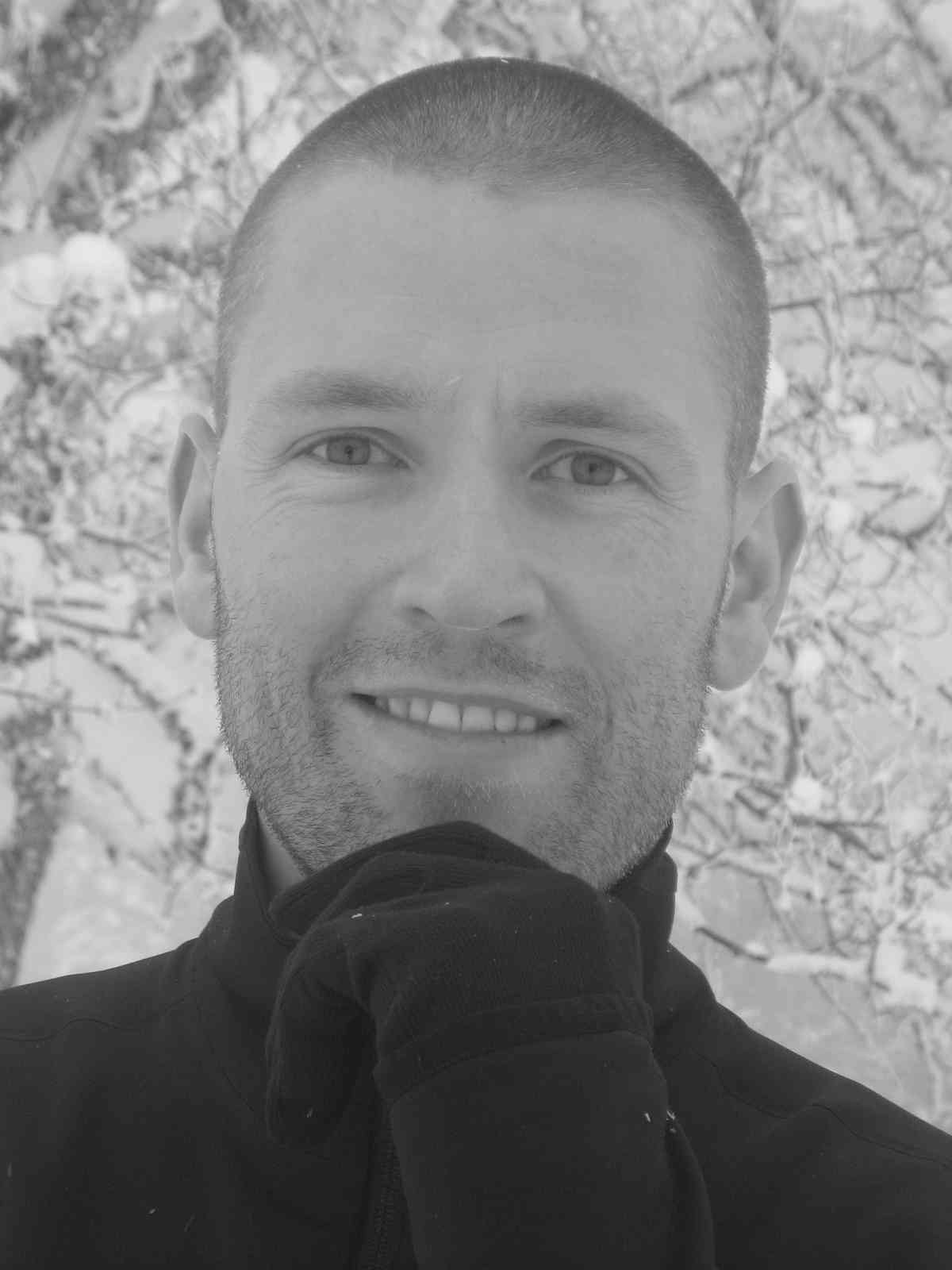}}]{Yves Wiaux}
  received the M.S. degree in physics and the Ph.D. degree in
  theoretical physics from the Universit\'e catholique de Louvain
  (UCL), Louvain-la-Neuve, Belgium, in 1999 and 2002, respectively.

  He was a Postdoctoral Researcher at the Signal Processing
  Laboratories of the Ecole Polytechnique F\'ed\'erale de Lausanne
  (EPFL), Switzerland, from 2003 to 2008. He was also a Postdoctoral
  Researcher of the Belgian National Science Foundation (F.R.S.-FNRS)
  at the Physics Department of UCL from 2005 to 2009. He is now a
  Ma\^itre Assistant of the University of Geneva (UniGE), Switzerland,
  with joint affiliation between the Institute of Electrical
  Engineering and the Institute of Bioengineering of EPFL, and the
  Department of Radiology and Medical Informatics of UniGE. His
  research lies at the intersection between complex data processing
  (including development on wavelets and compressed sensing) and
  applications in astrophysics (notably in cosmology and radio
  astronomy) and in biomedical sciences (notably in structural and
  diffusion MRI).
\end{IEEEbiography}

\begin{IEEEbiography}[{\includegraphics[width=1in,height=1.25in,clip,keepaspectratio]
 {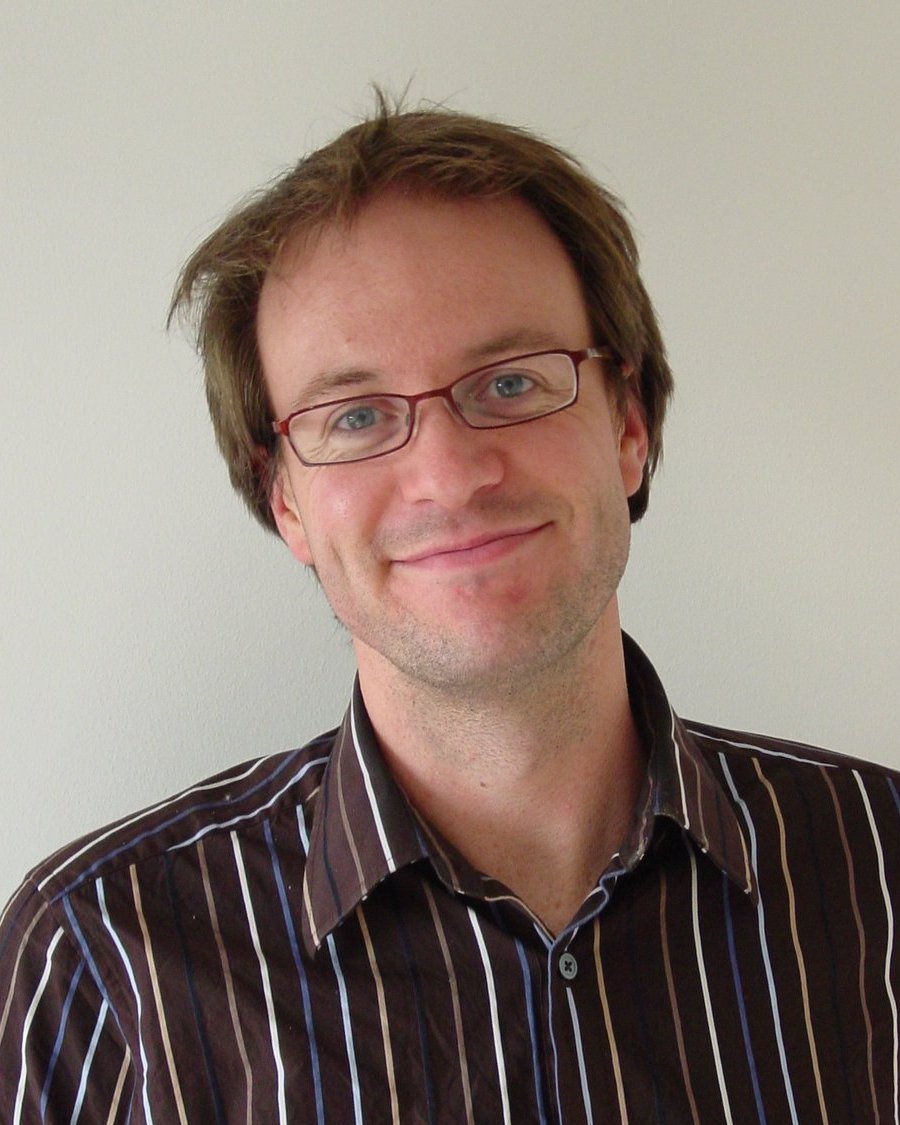}}]{Jason McEwen}
received a B.E.\ (Hons) degree in Electrical and Computer Engineering
from the University of Canterbury, New Zealand, in 2002 and a Ph.D.\
degree in Astrophysics from the University of Cambridge in 2007.

He held a Research Fellowship at Clare College, Cambridge, from 2007
to 2008, worked as a Quantitative Analyst from 2008 to 2010, and held
a position as a Postdoctoral Researcher at Ecole Polytechnique
F{\'e}d{\'e}rale de Lausanne (EPFL), Switzerland, from 2010 to
2011. From 2011 to 2012 he held a Leverhulme Trust Early Career
Fellowship at University College London (UCL), where he remains as a
Newton International Fellow, supported by the Royal Society and the
British Academy.  His research interests are focused on spherical
signal processing, including sampling theorems and wavelets on the
sphere, compressed sensing and Bayesian statistics, and applications
of these techniques to cosmology and radio interferometry.
\end{IEEEbiography}

\end{document}